\documentclass{aa}  
\usepackage{graphicx}
\usepackage{natbib}
\usepackage{amssymb}
\usepackage{amsmath}
\usepackage{amstext}
\usepackage{rotate}
\usepackage{rotating}
\usepackage{supertabular}
\usepackage{epsfig}
\usepackage{lscape}
\usepackage{enumerate}

\bibpunct{(}{)}{;}{a}{}{,}
\usepackage{txfonts}
\newcommand{\gsim}{\hbox{\rlap{\lower.55ex\hbox{$\sim$}} \kern-.3em
\raise.4ex \hbox{$>$}}}
\newcommand{\lsim}{\hbox{\rlap{\lower.55ex\hbox{$\sim$}} \kern-.3em
\raise.4ex \hbox{$<$}}}
\newcommand{\heiha}{He\,\textsc{i}$\lambda$6678/H$\alpha$}
\newcommand{\nha}{\textsc{[N\,ii]}$\lambda$6584/H$\alpha$}
\newcommand{\sha}{\textsc{[S\,ii]}$\lambda\lambda$6717,6731/H$\alpha$}

\newcommand{\ohb}{\textsc{[O\,iii]}$\lambda$5007/H$\beta$}
\newcommand{\oiii}{\textsc{[O\,iii]}$\lambda$5007}
\newcommand{\hb}{H$\beta$}
\newcommand{\ha}{H$\alpha$}
\newcommand{\nii}{\textsc{[N\,ii]}$\lambda$6584}
\newcommand{\sii}{\textsc{[S\,ii]}$\lambda\lambda$6717,6731}

\begin{document}
%
   \title{A study of the interplay between 
       ionized gas and  star clusters in the central
       region of \object{NGC~5253} with 2D spectroscopy\thanks{Based
         on observations  
       collected at the European Organisation for Astronomical
       Research in the Southern Hemisphere, Chile (ESO Programme
       078.B-0043).}}

   \author{A. Monreal-Ibero\inst{1}
          \and
           J. M. V\'{\i}lchez\inst{2}
          \and
           J. R. Walsh\inst{1}
          \and
           C. Mu\~noz-Tu\~n\'on\inst{3}
          }

   \offprints{A. Monreal-Ibero}

   \institute{European Organisation for Astronomical Research in the
              Southern Hemisphere, Karl-Schwarzschild-Strasse 2,
              D-85748 Garching bei M\"unchen, Germany\\ 
              \email{[amonreal,jwalsh]@eso.org}
          \and
           Instituto de Astrof\'{\i}sica de Andaluc\'{\i}a (CSIC), C/
              Camino Bajo de Hu\'etor, 50, 18008 Granada, Spain\\
              \email{jvm@@iaa.es}
          \and
           Instituto de Astrof\'{\i}sica de Canarias, C/
              V\'{\i}a L\'actea, s/n,  38205 La Laguna, Spain\\ 
            \email{cmt@iac.es}
             }

   \date{manuscript}

 
  \abstract
   {Starbursts are one of the main contributors to the
chemical enrichment of the interstellar medium. However,
mechanisms governing the interaction between the
recent star formation and the surrounding gas are not
fully understood. Because of their \emph{a priori} simplicity,
the subgroup of \ion{H}{ii} galaxies constitute an ideal
sample to study these mechanisms.} 
   {A detailed 2D study of the central
region of NGC~5253 has been performed to characterize the
stellar and ionized gas structure as well as the extinction
distribution, physical properties and kinematics of the ionized gas in
the central $\sim$210~pc$\times$130~pc.}  
   {We utilized optical integral field spectroscopy (IFS) data obtained
     with FLAMES.}
{A detailed extinction map for the ionized gas in \object{NGC~5253} 
shows that the largest extinction is associated with the prominent Giant 
\ion{H}{ii} region. There is an offset of $\sim$0\farcs5 between the
peak of the optical continuum and the extinction peak in agreement
with findings in the infrared. We found that stars suffer
less extinction than gas by a factor of $\sim$0.33.  
  The \textsc{[S\,ii]}$\lambda$6717/\textsc{[S\,ii]}$\lambda$6731 map
  shows an electron density ($N_e$) gradient declining from the
  peak of emission in \ha\ (790~cm$^{-3}$) outwards, while the argon line
  ratio traces areas
  with $N_e\sim 4200 - 6200$~cm$^{-3}$. 
  The area polluted with extra nitrogen, as deduced from the excess 
  \nha, extends up to distances of 3\farcs3 ($\sim$60~pc) 
  from the maximum pollution, which is offset by 
  $\sim$1\farcs5 from the peak of continuum emission.
  Wolf-Rayet features are
  distributed in an irregular pattern over a larger area 
  ($\sim100$~pc$\times100$~pc) and associated with young stellar
  clusters.
  We measured He$^+$ abundances over most of the field of view
  and values of He$^{++}$/H$^{+}\lsim0.0005$ in localized areas which 
  do not coincide, in general, with the areas presenting W-R emission
  or extra nitrogen. 
%
The line profiles are complex. Up to three emission components were
needed to reproduce them. One of them, associated with the giant
\ion{H}{ii} region, 
presents supersonic widths and  \nii\ and \sii\ emission lines
shifted up to $40$~km~s$^{-1}$ with respect to \ha. Similarly, one of
the narrow components presents offsets in 
the \nii\ line of  $\lsim20$~km~s$^{-1}$. This is the first time that
maps with such velocity offsets for a starburst galaxy have
been presented.  
The observables in the giant \ion{H}{ii} region fit with a scenario where
the two super stellar clusters (SSCs) produce an outflow that
encounters the previously quiescent gas.  
The south-west part of the FLAMES IFU field is consistent with
a more evolved stage where the  star clusters have already cleared out
their local environment.  
}  
 {}


   \titlerunning{The central region of NGC~5253 with FLAMES}
   \maketitle
%

\section{Introduction}

Starbursts are events characterized by star-formation rates much
higher than those found in 
gas-rich normal galaxies. They are considered one of the main
contributors to the chemical enrichment of the interstellar medium
(ISM) and can be found in galaxies covering a wide range of masses,
luminosities, metallicities and interaction stages such as blue
compact dwarfs, nuclei of spiral galaxies, or (Ultra)luminous Infrared
Galaxies \citep[see][and references therein]{con08}. 

A particularly interesting subset are the \ion{H}{ii}
galaxies, identified for the first time by \citet{har56}: gas-rich,
metal poor ($1/40\, Z_\odot\, \lsim\, Z\, \lsim\, 1/3\, 
Z_\odot$) dwarf systems characterized by the presence of
large ionized \ion{H}{ii} regions that dominate their optical spectra
\citep[see][for a review of these galaxies]{kun00}.
These systems are \emph{a priori} simple, which makes them the 
ideal laboratories to test the interplay between massive star formation and
the ISM. 

NGC~5253, an irregular galaxy located in the Centarus A /
M~83 galaxy complex \citep{kar07}, is a local example of an \ion{H}{ii}
galaxy. This galaxy is suffering a burst of star
formation which is believed to have been triggered by an encounter
with M~83  \citep{van80}. This is supported by the existence of the
\ion{H}{i} plume extending along the optical minor axis which is
best explained as tidal debris \citep{kob08}.

NGC~5253 constitutes an optimal target for the study of the
starburst phenomenon. On the one hand, its proximity
allows a linear spatial resolution to be achieved that is good enough 
to study the details of the interplay between the different components
(i.e. gas, 
dust and star clusters) in the central region. On the other hand, this system
has been observed in practically all spectral ranges from the X-ray to
the radio, and therefore a large amount of ancillary information is available. 

The basic characteristics of this galaxy are compiled in Table~
\ref{tabbasicdata}. Its stellar content has been widely studied and
more than 300 stellar clusters have been detected \citep{cre05}.
Multi-band photometry with the WFPC2 has revealed that
those in its central region present typical masses of $\sim
2-120\times10^{3}$~M$_\odot$ and are very young, with ages of $\sim1-12$~Myr
\citep[e.g.][]{har04}. In particular, HST-NICMOS images have revealed that the
nucleus of the galaxy is made out of two very massive
($\sim1-2\times10^6$~M$_\odot$) super stellar clusters (SSCs), with
ages of about $\sim3.5$~Myr, separated by $\sim0\farcs4$ \citep{alo04}, and
which are coincident with the double radio nebula detected at 1.3~cm
\citep{tur00}. Also, detection of spectral features characteristic of
Wolf-Rayet (W-R) stars in specific regions of the galaxy have
been reported \citep[e.g.][]{sch97}. Recently, seven supernova
remnants have been detected in the central region of this galaxy by
means of the [\ion{Fe}{ii}]$\lambda$1.644$\mu$m emission \citep{lab06}. 

NGC~5253 presents a filamentary structure in H$\alpha$ \citep[e.g.][]{mar98}
associated with extended diffuse emission in  X-ray which can be
explained as multiple superbubbles around its OBs 
associations and SSCs that are the results of the combined action of
stellar winds and supernovae \citep{str99,sum04}. 


\begin{table}
     \centering
     \caption[]{Basic data for NGC~5253 \label{basicdata}}
             \begin{tabular}{ccccccccc}
\hline
\hline
            \noalign{\smallskip}
Parameter & Value & Ref.\\
           \noalign{\smallskip}
           \hline
           \noalign{\smallskip}
Name               & NGC~5253                 & (a)\\
Other designations & ESO 445$-$ G004, Haro~10 & (a)\\
RA (J2000.0)       & 13h39m55.9s              & (a)\\
Dec(J2000.0)       & $-$31d38m24s             & (a)\\
$z$                & 0.001358                 & (a)\\
$D(Mpc)$           & 3.8                      & (b)\\
scale (pc/$^{\prime\prime}$) &  18.4          &    \\
$m_B$              & 10.78                    & (c)\\ 
$M_B$              & $-17.13$                 & (c)\\ 
$U-B$              & $-0.30$                  & (c)\\
$B-V$              & $0.50$                   & (c)\\
$V-R$              & $0.32$                   & (c)\\
$M_{HI} (M_\odot)$ & $1.4\times10^8$~M$_\odot$ & (d)\\
$Z/Z_\odot$        & $\sim$0.3$^{(\ast)}$ &  (e)\\
$\log(L_{fir}/L_\odot)$ & 8.95 & (f) \\
$\log(L_{ir}/L_\odot)$  & 9.21 & (f) \\
            \noalign{\smallskip}  
            \hline
         \end{tabular}
\begin{list}{}{}
\item[$^{(\ast)}$] We assumed $12+\log(\mathrm{O/H})_\odot = 8.66$ \citep{asp04}.
\item[$^{\mathrm{(a)}}$] NASA/IPAC Extragalactic Database (NED).
\item[$^{\mathrm{(b)}}$] \citet{sak04}.
\item[$^{\mathrm{(c)}}$] \citet{tay05}.
\item[$^{\mathrm{(d)}}$] \citet{kob08}.
\item[$^{\mathrm{(e)}}$] \citet{kob99}.
\item[$^{\mathrm{(f)}}$] \citet{san03}. Re-scaled to the distance
  adopted here.
\end{list} \label{tabbasicdata}
\end{table}

The measured metallicity of this galaxy is relatively low (see Table
\ref{tabbasicdata}) and presents a generally uniform distribution.
However, an increase in the abundance of nitrogen
 in the central region of $\sim2-3$ times the mean has been reported
 \citep{wal89,kob97,lop07}.  
No other elemental species appears to present spatial abundance fluctuations. The
reason for this nitrogen enhancement has not been fully clarified yet
although a connection with the W-R population has been suggested.

On account of their irregular structure, a proper characterization of the
physical properties of \ion{H}{ii} galaxies, necessary to explore the 
interplay of mechanisms acting between gas and
stars, requires high quality two-dimensional spectral information able to
produce a continuous mapping of the relevant quantities.
Such observations have traditionally been done in the optical and 
near-infrared by  mapping
the galaxy under study with a long-slit \citep[e.g.][]{vil98,wal89}. This
is, however, expensive in terms of 
telescope time and might be affected by some technical problems such as
misalignment of the slit or changes in the observing conditions with
time. The advent and popularization of integral 
field spectroscopy (IFS) facilities, able to record simultaneously the
spectra of an extended continuous field, overcomes these
difficulties. Nevertheless, work based on this technique 
devoted to the study of \ion{H}{ii} galaxies is still relatively rare
\citep[e.g.][]{lag09,bor09,jam09,keh08,gar08,izo06}. 

Here, we present IFS observations of the central area of NGC~5253
in order to study the mechanisms that govern the
interaction between the young stars and the surrounding ionized gas.  
The paper is organized as follows: section \ref{obsred} contains the
observational and technical details regarding the data reduction and
derivation of the required observables; section \ref{results} describes
the stellar and ionized gas structure as well as the extinction
distribution and the physical and kinematic properties of the ionized gas;
section \ref{discusion} discusses the evolutionary stage of the gas
surrounding the stellar clusters, focusing on the two most relevant
areas of the field of view (f.o.v.). Section \ref{summary} itemises
our results and conclusions.



\section{Observations, data reduction and line fitting \label{obsred}}

\subsection{Observations}

\begin{figure}[h!]
   \centering
\includegraphics[angle=0,scale=.80, clip=,bbllx=155, bblly=395,
bburx=450, bbury=685]{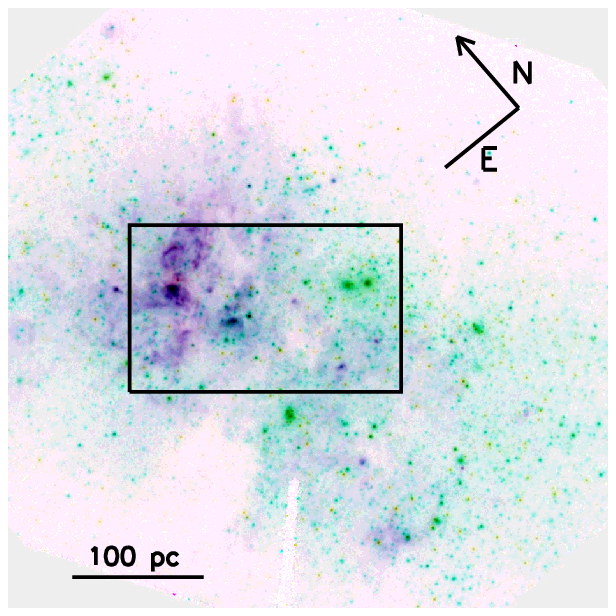}
   \caption[Covered field with FLAMES]{False colour image in filters
     F435W ($B$, blue), F658N ($H\alpha$, magenta), and
     F814W ($I$, green) for NGC~5253 using
     HST-ACS images (programme 10608, P.I.:Vacca) with the area covered
     by our FLAMES data over-plotted as a white rectangle. The orientation and
     scale for a distance of 3.8~Mpc are indicated. \label{apuntado}} 
 \end{figure}

Data were obtained with the \emph{Fibre Large Array Multi Element
  Spectrograph}, FLAMES \citep{pas02} at Kueyen, Telescope Unit 2
  of the 8~m VLT at ESO's observatory on Paranal, on February 10,
  2007. The central region of the galaxy was observed with the ARGUS
  Integral Field Unit (IFU) which has a field of view of $11\farcs5
  \times 7\farcs3$ with a sampling of
  0.52$^{\prime\prime}$/lens. In addition, ARGUS
    has 15 fibers that can simultaneously observe the sky and which
    were arranged forming a circle around the IFU. The precise covered area is
  shown in Figure   
  \ref{apuntado} which contains the FLAMES field of view over-plotted on
  an HST B, \ha, I colour image.

We utilized two different gratings in order to obtain information for the most
important emission lines in the optical spectral range. Data were taken
under photometric conditions and seeing ranged typically between
$0\farcs8$ and $1\farcs0$.   The covered spectral
range, resolving power, exposure time and airmass for each
configuration are shown in Table \ref{log_observaciones}. In addition
to the science frames, continuum and ThAr arc lamps exposures as well as
frames for the spectrophotometric standard star CD-32\,9927 were
obtained. 

\begin{table}
\centering
      \caption[]{Observation log \label{log_observaciones}}
              \begin{tabular}{ccccccccc}
            \hline
            \noalign{\smallskip}
Grating & Spectral range & Resolution & t$_{\mathrm{exp}}$ & Airmass\\
        &  (\AA)         &              & (s)        &      \\
            \noalign{\smallskip}
            \hline
            \noalign{\smallskip}
L682.2    & 6\,438--7\,184 & 13\,700 & $5\times1\,500$ &  1.75--1.13\\
L479.7    & 4\,501--5\,078 & 12\,000 & $5\times1\,500$ &  1.11--1.01\\
            \hline
         \end{tabular}

\end{table}

\subsection{Data reduction}

The basic reduction steps for the FLAMES data were performed with a
combination of the pipeline provided by ESO (version
1.0)\footnote{http://www.eso.org/projects/dfs/dfs-shared/web/vlt/vlt-instrument-pipelines.html.}
via \texttt{esorex}, version 2.0.2 and 
some IRAF\footnote{The Image Reduction and Analysis Facility \emph{IRAF} is
  distributed by the National Optical Astronomy Observatories which is
  operated by the association of Universities for Research in
  Astronomy, Inc. under cooperative agreement with the National
  Science Foundation.} routines. First of all we masked a bad column
in the raw data using the task \texttt{fixpix} within IRAF. Then, 
each individual frame was processed using the ESO pipeline in
order to perform bias subtraction, spectral tracing and
extraction, wavelength calibration and correction of fibre transmission.  

Uncertainties in the relative wavelength calibration were estimated by fitting 
a Gaussian to three isolated lines in every spectrum of the arc
exposure. The standard deviation of the central wavelength for a
certain line gives an idea of the associated error in that spectral
range. We were able to determine the centroid of the lines with an 
uncertainty of $\sim$0.005\AA, which translates into velocities of
$\sim$0.3~km~s$^{-1}$. The spectral resolution was very uniform over
the whole field-of-view with values of $0.178\pm0.004$ \AA{} and
$0.241\pm0.009$ \AA{}, FWHM for the blue and red configuration
respectively, which translates into $\sigma_{instru} \sim$
4.7~km~s$^{-1}$.

For the sky subtraction, we created a good signal-to-noise (S/N) spectrum
by averaging the spectra of the sky fibres in each individual frame.
This sky spectrum was subsequently subtracted from every spectrum. 
In several of the
sky fibres, the strongest emission lines, namely \oiii{} in the blue
frames and \ha{} in the red frames, could clearly be detected. We
attributed this effect to some cross-talk from the adjacent fibers. A direct
comparison of the flux in the sky and adjacent fibers showed that this
contribution was always \lsim0.6\% which is negligible
in terms of sky subtraction. However, in order to reduce this 
contamination to a minimum, we decided not to use these fibres in 
the creation of the high S/N sky spectra. 

Regarding the flux calibration, a spectrum for the calibration star
was created by co-adding all the fibers of the standard star
frames. Then, a sensitivity function was determined with the IRAF tasks
\texttt{standard} and \texttt{sensfunc} and science frames were 
calibrated with \texttt{calibrate}. Afterwards, frames corresponding 
to each configuration were combined
and cosmic rays rejected with the task \texttt{imcombine}.
As a last step, the data were reformatted into two easier-to-use 
data cubes, with two spatial and one spectral dimension, using the 
known position of the lenses within the array. 

\subsection{Line fitting and map creation \label{linefitting}}

In order to obtain the relevant emission line information, line
profiles were fitted using Gaussian functions. This procedure was done
in a semi-automatic way using the IDL based routine MPFITEXPR \footnote{See
http://purl.com/net/mpfit.} \citep{mar09}
which offers ample flexibility in case constraints on the parameters 
of the fit are included, such as lines in fixed ratio. 
The procedure was as follows.

As a first step, we fit all the lines by a single Gaussian. The
H$\alpha$+[N\,\textsc{ii}] complex was fitted simultaneously by one 
Gaussian per emission line plus a flat continuum first-degree
polynomial using a common 
  width for the three lines and fixing the separation in wavelength
  between the lines according to the redshift provided at
NED\footnote{See http://nedwww.ipac.caltech.edu/.} and the nitrogen
line ratio ($\lambda$6583/$\lambda$6548) to 3. 
The same procedure was repeated for the
[S\,\textsc{ii}]$\lambda\lambda$6717,6730 doublet, the
[Ar\,\textsc{iv}]$\lambda$4711 line (which was fitted jointly with the
[Fe\,\textsc{iii}]$\lambda$4701 and He\,\textsc{i}$\lambda$4713) and the
[Ar\,\textsc{iv}]$\lambda$4740 line (which was fitted jointly with the
[Fe\,\textsc{iii}]$\lambda$4734  line), but this time without any
restriction on the line ratios. Finally, \hb, \oiii, and
\ion{He}{i}$\lambda$6678  were individually fitted.

This single Gaussian fit gave a good measurement for the line fluxes and
the results of these fits are the used in all the forthcoming
analysis, with the exception of the kinematics. This latter analysis 
requires a more complex line fitting scheme, since several
lines showed signs of asymmetries and/or multiple 
components in their profiles for a large number of spaxels. In those
cases, multi-component fits were performed. Over the 
whole field of view we compared the measured flux from performing 
the fit with a single Gaussian to the fit by several components in 
the brightest emission lines (namely: \hb,
\oiii, \ha, \nii\ and \sii). Differences between the two sets of line
fluxes ranged
typically from 0\% to 15\%, depending on the spaxel and the emission
line, and translated into differences in the line ratios \lsim0.06~dex.  

In all the cases, MPFITEXPR estimated an error for the fit using the
standard deviation of the adjacent continuum. 
Those fits with a ratio between line flux and error less than 
three were automatically rejected. The remaining spectra were visually
inspected and classified as good or bad fits.

Finally, for each of the observables, we used the derived quantity
together with the position within the data-cube for each spaxel to
create an image suitable to be manipulated with standard astronomical
software. Hereafter, we will use both terms, \emph{map} and
\emph{image}, when referring to these.

\section{Results \label{results}}

\subsection{Stellar and ionized gas structure \label{morfologia}}

\begin{table}
\centering
      \caption[]{Main reference clusters.}
       \label{cumulos}
              \begin{tabular}{ccccccccc}
            \hline
            \noalign{\smallskip}
Name & FLAMES                     & \multicolumn{4}{c}{Other names}\\
     &  coord.($^{\prime\prime}$,$^{\prime\prime}$)  &
            C97$^{\mathrm{(a)}}$ &
            H04$^{\mathrm{(b)}}$ &
            K97$^{\mathrm{(c)}}$  & 
            AH04$^{\mathrm{(d)}}$\\
            \noalign{\smallskip}
            \hline
            \noalign{\smallskip}
\#1  & $(3.6,0.5)$  & N5253-5 & 1          & UV3 & C1+C2 \\
\#2  & $(1.0,-1.0)$ & N5253-4 & 4,8,24,25  & UV1 & $-$   \\
\#3  & $(-4.2,1.0)$ & N5253-3 & 3,5        & $-$ & C4+C5 \\
            \hline
         \end{tabular}
\begin{list}{}{}
\item[$^{\mathrm{(a)}}$] \citet{cal97}.
\item[$^{\mathrm{(b)}}$] \citet{har04}.
\item[$^{\mathrm{(c)}}$] \citet{kob97}.
\item[$^{\mathrm{(d)}}$] \citet{alo04}.
\end{list}
\end{table}

\begin{figure}[h!]
   \centering
\includegraphics[angle=0,width=0.48\textwidth,clip=,bbllx=28, bblly=18,
bburx=543, bbury=360]{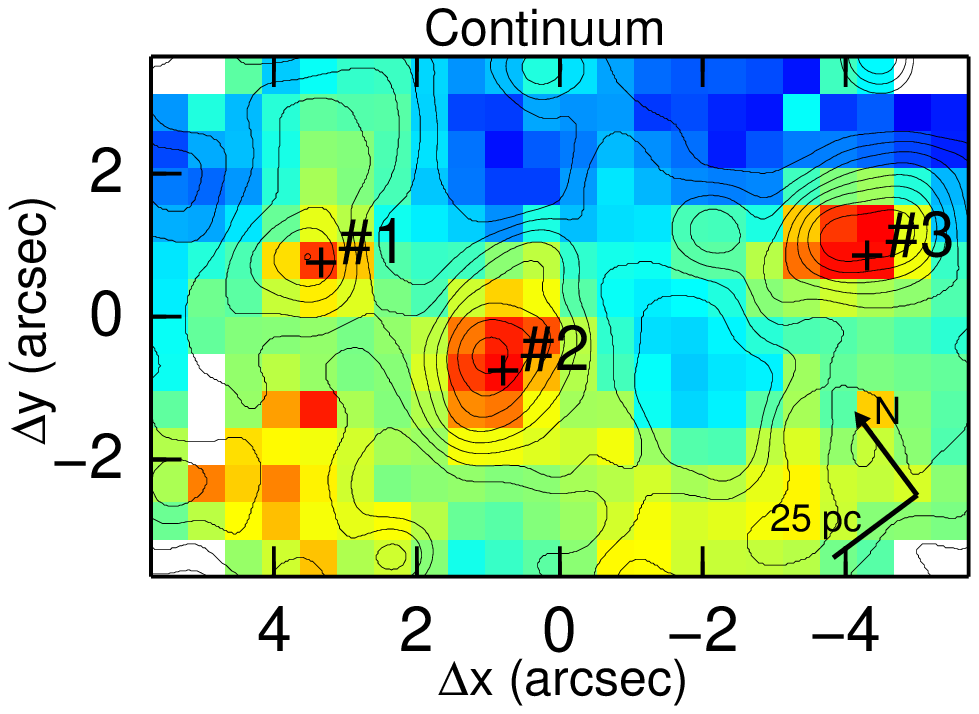}
\includegraphics[angle=0,width=0.48\textwidth,clip=,bbllx=28, bblly=18,
bburx=543, bbury=360]{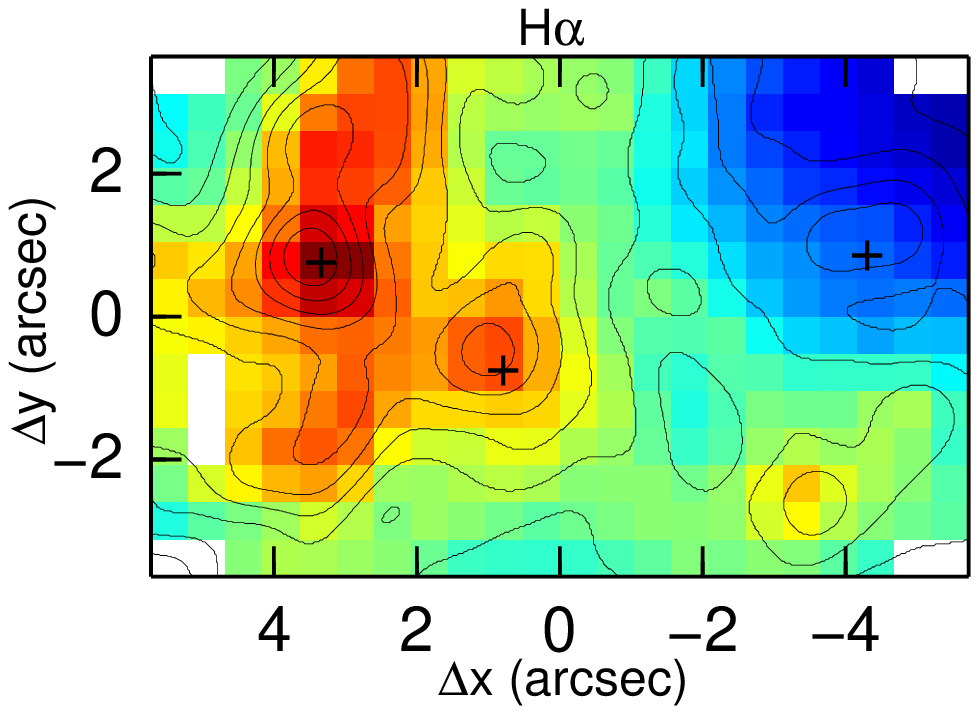}
   \caption[Covered field with FLAMES]{\emph{Top:} Stellar component
     distribution as traced by a 
     continuum map made from the average flux in the spectral range $6525 -
     6545$~\AA{} and  $6600 - 6620$~\AA. \emph{Bottom:} Ionized gas
     distribution as traced by the H$\alpha$ emission line.
     We have over-plotted contours corresponding to
     the HST-ACS images in the F814W (\emph{top}) and F658N
     (\emph{bottom}) filters (programme
     10609, P.I.: Vacca) convolved with a Gaussian of 0\farcs8 to
     simulate the seeing at Paranal. The position of the three main
     peaks of continuum emission are marked with crosses.
     Both maps are presented in logarithmic scale in order to emphasize 
     the relevant morphological features and cover
     a range of 2.0~dex for the \ha\ and 0.9~dex for the continuum
     map. Flux units are arbitrary.
  \label{estructura}}
 \end{figure}

Figure \ref{estructura} displays the stellar structure, as traced by a
continuum close to H$\alpha$, as well as the one for the ionized gas
(traced by the H$\alpha$ emission line). The over-plotted contours,
which represent the HST-ACS images in the F659N and F814W bands
convolved with a Gaussian to match the seeing at Paranal, show good
correspondence between the images created from the IFS data and the HST
images  (although obviously with poorer resolution for the ground-based
FLAMES data). A direct comparison of these maps shows how the stellar
and ionized gas structure differs.

The continuum image displays three main peaks of emission which will
be used through the paper as reference. We have
associated each of these peaks with one or more star clusters by
direct comparison with ACS images. Table \ref{cumulos} compiles
their positions within the FLAMES field of view together with the
names of the corresponding clusters according to several reference works. 
These clusters trace a sequence in age as we move towards the right 
(south-west) in the
FLAMES field of view. The clusters associated with peak \#1 are
very young \citep[$\sim3-8$~Myr, ][]{har04,alo04}, those associated with
peak \#2 display a 
range of ages from very young to intermediate age
\citep[$\sim6-170$~Myr, ][]{har04}, while the stars in the pair of
clusters associated with peak \#3 seem to have intermediate ages
\citep[70-113~Myr, ][]{har04}. 




The H$\alpha$ emission line reproduces the structure described by
\citet{cal97} using an HST WFPC2 image. Briefly, the central region of
NGC5253 is divided into two parts by a dust lane that crosses the galaxy
along the east-west direction (from
$\sim$[2\farcs0,-3\farcs0]\footnote{Hereafter, the 
different quoted positions will be refereed as
  [$^{\prime\prime}$,$^{\prime\prime}$] and using the FLAMES
  f.o.v. as reference.} to the Complex \#3). 
Most of the H$\alpha$ emission is located towards the north
of this lane where there is giant H\,\textsc{ii} region associated with the
Complex \#1. This region shows two tongue-shaped extensions towards the
upper and lower part of the FLAMES field of view (P.A. on the sky of
$\sim45^{\circ}$ and $\sim-135^{\circ}$, respectively) as well as a
extension  at P.A.$\sim155^{\circ}$ which contains the Complex \#2. 
Towards the south of the dust lane, the emission is dominated
by a peak at $\sim$[-3\farcs5,-2\farcs5] which could be associated with
cluster 17 in \citet{har04}.   



\subsection{Extinction structure \label{secextincion}}

Extinction  was derived assuming an intrinsic Balmer emission
line ratio of H$\alpha$/H$\beta$ = 2.87 \citep[][for an $T_e =
  10\,000$~K]{ost06} and using the extinction 
curve of \citet{flu94}. Since the H$\alpha$ and H$\beta$ emission lines
are separated by 
$\sim$1700~\AA{} and both of them had high signal-to-noise ratio in
only a single exposure, we decided
to obtain the extinction maps from the LR3 and LR6 exposures observed at 
the smallest airmass (1.009 and 1.160, respectively), thus minimizing any effect
due to differential atmospheric refraction.

   \begin{figure}[h!]
   \centering
\includegraphics[angle=0,scale=.43, clip=,bbllx=40, bblly=0,
bburx=600, bbury=355]{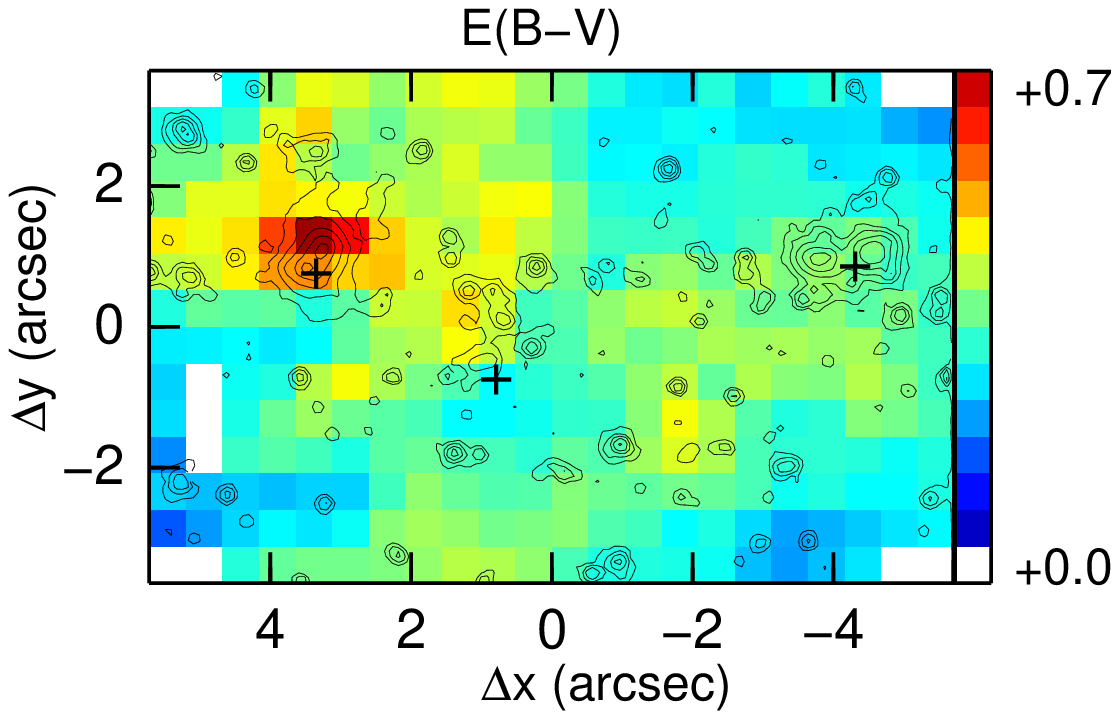}
   \caption[Reddening map.]{Reddening map obtained from the hydrogen
   recombination lines assuming H$\alpha$/H$\beta$ = 2.87 and the
   extinction law of \citet{flu94} and $E(B-V) = 
   A_V / 3.1$ \citep{rie85}. The position of the three main 
     peaks of continuum emission are shown for reference. We have
   over-plotted the F160W band NICMOS-HST image from \citet{alo02}
   with contours for comparison. 
  \label{mapa_ebv}}
 \end{figure}

We have not included any correction for an underlying
stellar population. We inspected carefully each individual spectrum
to look for the presence of the stellar absorption feature in
H$\beta$. Only in those spaxels associated with the area around the 
Complex \#3 ($\sim$9-10~spaxels or $\sim$1\farcs5$\times$1\farcs5 in
total) was such an absorption 
detected. We estimated the influence of this component by
  fitting both the absorption and the emission component in the most
  affected spaxel. Equivalent widths were $\sim$3 and  $\sim$9~\AA,
  respectively. In these spaxels,    
the absorption line was typically $\sim15-20$~times wider and with
about half - one third of the flux of the emission line.
This implies an underestimation of H$\beta$
emission line flux of about 10\%. For the particular area around
Complex \#3, this translates
into a real  extinction of $A_V \sim 0.9$~mag instead of the measured
$A_V \sim1.2$~mag. 


The corresponding reddening map was determined
assuming  $E(B-V) = A_V / 3.1$ \citep{rie85} and is presented in
Figure \ref{mapa_ebv}. This map shows that the extinction
is distributed in a non-uniform manner ranging from $E(B-V) = 0.16$ to
$0.64$ (mean 0.33, standard deviation, 0.07). Given that Galactic reddening for
NGC~5253 is 0.056 \citep{sch98}, nearly all of the extinction can be
considered intrinsic to the galaxy.

In general terms, the structure presented in this map
coincides with the one presented by 
\citet{cal97}. The dust lane mentioned in the previous section is
clearly visible here and it causes extinction of $A_V \sim
0.10-1.13$~mag. However, the larger measured extinction values are
associated with the giant \ion{H}{ii} region, in agreement with the
\ion{H}{i} distribution \citep{kob08}. Dust in this area forms an
S-shaped distribution with $A_V \sim 0.13-1.15$~mag in the arms.

\begin{figure}[h!]
 \centering
\includegraphics[angle=0,width=0.4\textwidth, clip=,bbllx=40, bblly=15,
bburx=685, bbury=410]{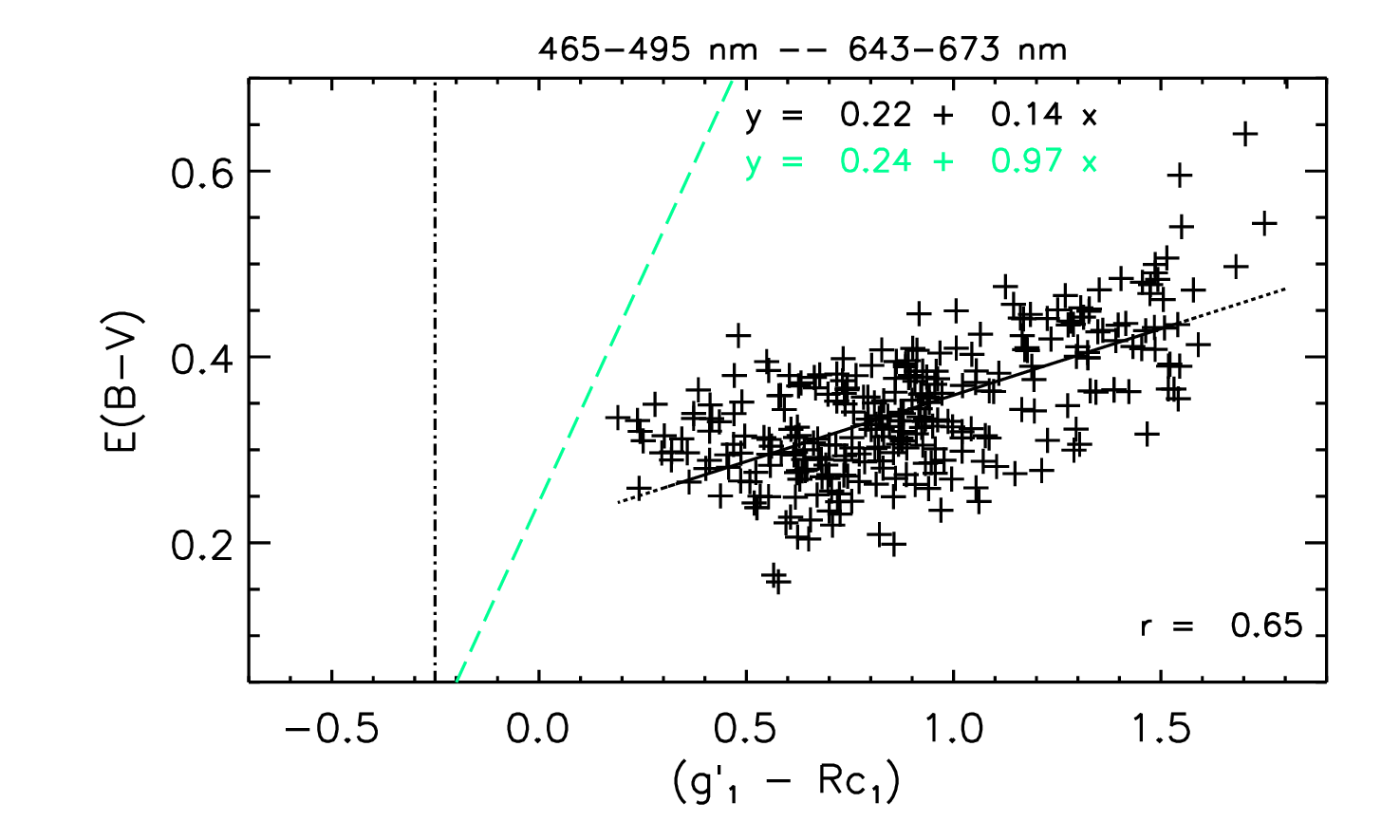}
\includegraphics[angle=0,width=0.4\textwidth, clip=,bbllx=40, bblly=15,
bburx=685, bbury=410]{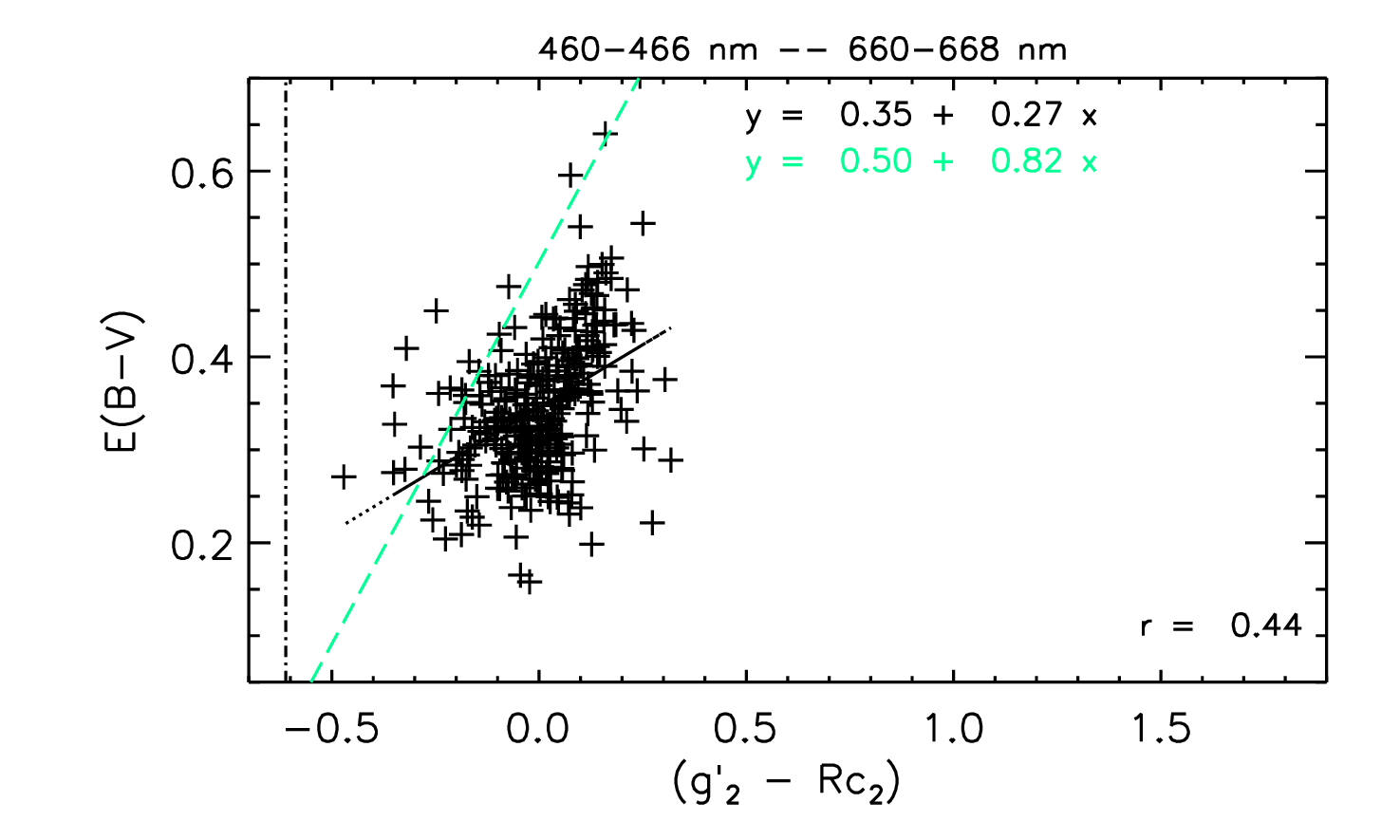}
   \caption[]{Relation between the derived reddening and a colour
   similar to $g^{\prime} - R_c$. The vertical line represent the
   $g^{\prime} - R_c$ colour expected for a Im type galaxy
   without any extinction while the green dashed line
     corresponds to the expected 
     relation if gas and stars were suffering the same amount of
     extinction (see text for details). The 1-degree polynomial fit to
     all the data appear as a continuous line. The corresponding fit is
   shown in the upper left corner while the Pearson's coefficient of
   correlation is indicated in the lower right corner.
   The simulated $g^{\prime} - R_c$ colours have been done by 
   integrating the flux in the spectral ranges of $4\,653 - 4\,953$~\AA\
   ($g^{\prime}_1$) and $6\,431 - 6\,731$~\AA\ ($R_{c1}$)   for the upper
   panel and  $4\,600 - 4\,660$~\AA\ ($g^{\prime}_2$) and $6\,600 - 6\,680$~\AA\
   ($R_{c2}$) for the lower one. 
   
  \label{colorvsebv}}
 \end{figure}

   \begin{figure}[h!]
   \centering
\includegraphics[angle=0,scale=.43, clip=,bbllx=40, bblly=0,
bburx=600, bbury=355]{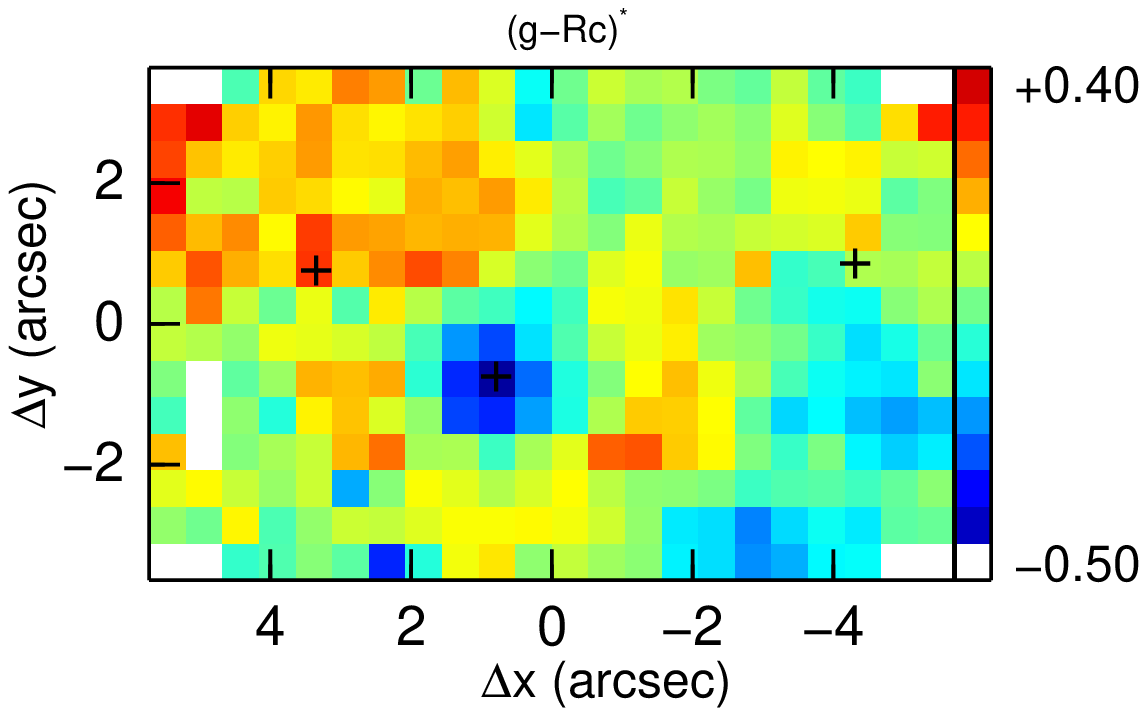}
   \caption[Reddening map.]{Line free colour map. The simulated filters
   have been defined as explained in Figure
   \ref{colorvsebv}.  The position of the three main 
     peaks of continuum emission are shown for reference. 
  \label{mapa_color}}
 \end{figure}

In order to explore the relation between the extinction
  suffered by the gas and by the stellar populations,
  our $E(B-V)$ measurements were compared with colours defined
  \emph{ad hoc}. For the covered spectral 
  range, it is not possible to exactly simulate any of the existing
  standard filters. It is possible, however, to create filters
  relatively similar to the $g^{\prime}$ and $R_c$ ones. We have
  simulated two set of filters.

In the first case, the flux was integrated over two large
wavelength ranges (465 -- 495~nm and 643 -- 673~nm) in order
to simulate broad filters. The relation between the reddening
derived for the ionized gas and the derived colour, hereafter
$(g^{\prime} _{1} - R_{c1})$, is shown in the upper panel of Figure
\ref{colorvsebv}, 
as would be observed with photometry. The first order polynomial fit
to the data and the Pearson correlation coefficient are included
on the plot. Also shown is the expected relation for an Im
galaxy with foreground reddening. This latter was derived for the
average of two Im templates (NGC 4449 and NGC 4485) from
\citet{ken92} applying a foreground screen of dust with
a standard Galactic reddening law \citep{car89} with R=3.1.
There is a strong difference between the expected
relation for a foreground  screen of dust and the measured values.
 On the one hand, colours are much redder. On the other hand,
  the slope of the  1-degree polynomial fit is much less steep than the
  expected one. Also, there is a very good correlation between the
  $E(B-V)$ and our synthetic  $(g^{\prime} - R_{c1})$ colour. All this
  can be attributed to the contamination of the gas emission lines,
  mainly H$\alpha$ and H$\beta$, in our filters.

In the second set, we restricted the spectral ranges for the simulated
  filters to a narrower wavelength range which was free from the
  contamination of the main emission lines. The map for this line-free
  colour is displayed in Figure \ref{mapa_color}. The structure
  resembles the one presented in Figure \ref{mapa_ebv} (i.e. dust
  lane, redder colours associated with the giant \ion{H}{ii} region),
  although there are differences, that can be attributed to differences
  in the properties of the stellar populations in the different clusters.
  The relation between the reddening and the 
  corresponding $(g^{\prime} _{2}- R_{c2})$ is shown in the lower panel
  of Figure \ref{colorvsebv}. This time colours are more similar to
  what is expected for a given stellar population suffering a certain
  amount of extinction.
However, for a given colour, stars do not reach the expected
  reddening if gas and stars were suffering the same extinction
  (i.e. data points are \emph{below} the green line).    
  The ratio between the slopes indicates that extinction in the stars is
  a factor 0.33 lower than the one for the ionized gas. 
  This is similar to what \citet{cal97} found using HST images who
  estimated that the extinction suffered by the stars is a factor 0.5 lower
  than for the ionized gas and can be explained if the dust
  has a larger covering factor for the ionized gas than for the stars
  \citep{cal94}.  


In general, our $E(B-V)$ measurements agree with previous
  ones using the same emission lines in specific areas
  \citep[e.g.][]{gon87,lop07} or with poorer spatial resolution
  \citep{wal89}. However there are discrepancies when comparing with
  the estimation of the extinction at other wavelengths. In particular,  
the peak of extinction ($A_V = 2.1$~mag, according to the Balmer line
ratio) is offset by $\sim0\farcs5$ from the peak of 
continuum emission. \citet{alo04} showed how in the central
area of NGC~5253 there are two massive star clusters, C1 and C2. While
C1 is the dominant source in the optical, coincident with our peak in
the continuum map, the more massive and extinguished C2 is the dominant
source in the infrared. The contours for the NICMOS $F160W$ image
in Figure \ref{mapa_ebv} show the good correspondence between our 
maximum of extinction and C2. Measurements in the near and 
mid-infrared suggest extinctions of $A_V \sim 17$~mag for this cluster
\citep{tur03,alo04,mar05}. The discrepancy between these two values
indicates that a foreground screen model is not the appropriate one to
explain the distribution of the dust in the giant \ion{H}{ii} region.


\subsection{Electron density distribution \label{secdensidad}}

   \begin{figure}
   \centering
\includegraphics[angle=0,scale=.43, clip=,bbllx=40, bblly=0,
bburx=600, bbury=355]{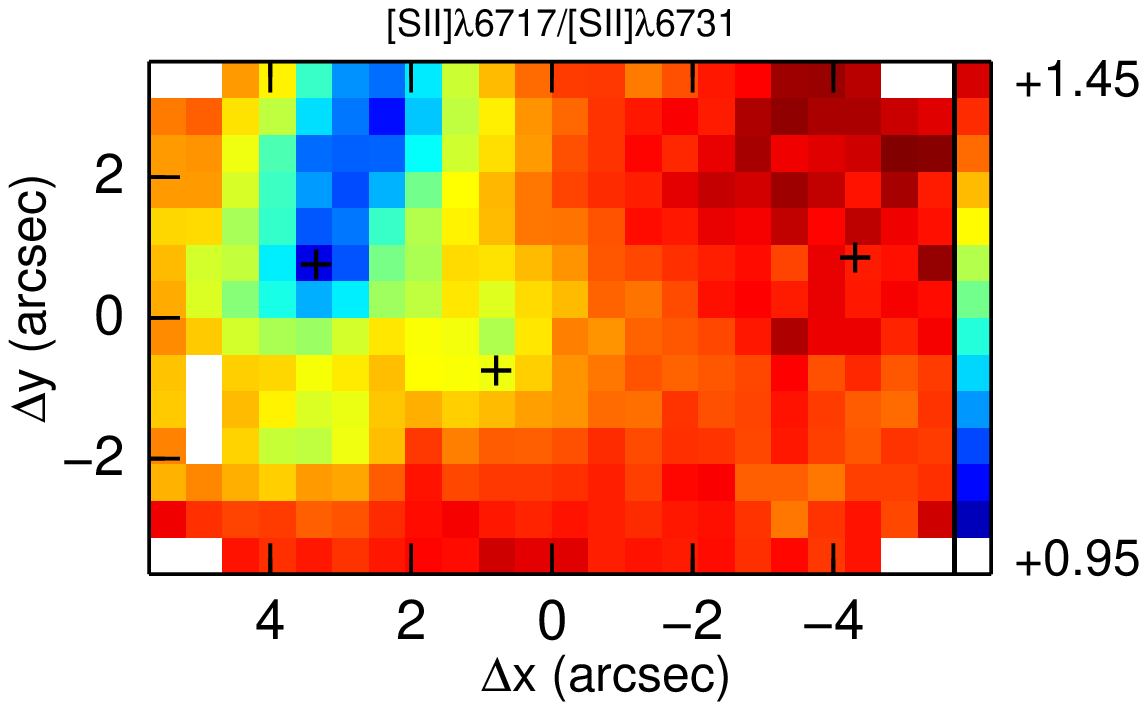}
\includegraphics[angle=0,scale=.43, clip=,bbllx=40, bblly=0,
bburx=600, bbury=355]{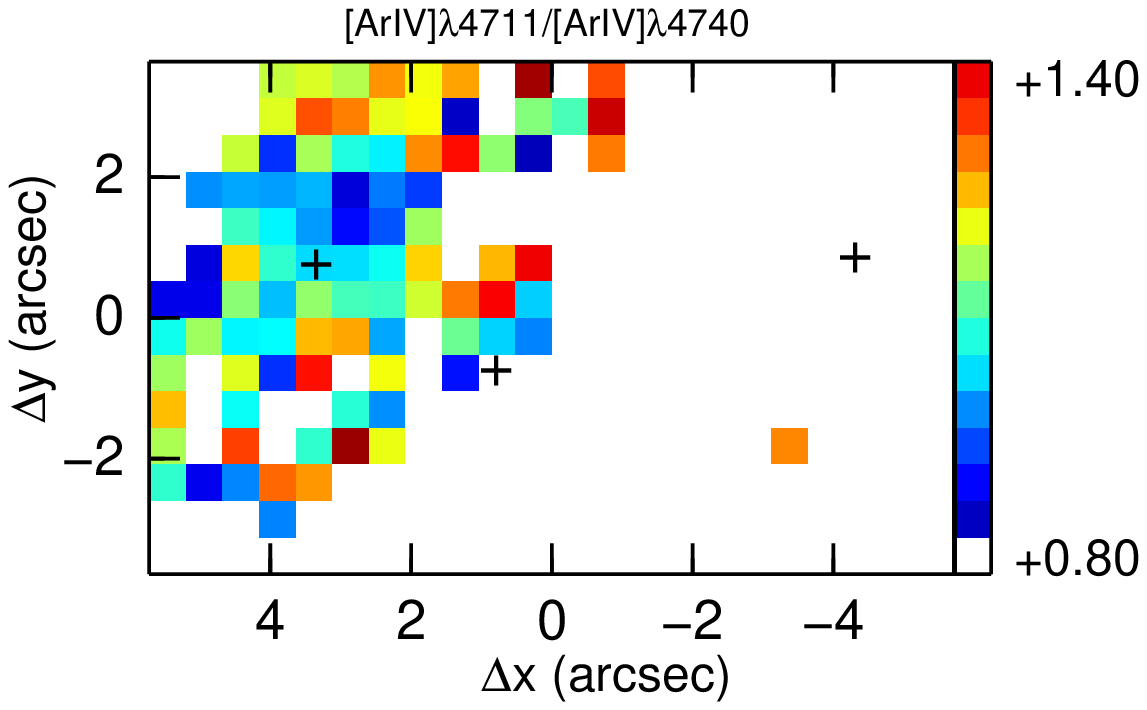}
   \caption[Line ratio maps for the electron density diagnostic.]
  {Maps for the line ratios sensitive to the electron
  density. The position of the three main 
     peaks of continuum emission are shown as crosses
     for reference. The displayed ranges in the line ratios imply
       electron densities of $<$10-790~cm$^{-3}$ and
       190-8750~cm$^{-3}$ for the sulphur and argon line ratio, respectively.
\label{mapas2as2r}} 
 \end{figure}

Electron density ($N_e$) can be determined from the
\textsc{[S\,ii]}$\lambda$6717/\textsc{[S\,ii]}$\lambda$6731 and
[Ar\textsc{\,iv}]$\lambda$4711/[Ar\textsc{\,iv}]$\lambda$4740
line ratios; values of 1.25 and
1.00 respectively were measured in a spectrum made from the
  sum of all our spaxels. Hereafter, we will refer to this spectrum as the
\emph{integrated spectrum}. 
Electron densities were determined assuming an electron
temperature of 11\,650 K, the average of the values given in
\citet{lop07}, and using the task
\texttt{temden}, based on the \texttt{fivel} program \citep{sha95}
included in the IRAF package \texttt{nebular}. Derived values of $N_e$
for the two line ratios were
180~cm~$^{-3}$  and 4520~cm~$^{-3}$, respectively.
Differences between the electron densities derived from
  the argon and sulphur lines are usually found in ionized gaseous
  nebulae \citep[see][]{wan04} and are understood in terms of the
  ionization structure of the nebulae under study: [\ion{Ar}{iv}] lines
  normally come from inner regions of higher ionization degree than
  [\ion{S}{ii}] lines. Typically, for giant Galactic and extragalactic
  \ion{H}{ii} regions, derived $N_e$ from these two line ratios differ
  in a factor of $\sim$5 \cite[e.g. ][]{est02,tsa03} which is much
  lower than what we find for the \emph{integrated} spectrum of
  NGC~5253 ($\sim$25). However, when only the giant
  \ion{H}{ii} region is taken 
  into account (i.e. the area of $\sim$90 spaxels where the argon
  lines are detected) the 
  difference between the densities derived from the 
  argon and the sulphur lines ($\sim$10, see typical
  values for the densities below) is more similar to
  those found for other \ion{H}{ii} regions. 
  
Maps for both ratios are shown in Figure
\ref{mapas2as2r}. According to the sulphur line ratio - detected
over the whole field - densities range from very low values, of the
order of the low density limit, in a region of about $5^{\prime\prime}\times
5^{\prime\prime}$ in the upper right corner of the field just above
the Complex \#3, to 790~cm~$^{-3}$ at the
peak of the emission in the cluster associated with the H\,\textsc{ii}
region, with a mean (median) over the field of view of $\sim130\,
(90)$~cm~$^{-3}$. The rest of the H\,\textsc{ii} regions still present
high densities (of about 400~cm~$^{-3}$ as a whole, 480~cm~$^{-3}$ in
the H\,\textsc{ii}-2, \citep[i.e. the upper area of the
  giant \ion{H}{ii} region,][]{kob97}. This agrees well with the value
estimated from long-slit measurements \citep{lop07}.
The tail and the region associated with  the cluster UV-1 present
intermediate values (of about 200~cm~$^{-3}$). 


The argon line ratio is used to sample the densest regions. The map for
this ratio was somewhat noisier and allowed an
estimation of the electron density only in the giant H\,\textsc{ii}
region. The densities derived from this line ratio are comparatively
higher, with a mean (median) of 3400~(3150)~cm~$^{-3}$. As
it happened in the case of the extinction, 
the peak of electron density according to this line ratio is offset
by $\sim0\farcs7-1\farcs1$ towards the north-west with respect of the
peak of continuum emission at Complex \#1.

We created higher S/N ratio spectra by co-addition of
$3\times3$~spaxel apertures associated with 
certain characteristic regions (i.e. the clusters at the core of 
the systems, C1+C2, and the regions \ion{H}{ii}-2, \ion{H}{ii}-1 and
UV-1 of \citet{kob97}). 
The largest values are measured around the 
core (i.e. C1+C2) where the [Ar~IV] electron density can be as high as
6200~cm~$^{-3}$. As we move further
away from this region, the measured electron density becomes
lower. Thus, \ion{H}{ii}-2 presents similar, although slightly lower,
densities ($\sim6100$~cm~$^{-3}$), followed by \ion{H}{ii}-1 with
$\sim4200$~cm~$^{-3}$ and UV1 with $\sim3300$~cm~$^{-3}$. These 
values agree,
within the errors, with those reported in \citet{lop07} for
similar apertures.

An interesting point arises when the different density values
derived for the integrated spectrum and for each individual
spaxel/aperture  are compared (180~cm$^{-3}$, and up to 790~cm$^{-3}$,
respectively when using the sulphur line ratio). The covered
f.o.v. ($\sim$210~pc$\times$135~pc) is comparable to the linear scales
that one can resolve from the ground at distances of $\sim$40~Mpc (or
$z\sim0.01$). Such a comparison illustrates how aperture effects can
cause important underestimation of the electron density in the
\ion{H}{ii} regions in starbursts at such distances, or further away.






\subsection{Ionization structure, excitation sources and nitrogen
  enhancement \label{secionistruc}}

\begin{figure}[h!]
   \centering
\includegraphics[angle=0,scale=.43, clip=,bbllx=28, bblly=0,
bburx=600, bbury=355]{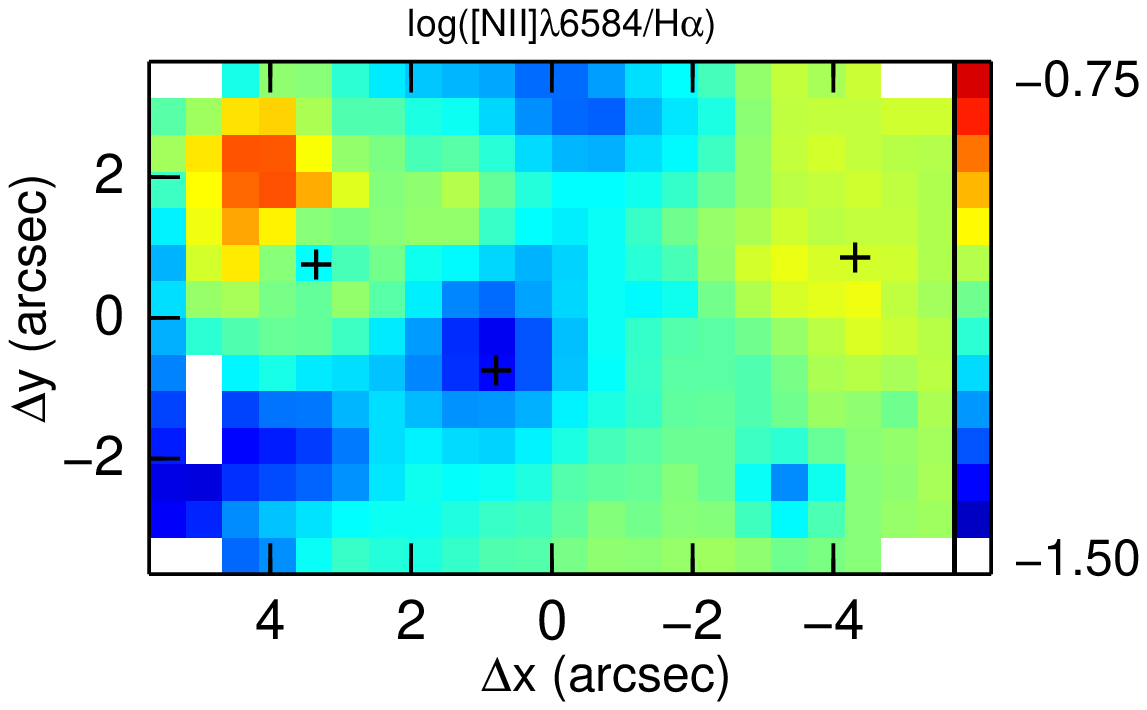}
\includegraphics[angle=0,scale=.43, clip=,bbllx=28, bblly=0,
bburx=600, bbury=355]{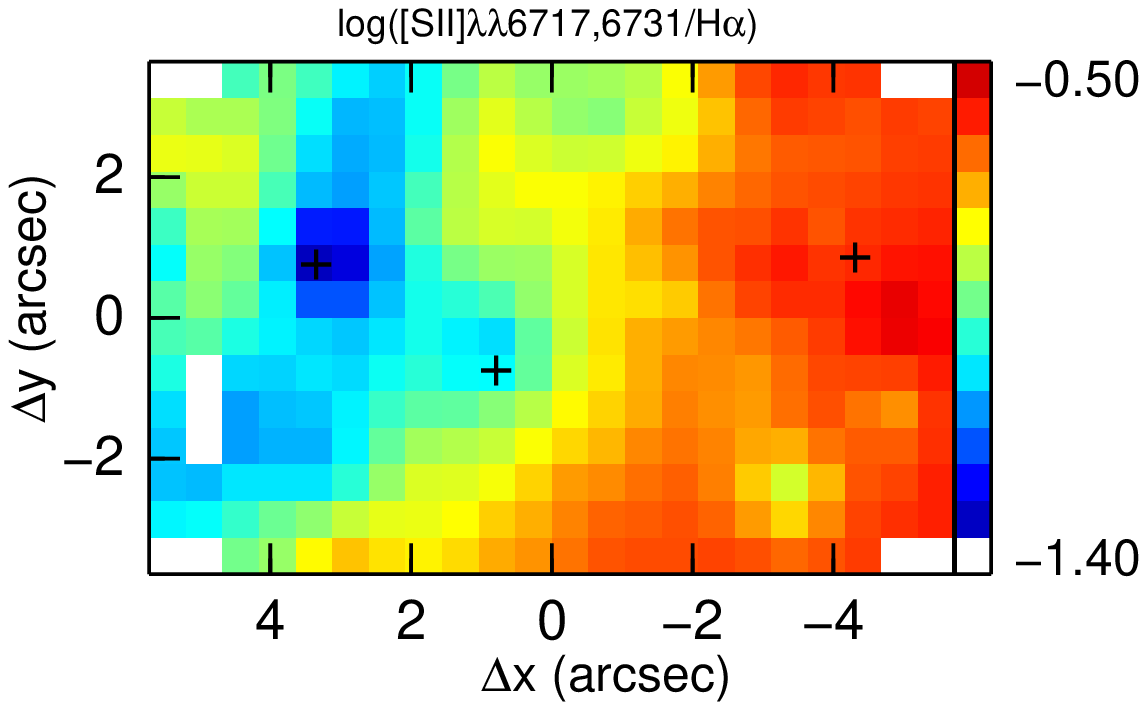}
\includegraphics[angle=0,scale=.43, clip=,bbllx=28, bblly=0,
bburx=600, bbury=355]{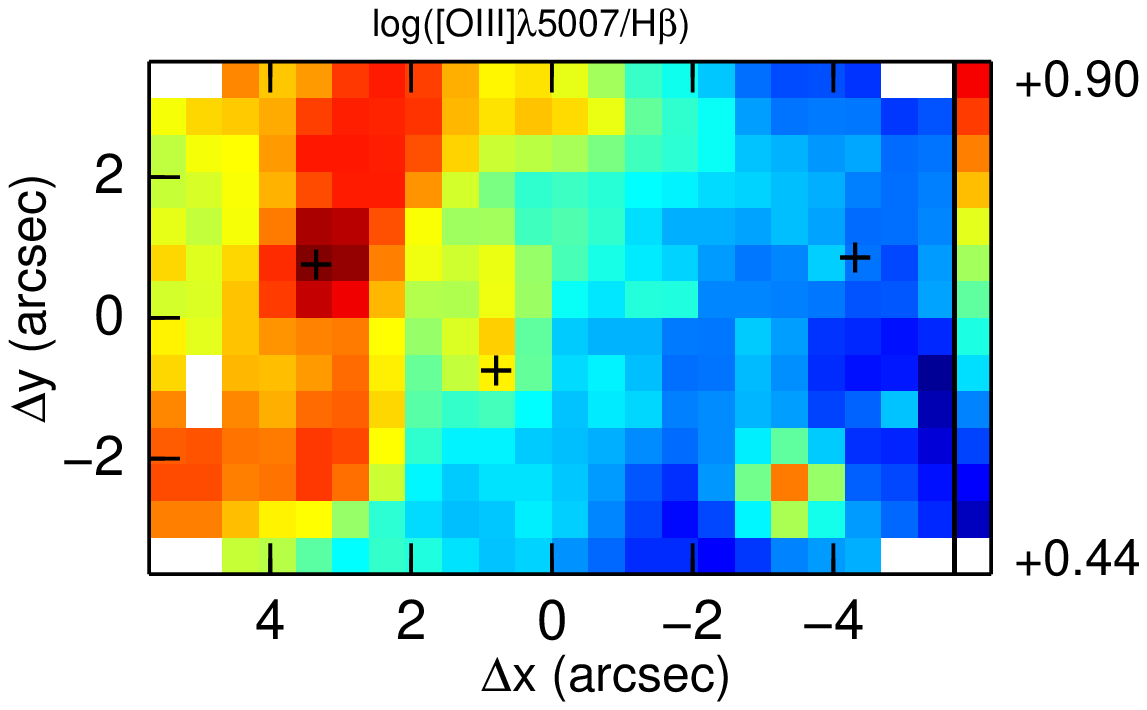}
   \caption[Emission line ratio maps. $\log$]{Emission line
   ratio maps. \emph{Up:} $\log$
   (\textsc{[N\,ii]}$\lambda$6584/H$\alpha$.
 \emph{Middle:} $\log$ (\textsc{[S\,ii]}$\lambda\lambda$6717,6731/H$\alpha$).
 \emph{Bottom:} $\log$ (\textsc{[O\,iii]}$\lambda$5007/H$\beta$).
 The position of the three main 
     peaks of continuum emission are shown for reference.
  \label{mapascocientes}}
 \end{figure}

The ionization structure of the interstellar medium can be studied by
means of diagnostic diagrams. Different areas of
a given diagram are explained by different ionization 
mechanisms. In the optical spectral range, the most widely
used are probably those proposed by \citet{bal81} and later reviewed
by \citet{vei87}, the so-called BPT diagrams. 
In Figure \ref{mapascocientes} the maps for the three available line ratios
involved in these diagrams - namely
[N\,\textsc{ii}]$\lambda$6584/H$\alpha$,
[S\,\textsc{ii}]$\lambda\lambda$6717,6731/H$\alpha$,
[O\,\textsc{iii}]$\lambda$5007/H$\beta$ - are shown on a logarithmic
scale. This figure
shows that the ionization structure in the central 
region of this galaxy is complex. Not only do the line ratios not show
a uniform distribution, but the structure changes depending on the
particular line ratio.

   \begin{figure}[h!]
   \centering
\includegraphics[angle=0,width=0.4\textwidth, clip=,bbllx=25, bblly=12,
bburx=452, bbury=643]{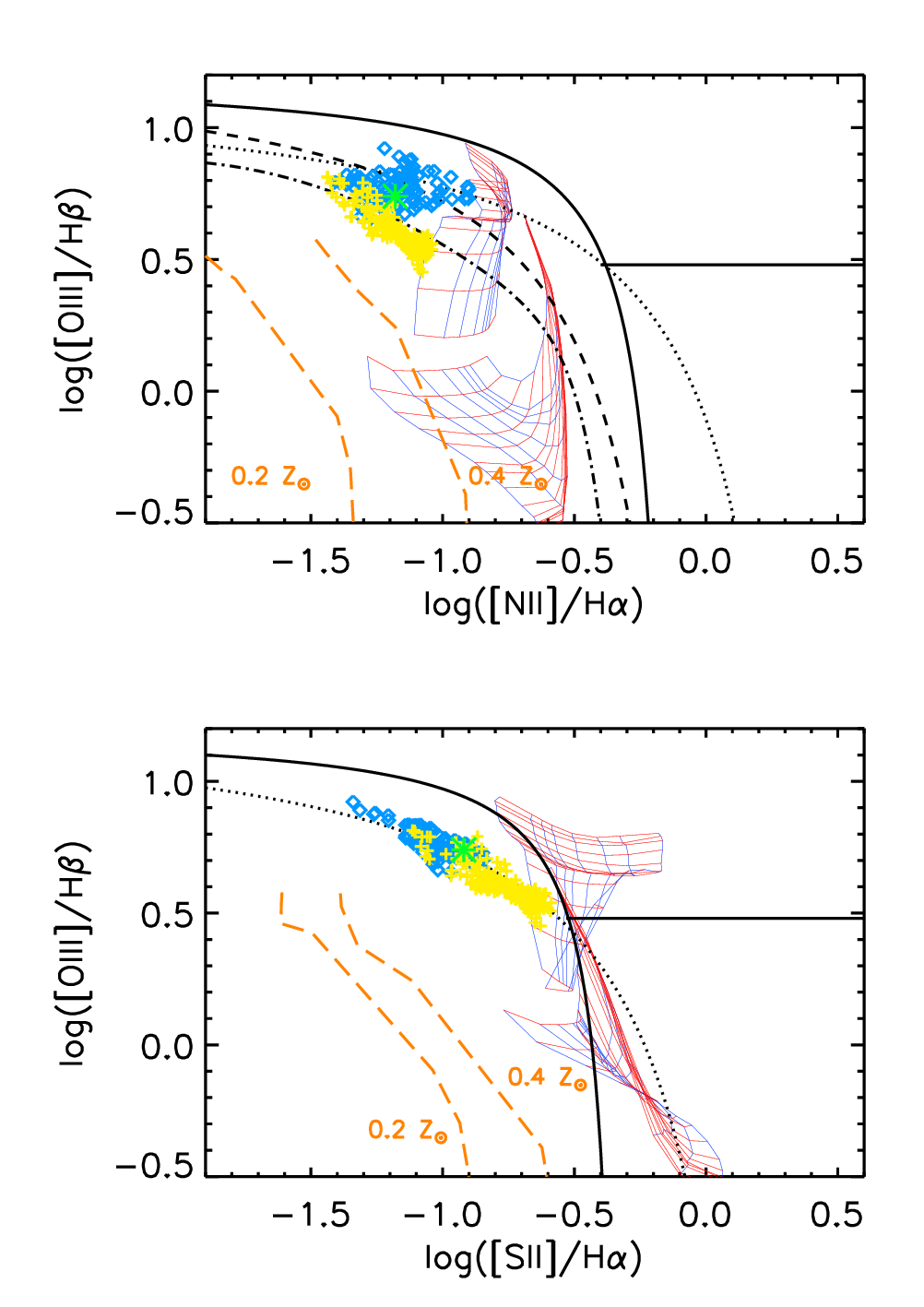}
   \caption[]{Position of the individual spaxels in NGC~5253 in the
     diagnostic diagrams 
     proposed by \citet{vei87}. 
   Data points above and below the 3-$\sigma$ level of the
   first-degree polynomial fit in the \nha\ vs. \sha\ diagram (see
   text for details) have been represented with blue diamonds and
   yellow crosses respectively. The solid curves show
     the empirical borders found by Veilleux \& Osterbrock
     between ionization 
     caused by different mechanisms, while the dotted lines show the
     theoretical borders proposed by \citet{kew01a} to delimit the area
     where the line ratios can be explained by star formation. Black
     dashed and dot-dashed lines show the revised borders by 
     \citet{kau03} and \citet{sta06}, respectively. Solid horizontal
     lines show the border between classical Seyfert galaxies (above
     the line) and 
     LINERs (below the line), at \ohb=3.  Long-dashed lines in orange represent
     the predictions from models of photo-ionization by stars
     presented in \citet{dop06} for $R=0$ and ages of $2-6$~Myr. The
     grids represent the 
     predictions for the shock models presented in \citet{all08}, with
     and without precursor,
     assuming the abundances of the LMC and $N_e = 1$~cm$^{-3}$. Lines
     of constant magnetic parameter (blue) cover a
     range of $B=0.0001-10$~$\mu$G, while lines of constant shock
     velocity (red) cover a range of $v_s =
     150-500$~km~s$^{-1}$. Green asterisks mark the derived 
     values for the integrated spectrum of the whole IFU field.
       \label{diagdiag}}
 \end{figure}

Both the \ohb{} and the \sha{} line ratios display a gradient away from 
the peak of emission at Complex \#1 and
with a structure that follows that of the ionized gas. Thus the
\sha{} (\ohb) ratio is smallest (largest) at Complex \#1,
presents somewhat intermediate values in the two tongue-shaped
extensions and the Complex \#2 
and is relatively high (low) in the rest of the field, with a
secondary minimum (maximum) at $\sim[-3\farcs5,-2\farcs5]$, the position of a
secondary peak in the H$\alpha$ emission. This is coincident with the
structure presented in \citep{cal04}. 

The \nha{} line ratio, however, display a different structure. While in
the right half of the FLAMES field of view, the behavior is quite similar
to the one observed for the \sha{} ratio (i.e. values relatively high,
local minimum at $\sim[-3\farcs5,-2\farcs5]$), the left half,
dominated by the giant \ion{H}{ii} region, displays a completely
different pattern. The lowest values are associated with Complex
\#2 and the southern extension, rather than Complex \#1,
and the \nha{} line ratios are highest at $\sim[4\farcs0,2\farcs0]$. 

In Figure \ref{diagdiag} we show the position of each spaxel of the
FLAMES field of view, as well as for the integrated spectrum, after
co-adding all the spaxels in the BPT diagnostic diagrams together with the 
borders that separate \ion{H}{ii} region-like ionization from
ionization by other mechanisms according to several authors
\citep{vei87,kew01a,kau03,sta06}.  We also show the
  predictions for models of photo-ionization caused by stars
  \citep{dop06} that take into account the effect of the stellar
  winds on the dynamical evolution of the region. In these models, the
  ionization parameter is replaced by a new variable $R$ that depends
  on the mass of the ionizing cluster and the pressure of the
  interstellar medium ($R$ = (M$_{Cl}$/M$_\odot$)/(P$_o$/k), with
  P$_o$/k measured in  cm$^{−3}$~K). Also, the predictions for
  shocks models for a LMC metallicity are included. Given the relatively
    low metallicity of NGC~5253, these are the most appropriate
    ones. They were calculated assuming a $N_e =1$~cm$^{-3}$ and cover
  an ample range of magnetic parameters, $B$, and shock velocities,
  $v_s$, \citep[see][for details]{all08}. 


As demonstrated for the electron density, these diagrams illustrate
very clearly how resolution effects can influence the measured
line ratios. Values derived for individual spaxels cover a range of 
$\sim$0.5, 1.0 and 0.5~dex for the \nha, \sha, and \ohb\ line ratios
respectively, 
with mean values similar to the integrated values ($-1.15$, $\-0.92$,
and 0.74). This is particularly relevant when interpreting the
ionization mechanisms in galaxies at larger distances where the 
spectrum can sample a region with a range in ionization properties.
This loss of spatial resolution thus 'smears' the
determination of the ionization mechanism by a set of line ratios. 
Even if this given set of line ratios is typical of
photoionization caused by stars, it is not possible to exclude some
contribution due to other mechanisms at scales unresolved by the 
particular observations. 

Regarding the individual measurements, although all line ratios are
within the typical values expected for an \ion{H}{ii} region-like
ionization, two differences between these diagrams arise. The first
one is that the diagram involving the \sha\ line 
ratio indicates a somewhat higher ionization degree than the one involving
the \nha\ line ratio. That is: values for the diagram
  involving the \sha\ line ratio
  are at the limit of what can be explained by pure
photo-ionization in an \ion{H}{ii} region according to the
\citet{kew01a} theoretical borders. On the contrary, most of the data
points in the diagram involving the \nha\ are clearly in the area
associated to photoionization caused by stars. A comparison with the
predictions of the models for metallicities similar to the one of
\object{NGC~5253} shows how the measured line ratios present
  intermediate values between those predicted by ionization caused by shocks
  and those by pure stellar photoionization.
That this is exactly what one would
expect if shocks caused by the mechanical input from stellar winds or
supernovae within the starburst were contributing to the observed
spectra.  Also, this comparison supports previous studies that show
how models of photoionization caused by stars underpredict the \ohb\ line
ratios, specially in the low-metallicity cases \citep{bri08,dop06}.


   \begin{figure}[h!]
   \centering
\includegraphics[angle=0,width=0.45\textwidth, clip=,bbllx=25, bblly=15,
bburx=700, bbury=540]{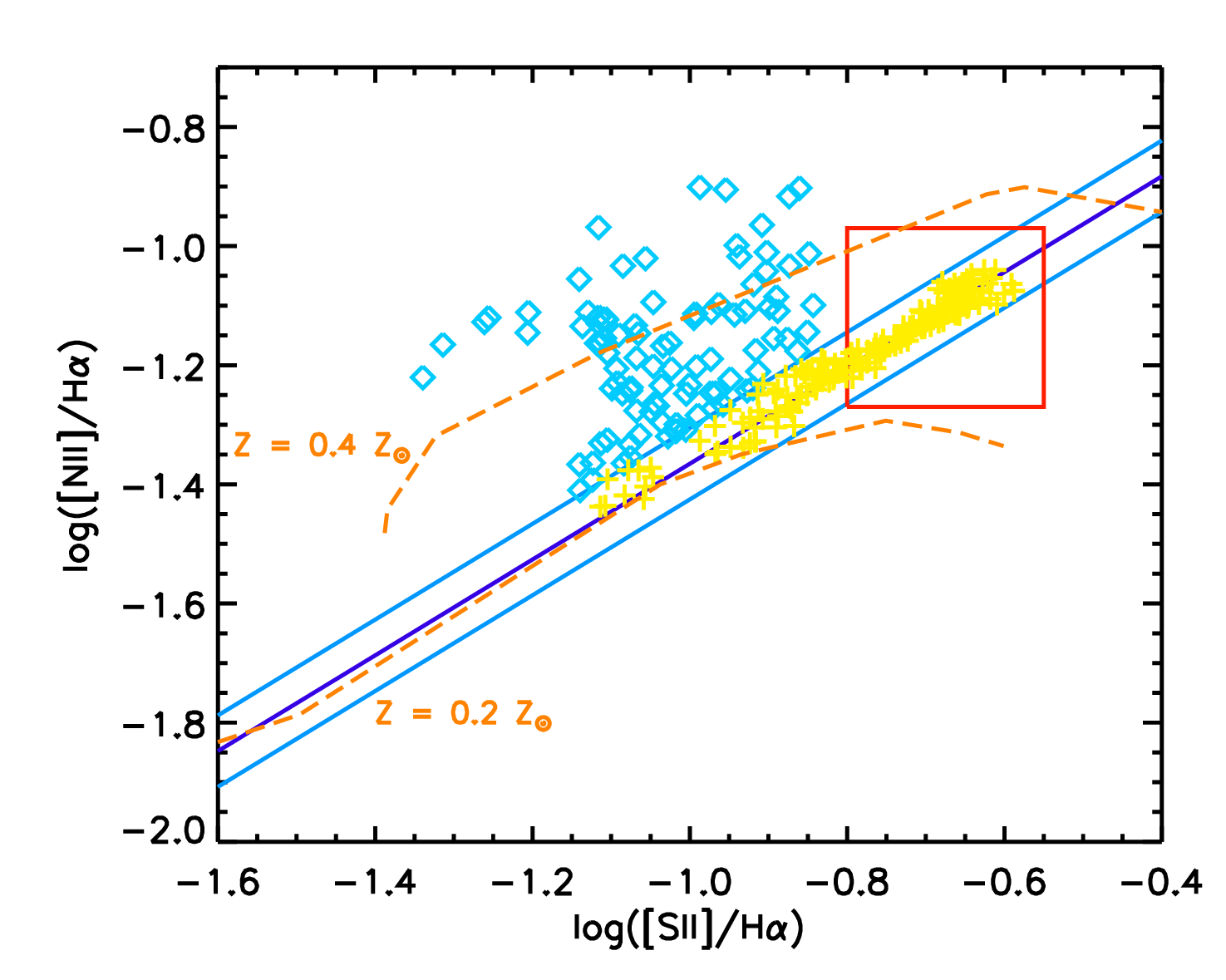}
   \caption[]{\sha\ vs. \nha\ diagram. The red box
   includes the data points utilized to determine the 
   first-degree polynomial fit (dark blue continuous line) between the
   \sha\ and \nha\ line ratios. 
   Data points above and below the 3-$\sigma$ level (light blue
   continuous line) have been represented with blue diamonds
   and yellow crosses respectively.  
   The expected line ratios  for \ion{H}{ii} regions for $Z=0.2 Z_\odot$ and
   $Z=0.4 Z_\odot$ with ages in the range of 2-6~Myr and $R=0$ according to the
   models presented in \citet{dop06} are over-plotted with dashed oranges lines
   for comparison.
  \label{s2havsn2gha}}
 \end{figure}

 \begin{figure}[h!]
 \centering
\includegraphics[angle=0,width=0.42\textwidth, clip=,bbllx=20, bblly=-10,
bburx=485, bbury=312]{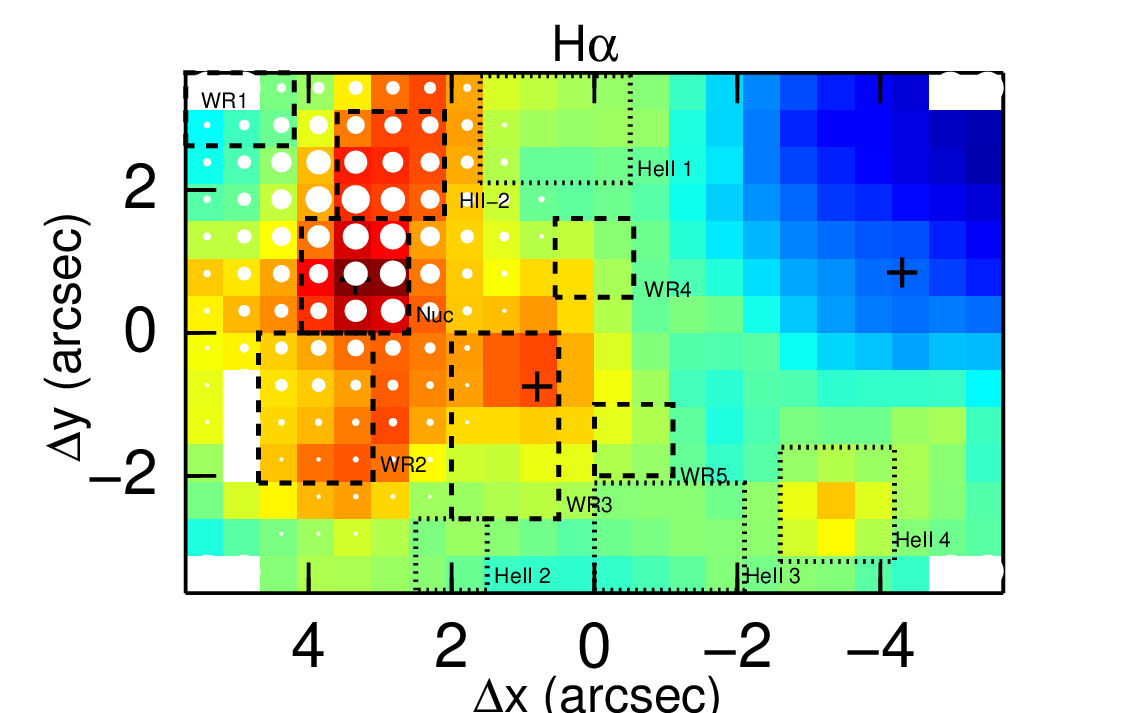}
   \caption[]{Spaxels with \nha\ line ratio above the 3-$\sigma$
   level of the fit presented in Figure \ref{s2havsn2gha} on
     top of our \ha\ emission line map. The size of
   the white circles is proportional to 
   the excess in the \nha\ line ratio. Dashed and dotted lines delimit the
   apertures utilized to extract the spectra with Wolf-Rayet and nebular
   \ion{He}{ii} features displayed in Figures \ref{especwr} and
   \ref{especheiineb} (see sections \ref{secwr} and
   \ref{seche}). The position of the three peaks of continuum emission in
    the map are shown by crosses for reference. 
  \label{nitroalto}}
 \end{figure}

The second diference is the distribution of the
data points in these diagrams. While the data points in the \sha{} vs. \ohb{}
diagram form a sequence, data points in the \nha{} vs. \ohb{} diagram are
distributed in two groups: a sequence similar to the one in the
\sha{} vs. \ohb{} diagram and a cloud of data points above that
sequence with larger \nha. 
This result can be interpreted either by local variations in
  the relative abundances or by changes in the ionization
  parameter. Here we will 
  explore the first option, which is the most accepted explanation
  \citep[e.g.][and references therein]{lop07} and is supported by the
  relatively constant ionization parameter found in specific
  areas via long-slit \citep[$\log(U)\sim$-3,][]{kob97}.
Long-slit measurements in specific areas of this field have shown
out how this galaxy present some regions with an over-abundance of
nitrogen \citep[e.g.][]{wal89,kob97}. For our measured line
ratios and using expression (22) in \citet{per09}, we measure a range
in $\log (N/O)$ of $-0.70$ to $-1.46$.
Here we will assume that this over-abundance is the cause of our
excess in the \nha\ line ratio and will use this excess to
precisely delimit the area presenting 
this over-abundance. To this aim, we placed the information of each of 
the spaxels in the \sha{} 
vs. \nha{} diagram, which better separates the two different groups
described above. This is presented in Figure \ref{s2havsn2gha}.
We have assumed that in the so-called \emph{un-polluted} areas, the 
\nha\ and \sha\ follow a linear relation. This is a reasonable
assumption since the \nii/\sii\ line ratio has a low dependence with
the abundance and the properties of the ionizing radiation field
\citep{kew02b}. This \emph{standard} relation was determined by
fitting a first-degree polynomial to the data points with
\sha$>-0.8$ (indicated in Figure \ref{s2havsn2gha} with a red box). 
Those spaxels whose \nha\ line ratio was in excess of more than
3-$\sigma$ from the relation determined by this fit, have been
identified as having an [N~II]/\ha excess, and are identified by
diamonds in Figure \ref{s2havsn2gha}. As can be seen from this 
figure, there are a number of spaxels where this excess is much 
above the \emph{standard} relation.

The data points thus identified with \nha\ excess are shown as a map
in Figure \ref{nitroalto} where the location and magnitude of the \nha\
excess is indicated by white circles, whose size is proportional to the 
size of the \nha\ excess. This figure can be interpreted as a
snapshot in the pollution process of the interstellar medium by the
SSCs in the central area of \object{NGC~5253}.  
The  pollution is affecting almost the whole giant \ion{H}{ii}
region. 
The largest values are found at $\sim1\farcs5$ towards
the north-west of the peak at Complex~\#1. Then, the quantity of extra
nitrogen decreases outwards following the two tongue-shaped extensions
towards the north-west and south-east. This is consistent with the HST
observations of \citet{kob97} who found nitrogen enrichment in their
\ion{H}{ii}-1 and \ion{H}{ii}-2, while the N/O ratio in UV-1 (see Table
\ref{cumulos}) was typical for metal-poor galaxies.


   \begin{figure*}
\begin{center}
\includegraphics[angle=0,width=0.45\textwidth, clip=,bbllx=65, bblly=68,
bburx=740, bbury=375]{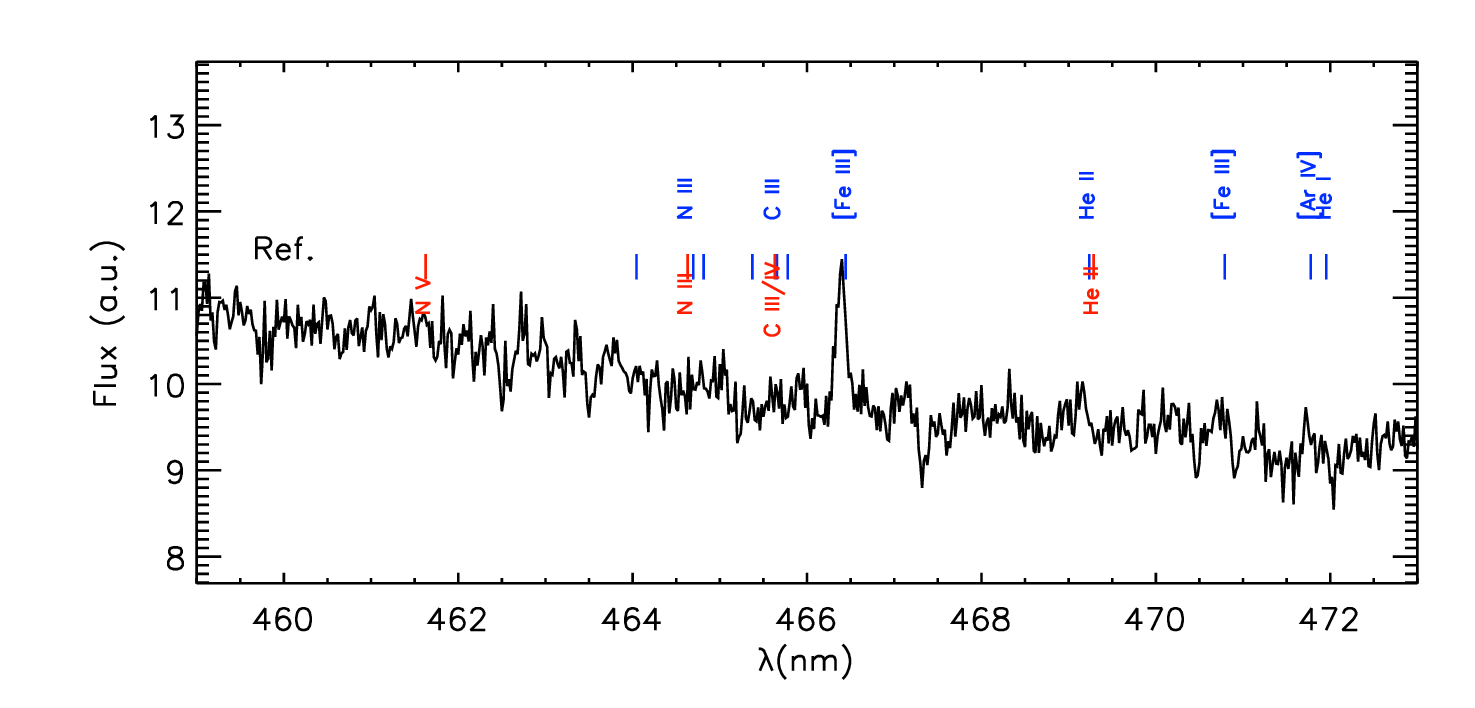}
\includegraphics[angle=0,width=0.45\textwidth, clip=,bbllx=65, bblly=68,
bburx=740, bbury=375]{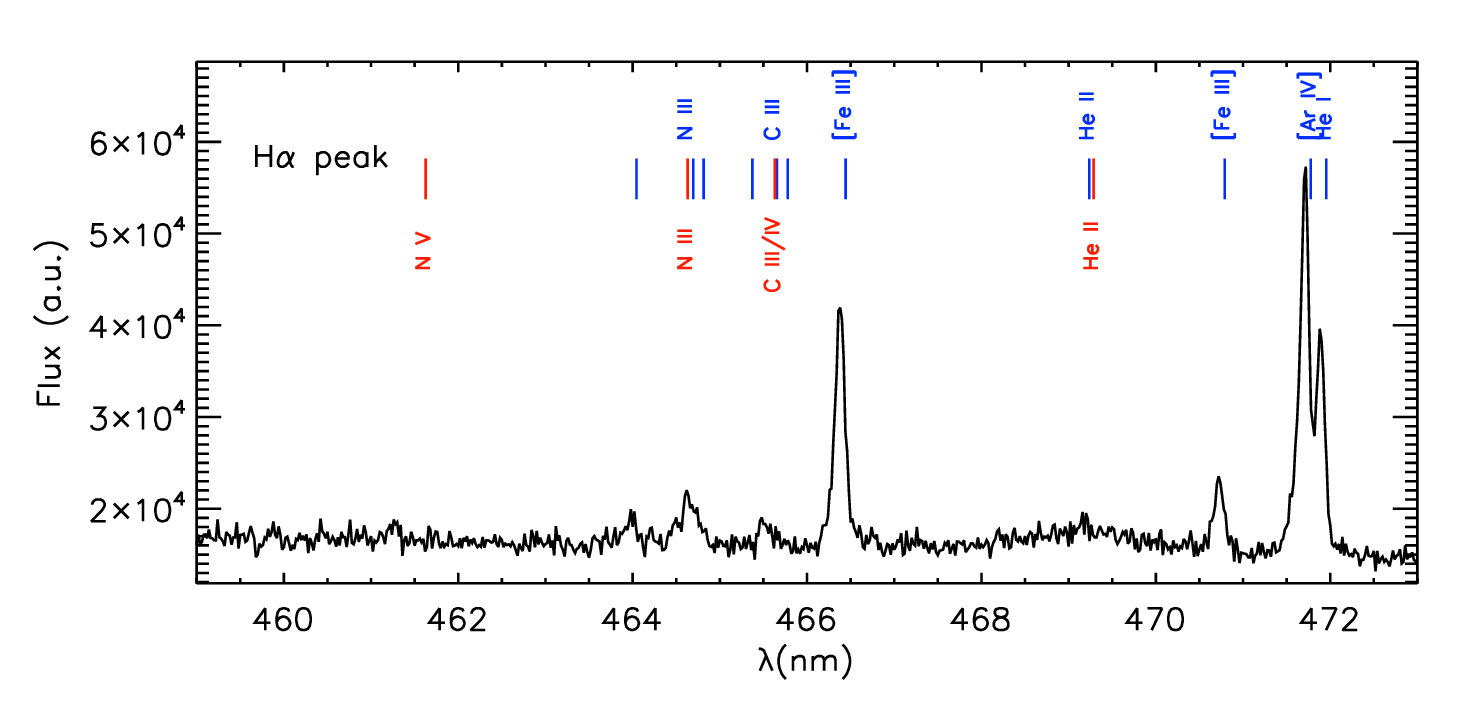}\\
\includegraphics[angle=0,width=0.45\textwidth, clip=,bbllx=65, bblly=68,
bburx=740, bbury=375]{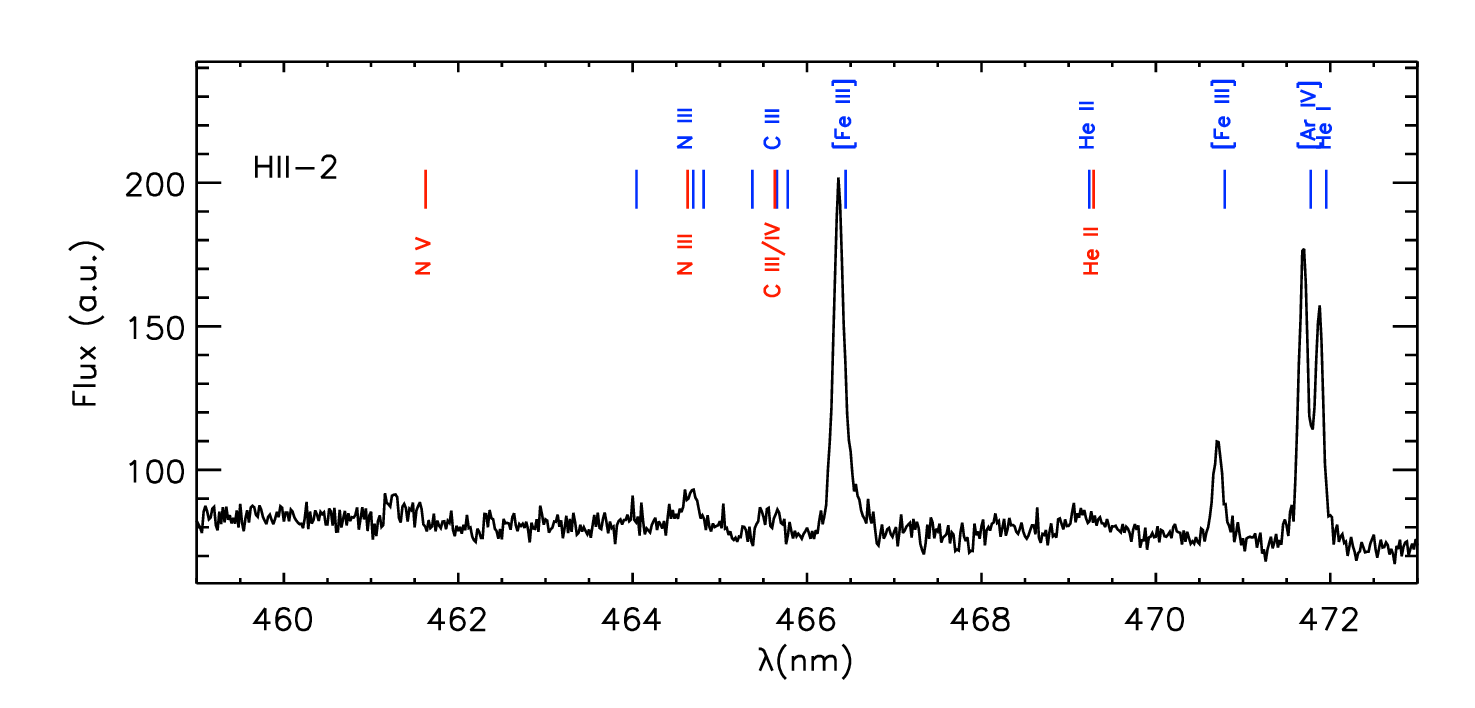}
\includegraphics[angle=0,width=0.45\textwidth, clip=,bbllx=65, bblly=68,
bburx=740, bbury=375]{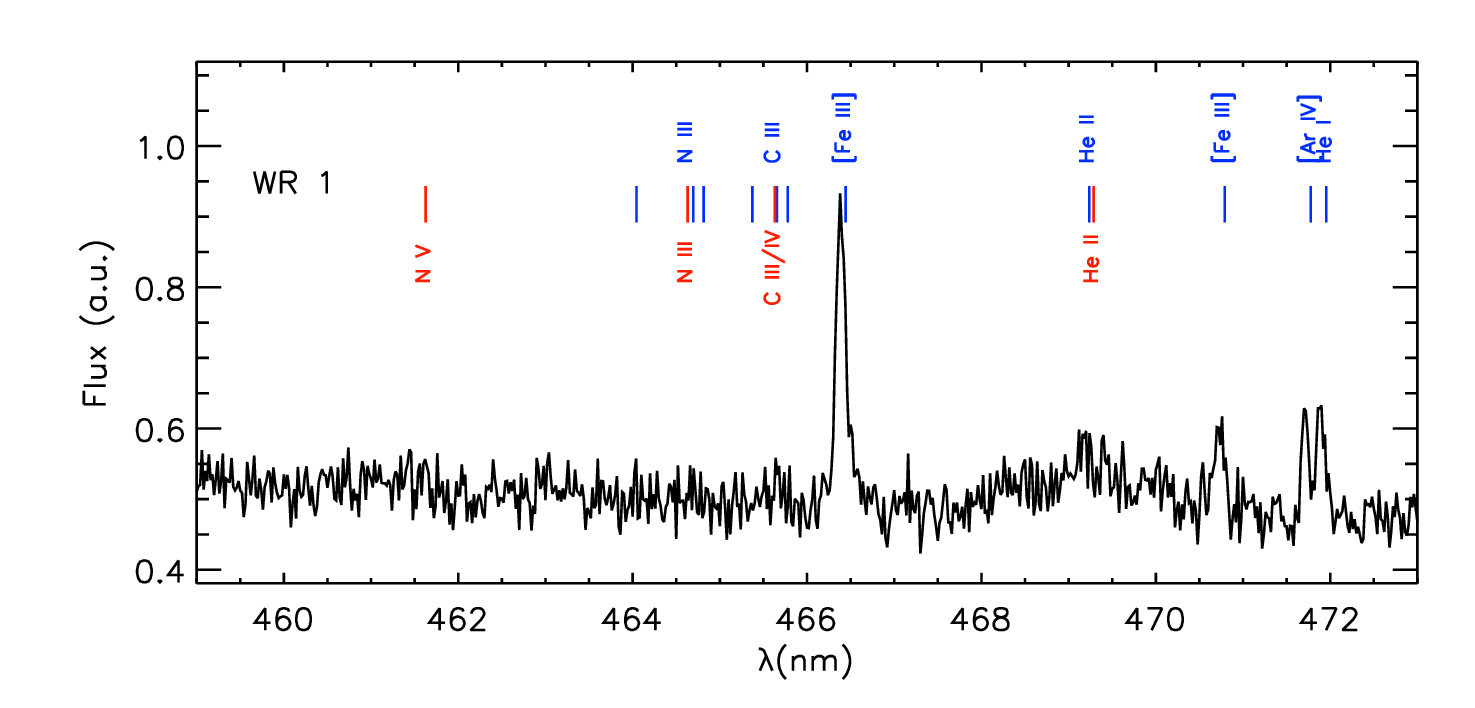}\\
\includegraphics[angle=0,width=0.45\textwidth, clip=,bbllx=65, bblly=68,
bburx=740, bbury=375]{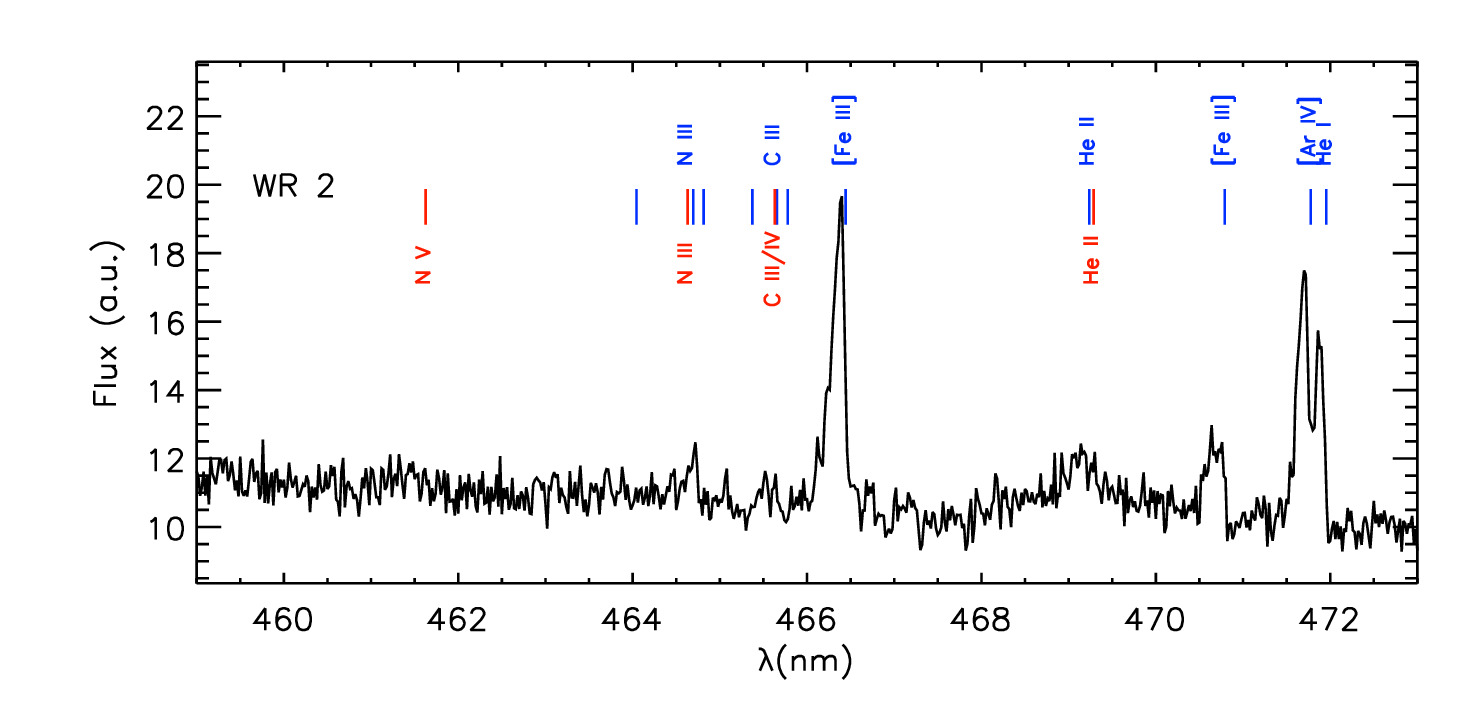}
\includegraphics[angle=0,width=0.45\textwidth, clip=,bbllx=65, bblly=68,
bburx=740, bbury=375]{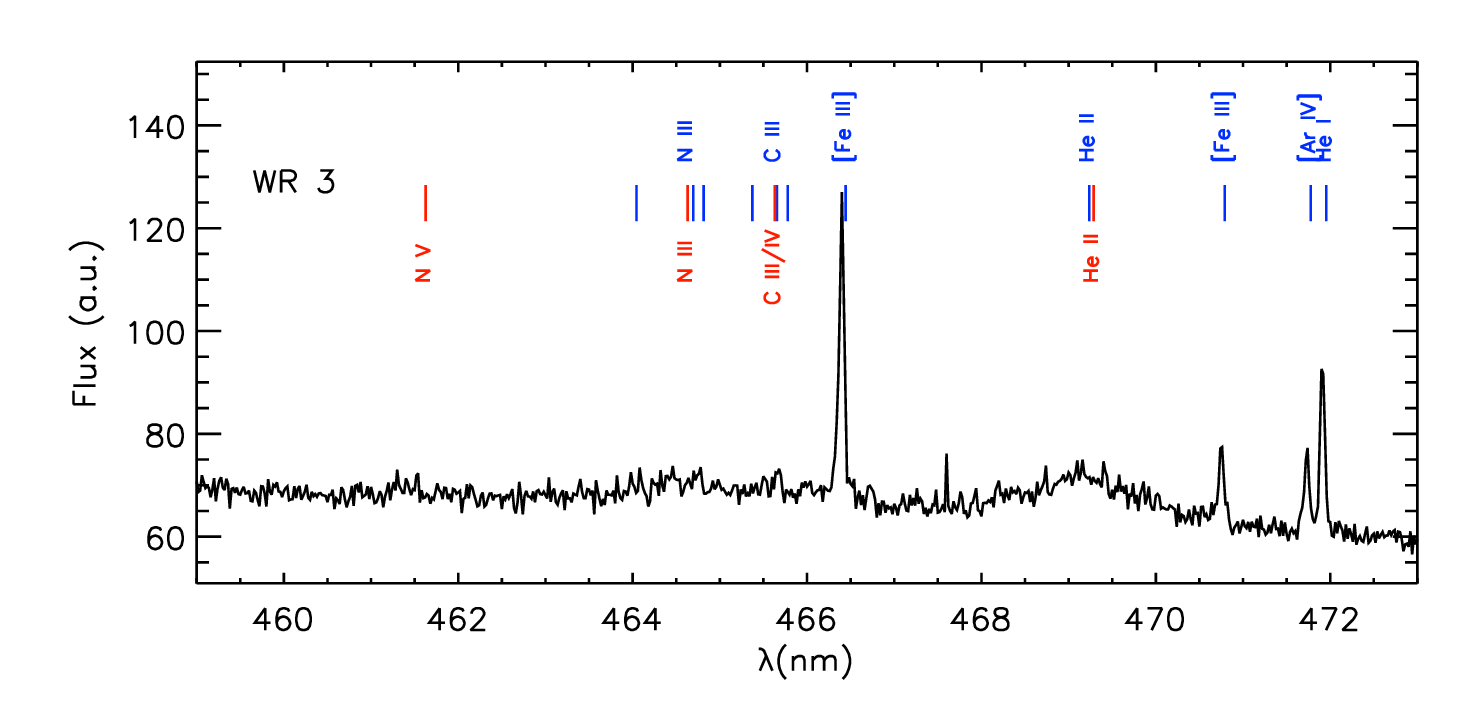}
\includegraphics[angle=0,width=0.45\textwidth, clip=,bbllx=65, bblly=68,
bburx=740, bbury=375]{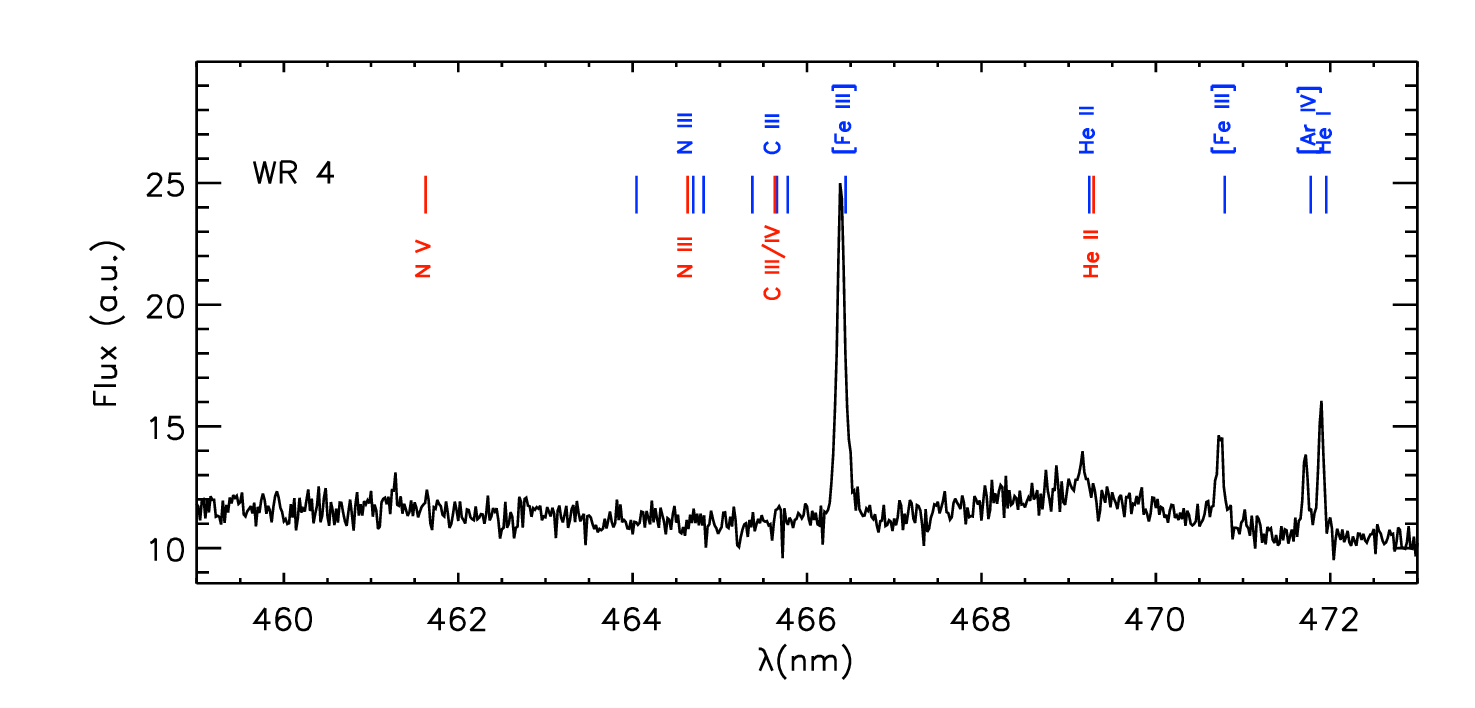}
\includegraphics[angle=0,width=0.45\textwidth, clip=,bbllx=65, bblly=68,
bburx=740, bbury=375]{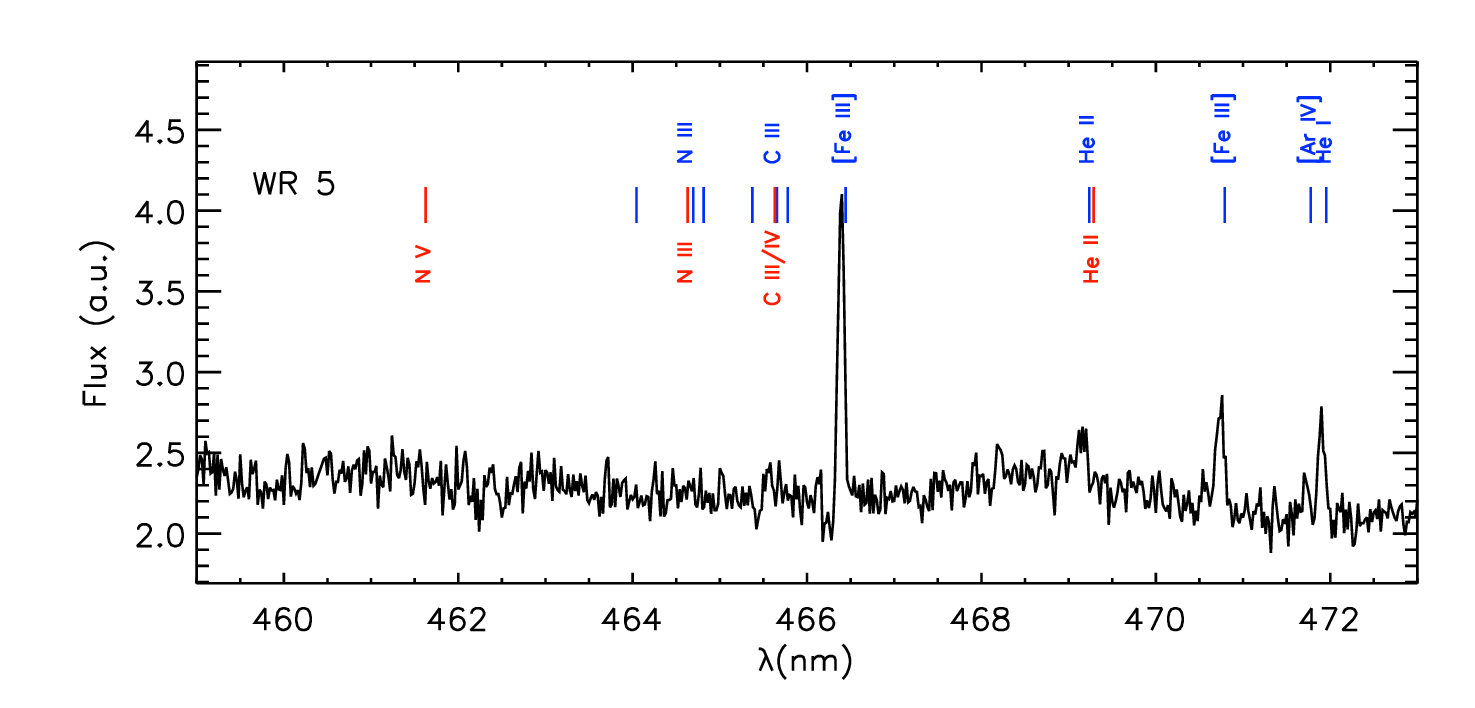}
\end{center}
   \caption[Spectra with Wolf-Rayet features]
  {Spectra showing Wolf-Rayet features. The first spectrum has been
  extracted from a low surface brightness area in the upper right
  corner of the FLAMES field of view and is presented here as
  reference. The positions of the nebular emission lines have been
  indicated with blue ticks and labels, while 
  those corresponding to Wolf-Rayet features appear in red. The position 
  and extent of the regions are indicated on Figure \ref{nitroalto}. \label{especwr}}
 \end{figure*}

\subsection{Wolf-Rayet features \label{secwr}} 

Wolf-Rayet (W-R) stars are very bright objects with strong broad
emission lines in their spectra. They are classified as WN (those
with strong lines of helium and nitrogen) and WC (those with strong
lines of helium, carbon and oxygen) and are understood as the result of the
evolution of massive O stars. As they evolve, they loose a significant
amount of their 
mass via stellar winds showing the products of the CNO-burning
first - identified as WN stars - and the He-burning afterwards - 
as WC stars  
\citep{con76}. The presence of W-R stars can be recognized via the W-R
bumps around 
$\lambda$4650~\AA{} (i.e. the \emph{blue bump}, characteristic of WN stars) and
$\lambda$5808~\AA{} (i.e. the \emph{red bump},  characteristic of WC
stars, but not covered by the present data).

\begin{table*}
\centering
      \caption[]{Ages of the clusters associated with the W-R regions
        according to different indicators.}
       \label{compacumu}
              \begin{tabular}{ccccccccc}
            \hline
            \noalign{\smallskip}
 & \multicolumn{2}{c}{\citet{har04}}  &
            \multicolumn{4}{c}{This work} \\
Region & Clusters & Age  &
         $\log$(WR/(WR+O))  & Age$_{\rm WR/O}$  &
         EW(\hb)            & Age$_{\rm H\beta}$  \\ 
 & Number & (Myr) & & (Myr) & (\AA) & (Myr)      \\
            \noalign{\smallskip}
            \hline
            \noalign{\smallskip}
Nucleus        & 1       & 3      & -1.99 & 2.9     & 245 & 2.7\\
\ion{H}{ii}-2  & 20      & 3      & -1.93 & 2.9     & 320 & 2.4\\
W-R~1          & 13      & 4      & -1.34 & 3.2/4.8 & 131 & 3.4\\
W-R~2          & 23      & 5      & -1.64 & 3.0/5.5 & 180 & 3.2\\
W-R~3          & 4,24,25 & 1-5    & -1.55 & 3.0/5.3 & 128 & 3.4\\
W-R~4          & 6,21    & 4      & -1.35 & 3.2/4.8 & 131 & 3.4\\
W-R~5          & \ldots  & \ldots & -1.17 & 3.4     &  86 & 4.7\\
            \hline
         \end{tabular}
\end{table*}

\citet{sch97,sch99} carried out a thorough search and
characterization of the W-R population in NGC~5253. They detected W-R
features (both WN and WC) at the peak of emission in the optical (our Complex 
\#1) and the ultraviolet (our Complex \#2). 
Due to the spatial coincidence of these detections and the N-enriched regions
found by \citet{wal89} and \citet{kob97} - at least in the case of the
nucleus - they suggested that these W-R stars could be the cause of
this enhancement.  
This is supported observationally by similar findings in other W-R galaxies.
For example, a recent survey using Sloan data of
W-R galaxies in the low-redshift Universe has shown that galaxies
belonging to this group present an elevated N/O ratio, in comparison with
similar non-W-R galaxies \citep{bri08}. Other suggested possibilities
to cause the enrichment in nitrogen include
planetary nebulae, O star winds, He-deficient 
W-R star winds, and luminous blue variables \citep{kob97}.

Here, we characterize the W-R population in NGC~5253 and 
explore the hypothesis of W-R stars as the cause of the
nitrogen enhancement by using the 2D spectral information provided by
the present data. In the previous section we have delimited very 
precisely the area that presents nitrogen enhancement. In a same
manner, it is possible to look for and localize the areas that present W-R
emission. Note that due to the continuous sampling of the present data
this can be done in a completely unbiased way.

We visually inspected each spectrum looking for the more
prominent W-R features in the \emph{blue bump}
(i.e. \ion{N}{iii}\,$\lambda$4640 and \ion{He}{ii}\,$\lambda$4686). 
The areas where these features have been found are marked in Figure
\ref{nitroalto} with dashed lines. 
The co-added and extracted spectra of each individual area appear in
Figure \ref{especwr} together with a reference spectrum, free of
W-R features, made by
co-adding 20 spaxels in the upper right corner of the FLAMES field
of view.

We confirm the detection of W-R features associated with the nucleus
and the brightest cluster in the ultraviolet, UV-1 (our W-R~3). 
In the same manner, we also detect a broad \ion{He}{ii} line
associated with the north-west and south-east extensions (i.e. \ion{H}{ii}-2
and W-R 2, respectively).  In addition, there are three more
areas which present W-R features, called W-R~1, W-R~4 and W-R~5, relatively far
($\sim58-83$~pc) from the main area of activity. 
Interestingly, two of these regions (W-R~4 and
W-R~5) present a narrow nebular \ion{He}{ii} on top of the broad W-R feature.

The short phase of W-R stars during star evolution 
makes their detection a very precise method for estimating the age of a
given stellar population. According to \citet{lei99}, typically an
instantaneous starburst shows these features at ages of $\sim3-6$~Myr
for metallicites of $Z = 0.004 - 0.008$, similar to the one in
\object{NGC~5253}. Thus, very young stellar clusters must be associated with the
areas that display these W-R features. We compared the positions of our
detections with the catalogue of clusters given by \citet{har04} and
compiled in  Table \ref{compacumu}.
All our regions, except W-R~5, are associated with one (or
several) young (i.e $<5$~Myr) star cluster(s). Regarding W-R~5,
\citet{har04} do not report any cluster associated with that
area. 

%

Cluster ages in \citet{har04} were estimated using both  broadband
  photometry and the \ha\ equivalent width. We
  estimated the ages by means of two indicators: the ratio between 
  the number of W-R and O stars; and the \hb\ equivalent width. 
The ratio between the number of W-R and O stars was estimated from
F(bb)/F(H$\beta$), where F(bb) and F(H$\beta$) are the flux in the
\emph{blue bump} (measured with \texttt{splot}) and in H$\beta$
respectively and using the relation proposed by
\cite{sch98}. Uncertainties are large, mainly due to the difficulty to
define the continuum and to avoid the contamination of the nebular
emission lines when measuring F(bb) but indicates a range in 
$\log$(WR/(WR+O)) of $\sim-2.0$ to $-1.2$ (see Table \ref{compacumu}, column 4).
There is a higher proportion of W-R stars in W-R~5, W-R~1 and W-R~4
and somewhat lower in those areas associated with the giant
\ion{H}{ii} region. The \hb\ equivalent widths
  are extremely high, consistent again with the expected youth of the
  stellar population. Predicted ages from these two age tracers are
  reported in Table \ref{compacumu}, columns 5 and 7. They  were estimated by
  using STARBURST99 \citep{lei99}, assuming an instantaneous burst of
  $Z=0.008$, an upper mass limit of $M_{up}$=100~M$_\odot$ and a
  Salpeter-type Initial Mass Function. The two age tracers give
  consistent age predictions and in agreement with those reported in
  \citet{har04}.   

The distribution of the W-R features, in an area of about
100~pc$\times$100~pc, much larger than the one polluted with nitrogen,
suggests that \emph{all} the detected W-R stars are not, in general, the cause of
this pollution.
Since the N-enrichment appears to be associated with the pair of
clusters in the core and, given the position of maximum,
most probably with the obscured SSC C2, the best W-R star candidates
to be the cause of this enrichment are those corresponding to our
\emph{Nuc} aperture, and perhaps also the \ion{H}{ii}-2 and W-R~2 regions. 


\begin{figure*}
\begin{center}
\includegraphics[angle=0,width=0.45\textwidth, clip=,bbllx=65, bblly=68,
bburx=740, bbury=375]{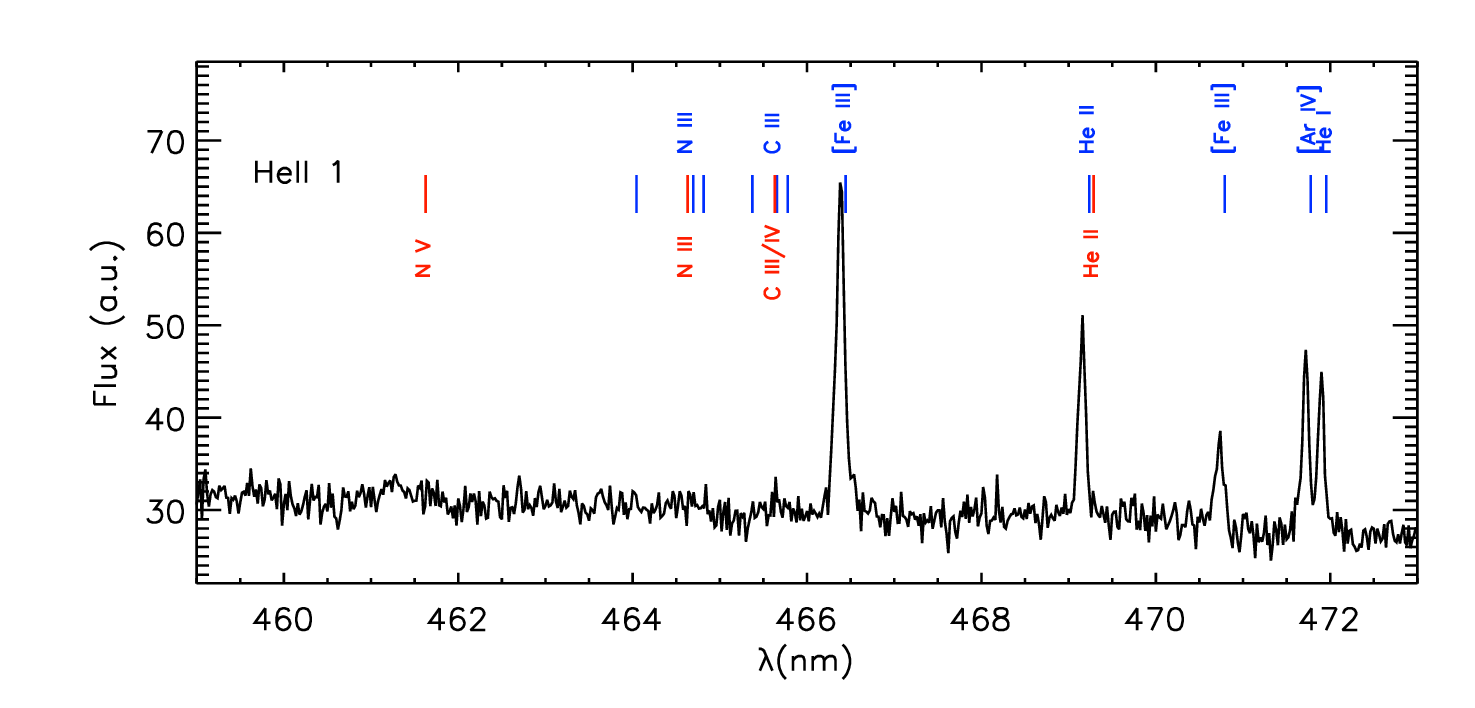}
\includegraphics[angle=0,width=0.45\textwidth, clip=,bbllx=65, bblly=68,
bburx=740, bbury=375]{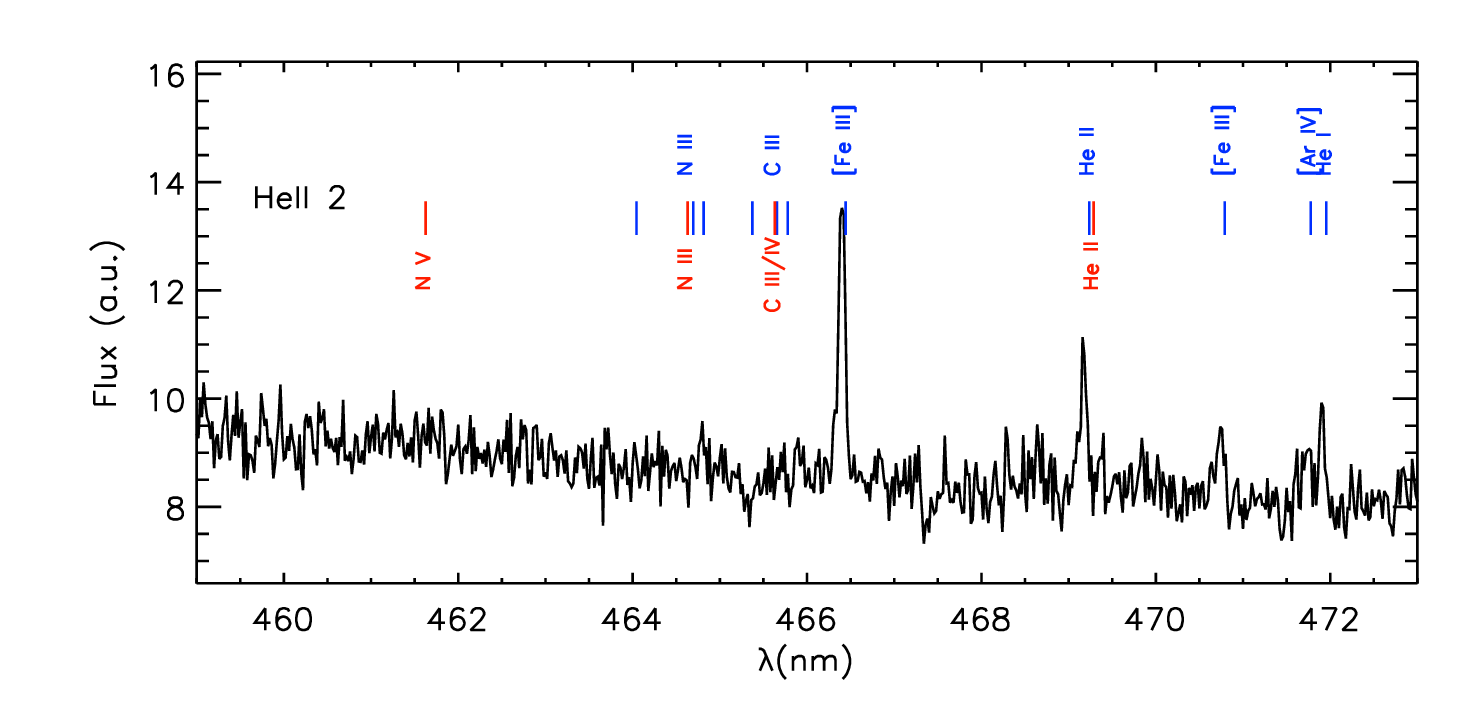}
\includegraphics[angle=0,width=0.45\textwidth, clip=,bbllx=65, bblly=68,
bburx=740, bbury=375]{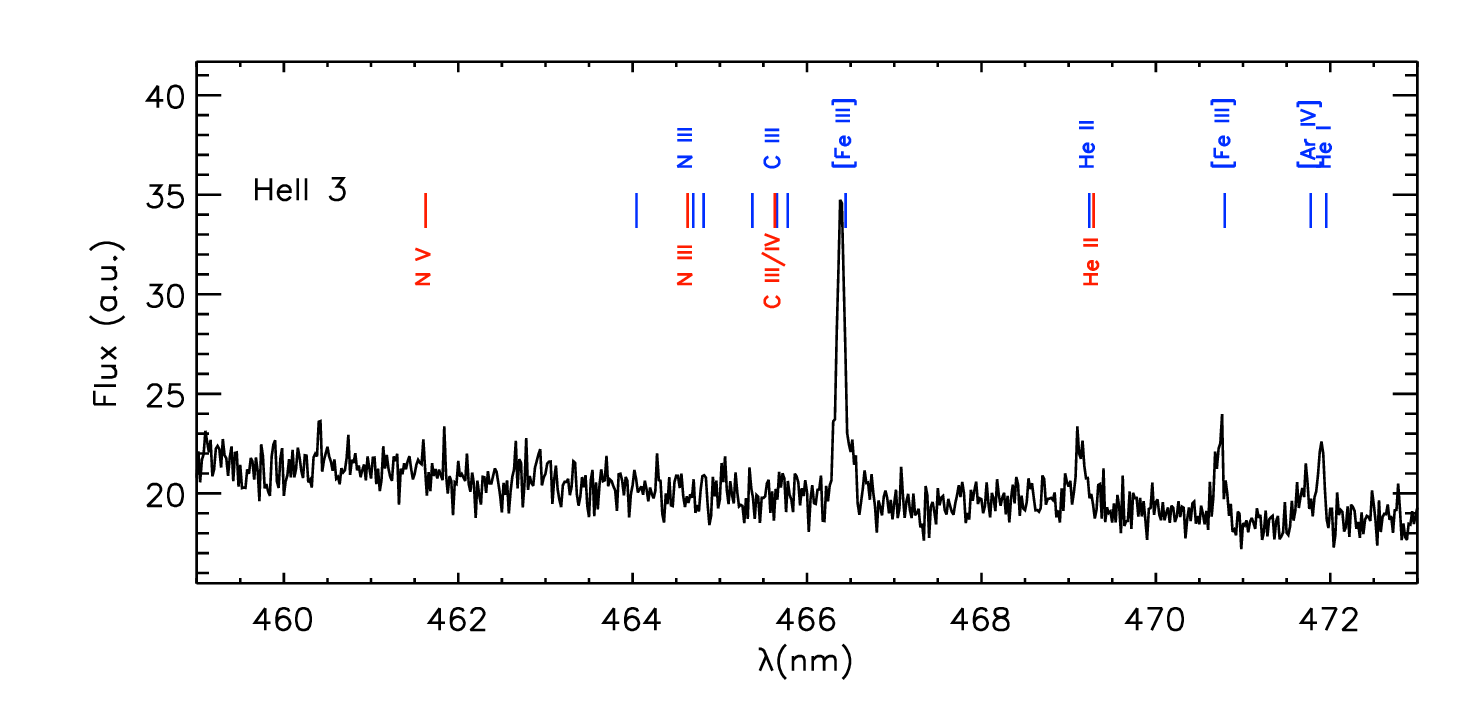}
\includegraphics[angle=0,width=0.45\textwidth, clip=,bbllx=65, bblly=68,
bburx=740, bbury=375]{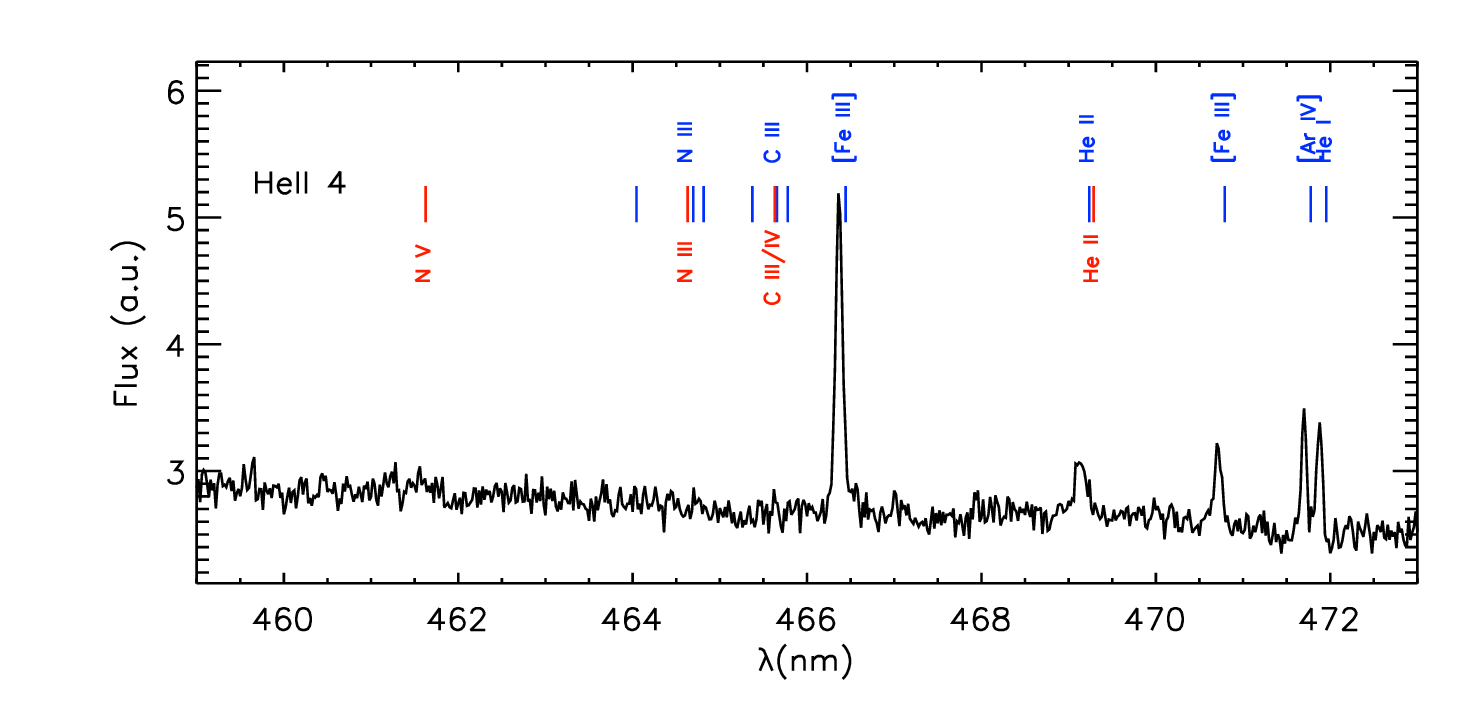}
\end{center}
   \caption[Spectra showing nebular \ion{He}{ii}]
  {Spectra showing nebular \ion{He}{ii}, but no W-R features. The
  positions of the nebular 
  emission lines have been indicated with blue ticks and labels, while
  those corresponding to Wolf-Rayet features appear in red. \label{especheiineb}}
 \end{figure*}

\subsection{Nebular \ion{He}{ii} and helium abundance \label{seche}}

The hypothesis that the W-R population is the cause of the nitrogen
enrichment in NGC~5253 requires an enhancement of the helium abundance
too \citep[e.g.][]{sch96}. This is nicely illustrated in \citet{kob97}
where different linear relations between the nitrogen and helium
abundances (N/H and He/H) are presented according to different
scenarios of nitrogen enrichment (W-Rs, PNe, etc.). The only
  scenario able to explain an extra quantity of nitrogen in the ISM
  without any extra helium counterpart would be the one where this
  nitrogen is caused during the late O-star wind phase.

\begin{figure}
   \centering
\includegraphics[angle=0,scale=.43, clip=,bbllx=40, bblly=0,
bburx=600, bbury=355]{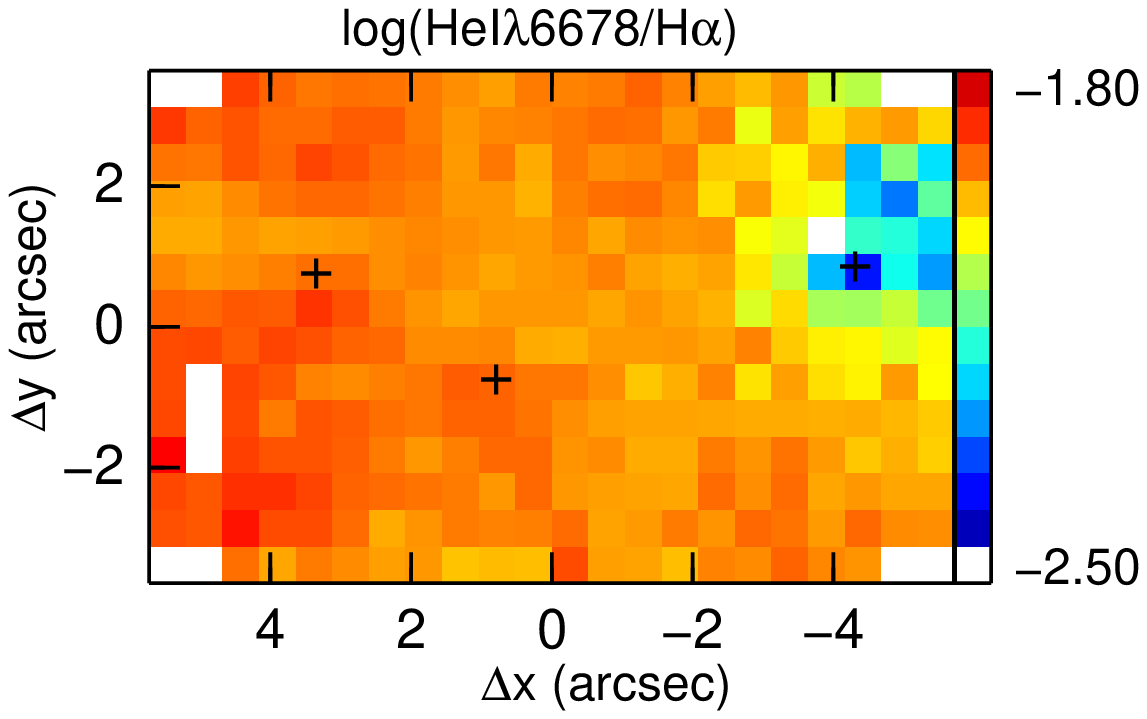}
   \caption[$\log(\ion{He}{i}\lambda$6678/H$\alpha$) line ratio map]{$\log(\ion{He}{i}\lambda$6678/H$\alpha$) line ratio map.  The position of the three
     peaks in the continuum map are shown for
     reference.\label{mapa_he1l6678ha}} 
 \end{figure}

As in previous sections, we can measure at each spaxel
the total helium abundance and compare it with that for nitrogen.  
Since lines like [\ion{O}{ii}]$\lambda\lambda$3726,3728
did not fall in the covered spectral range, we did not determine the
absolute nitrogen abundance. Instead, we used the mean of the
abundances determined by \citet{kob97} for their \ion{H}{ii}-1 and
\ion{H}{ii}-2 (N/H$\sim2.0\times10^{-5}$) to estimate how much helium
would be needed in the enriched areas, if the extra nitrogen were caused by
W-Rs (i.e. He/H$\sim0.12$). For the non N-enriched areas, we can use 
the measurement at UV-1 (N/H$\sim0.7\times10^{-5}$) which requires
He/H$\sim0.09$. Helium abundance can be determined as:

\begin{equation}
He/H = \mathrm{icf} \times (He^{+}/H^{+} + He^{++} / H^{+})
\end{equation}

\noindent
where icf is a correction factor due to the presence of neutral
helium. We assumed $icf\sim1.0$, which is consistent with
the predictions of photoionization models for our measured \ohb\ line
ratios \citep{hol02}.
Since the \ion{He}{i}$\lambda$6678 was detected in every spaxel of the
FLAMES field of view and with good S/N, for the purpose of this
work, we determined $y^{+} =
\rm{He}^{+}/\rm{H}^{+}$ from the \heiha\ line ratio using the expression

\begin{equation}
y^{+} = 2.58 t^{0.25} (2.87\,\ion{He}{i}\lambda6678/\mathrm{H}\alpha)
\end{equation}

\noindent
where $t$ is the electron temperature in units of $10^4$~K
\citep{pag92}. As in section \ref{secdensidad}, we assumed $T_e = 11\,650$~K.
Figure \ref{mapa_he1l6678ha} shows the 2D structure of the \heiha\
line ratio. It is relatively uniform with the exception 
of some spaxels in the upper right corner, close to Complex
\#3. This area, relatively far from the main photo-ionization
source and with low \ohb\ line ratio would be the only region where
one can expect a substantial contribution of neutral helium.

\begin{figure}
   \centering
\includegraphics[angle=0,width=0.42\textwidth, clip=,bbllx=40, bblly=10,
bburx=550, bbury=400]{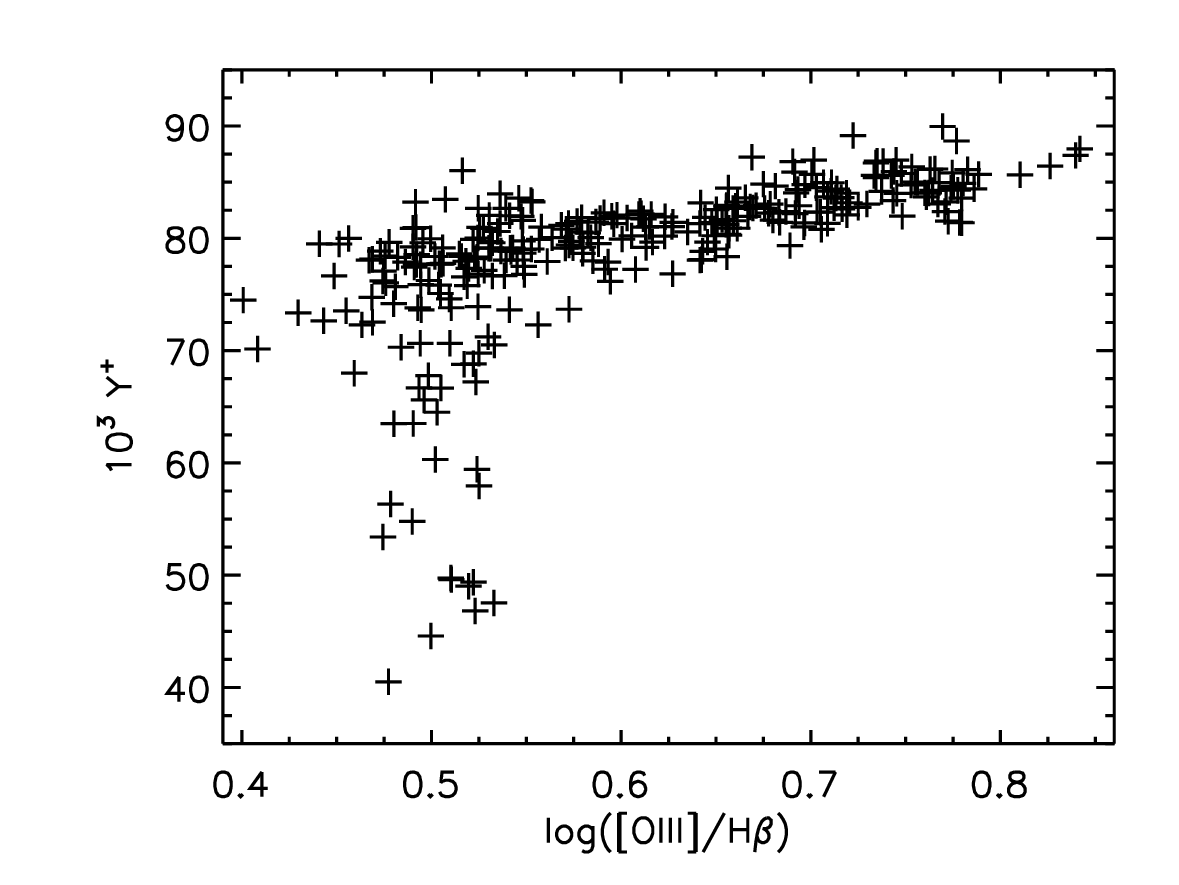}
   \caption[He$^+$ abundance vs. \ohb.]{He$^+$ abundance vs. \ohb\ for
     each individual spaxel.\label{ymasvso3ghb}} 
 \end{figure}

Figure \ref{ymasvso3ghb} presents the derived He$^+$ abundances vs. 
the \ohb\ line ratio. With the exception of the data corresponding to
the right upper corner of the FLAMES f.o.v., He$^+$/H$^+$ range between
0.075 and 0.090, being higher in the higher excitation zones (i.e. the giant
\ion{H}{ii} region). These values are in agreement with previous
measurements of  He$^+$/H$^+$ in specific areas \citep{pag92,kob97,wal89}.
They are consistent with a scenario without extra N-enrichment
and still far, by a factor $\sim1.3-1.7$, from the
required $\sim$0.12 \emph{total} helium abundance in the W-R
scenario, in particular in the areas enriched with nitrogen. 

   \begin{figure}
   \centering
\includegraphics[angle=0,width=0.42\textwidth, clip=,bbllx=40, bblly=10,
bburx=550, bbury=400]{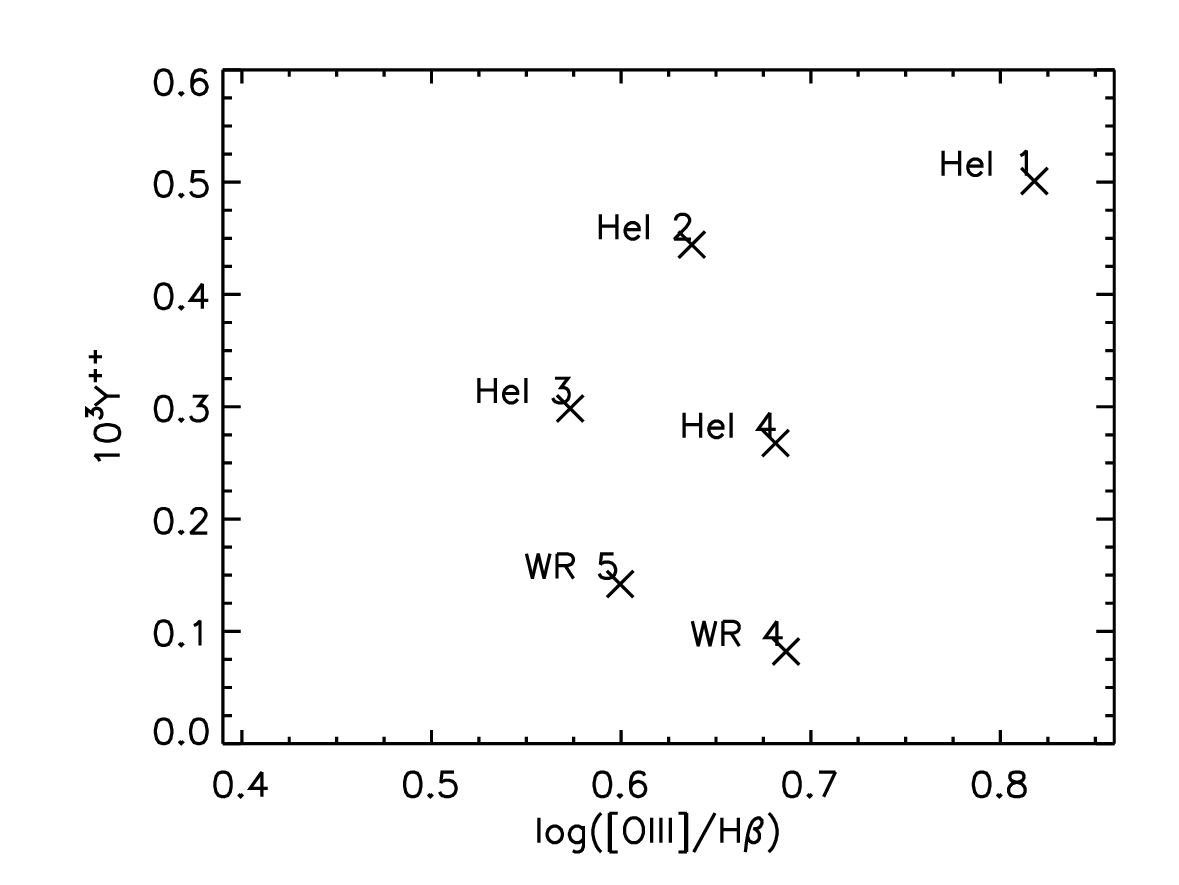}
   \caption[He$^{++}$ abundance vs. \ohb.]
  {He$^{++}$ abundance vs. \ohb\ for each of the extracted spectra
  with \ion{He}{ii}$\lambda$4686 detection (dotted regions in Figure 
  \ref{nitroalto}). \label{ymasmasvso3ghb}} 
 \end{figure}

What about the He$^{++}$/H$^+$, whose abundance can be determined via the
nebular \ion{He}{ii}$\lambda$4686 emission line? 
This line turned out to be rather
elusive. \citet{cam86} mentioned a possible detection in their regions B
and C.  This result, however, has not been confirmed afterwards
\citep[see][and references therein]{lop07}.
As with the W-R features, we looked for the \ion{He}{ii} nebular
line in each individual spectrum. Those that presented a spatial
continuity were taken to define an area and were co-added before extracting. 
The selected areas are marked in Figure \ref{nitroalto} with dotted lines.
The co-added and extracted spectra of each individual region appear in
Figure \ref{especheiineb}.
In addition to these regions, as mentioned in section \ref{secwr}, 
nebular \ion{He}{ii} in the W-R~4 and W-R~5 has also been detected.

These detections are, in general, neither associated with the area of
nitrogen enhancement nor with those presenting W-Rs features.
This lack of coincidence seems difficult to reconcile with a scenario
where this enhancement, and the existence of He$^{++}$, share a common
origin. Moreover, for the purpose of this work, we 
estimated the $\ion{He}{ii}$ abundances in this areas using: 

\begin{equation}
y^{++} = 0.084 t^{0.14} (\ion{He}{ii}\lambda4686/\mathrm{H}\beta)
\end{equation}


\noindent
from \citet{pag92}. Derived values for the individual regions are
shown Figure \ref{ymasmasvso3ghb} as a function of the \ohb\ line
ratio. They range between 0.0001 and 0.0005. Although
  uncertainties are large, up to 0.0006, due to the weakness of the
  \ion{He}{ii}$\lambda$4686 emission line, these values are clearly
  very far from the values of $\sim0.030 - 0.050$ required to bring the 
  helium abundance up to $\sim$0.12 in the W-R enrichment scenario. 
  Given the depth and continuous
  mapping of the present data, we can clearly exclude the possibility
  of further detections of larger quantities of He$^{++}$ based on
  optical observations. Thus the present data support 
the scenario suggested by \citet{kob97} where the N-enrichment should 
arise during the late O-star wind phase. 
In view of the extra-nitrogen distribution and the extinction map,
the only place where these larger quantities of He$^{++}$  could be
found (if they existed) is in Complex \#1. However, they should be
highly extinguished and would required a search for emission lines at
longer wavelengths as for example \ion{He}{ii} (7-10) at 21\,891~\AA~
\citep{hor99}.  

Since the nebular \ion{He}{ii} is not associated with the area
  showing N-enhancement, there is still the open question as to its
  origin. \citet{gar91} explored the different mechanisms capable of
  producing this emission in extragalactic \ion{H}{ii} regions. The
  first suggestion is photoionization by fast shocks. However, we have
  seen in section \ref{secionistruc}, that shocks do not appear to play
  a dominant role in the central parts of \object{NGC~5253}. Moreover, the
  measured $\log$(\ion{He}{II}$\lambda$4686/\hb) are $\sim-3.0$ to $-2.1$, much
  lower than those predicted by shocks models with $N_e=1$cm$^{-3}$
  and LMC abundances \citep[$\sim-1.4$ to $-0.4$,][]{all08}.

Another possibility discussed by \citet{gar91} is hot
  ($T\lsim70\,000$~K) 
  stellar ionizing continua. This looks like a plausible explanation for those
  cases where we had detected nebular \ion{He}{ii}$\lambda$4686 on top of the
  \emph{blue bump} (WR~4 and WR~5).

The last option would be photoionization caused by
  X-rays. The only point sources detected by \citet{sum04} that fall in
  our f.o.v. are sources 17, 18, and 19. This last source appears to be
  associated with the Complex \#1 and thus, not related to this
  discussion (since no nebular \ion{He}{ii} was detected for this region).
  Sources 18 and 17 could, however, be associated with \ion{He}{ii}-1 and
  \ion{He}{ii}-4 detections, respectively. In particular, the latter region
  coincides with the secondary peak of emission in the \ha\ image and is
  associated with cluster 17 of the sample catalogued by
  \citet{har04}. No satisfactory explanation was found
    for the cause of the ionization at \ion{He}{ii}-2 and \ion{He}{ii}-3.

\subsection{Kinematics of the ionized gas \label{seccinematica}}

\begin{figure*}[htb!]
 \centering
\includegraphics[angle=0,width=0.27\textwidth, clip=,bbllx=40, bblly=40,
bburx=525, bbury=760]{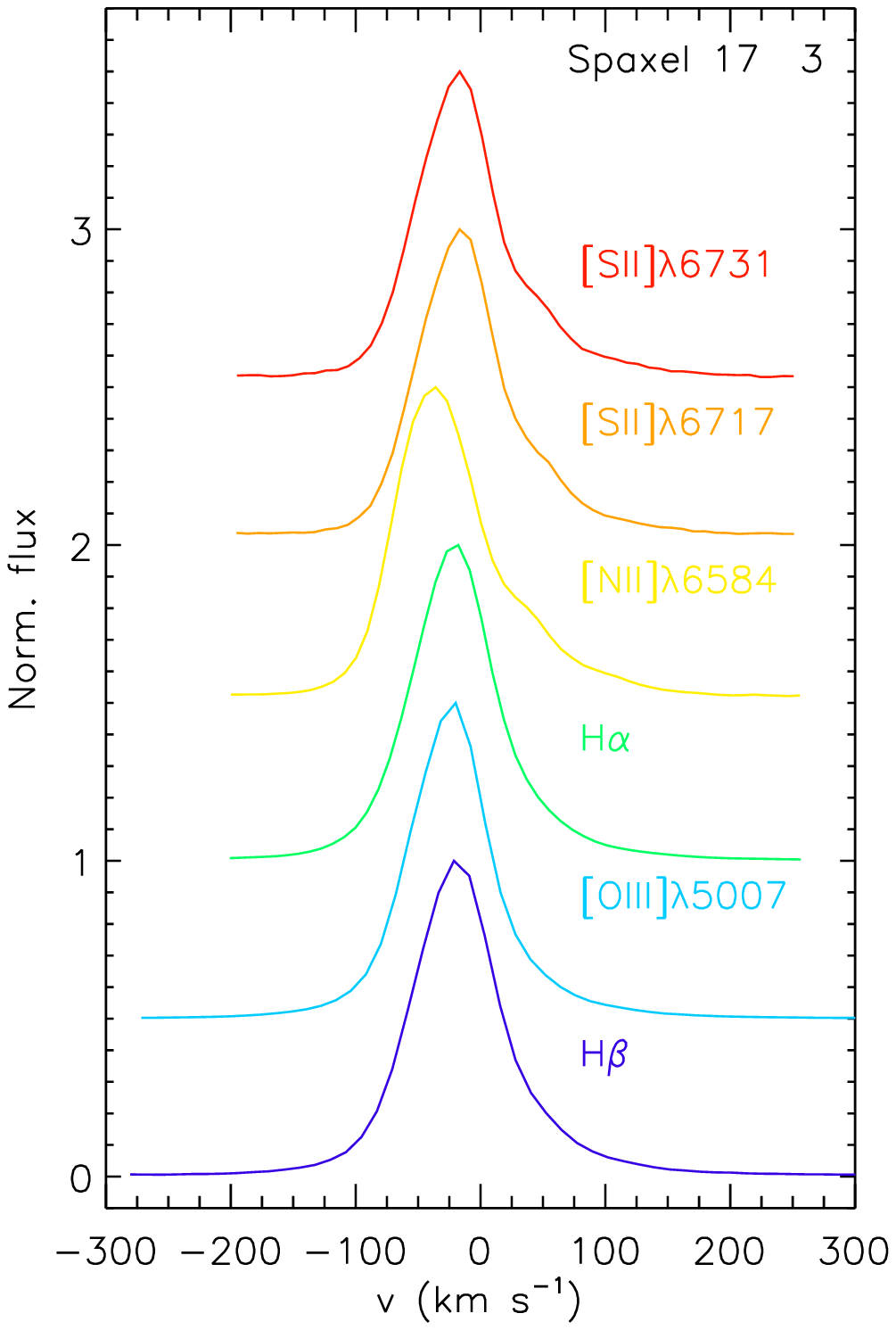}
\includegraphics[angle=0,width=0.27\textwidth, clip=,bbllx=40, bblly=40,
bburx=525, bbury=760]{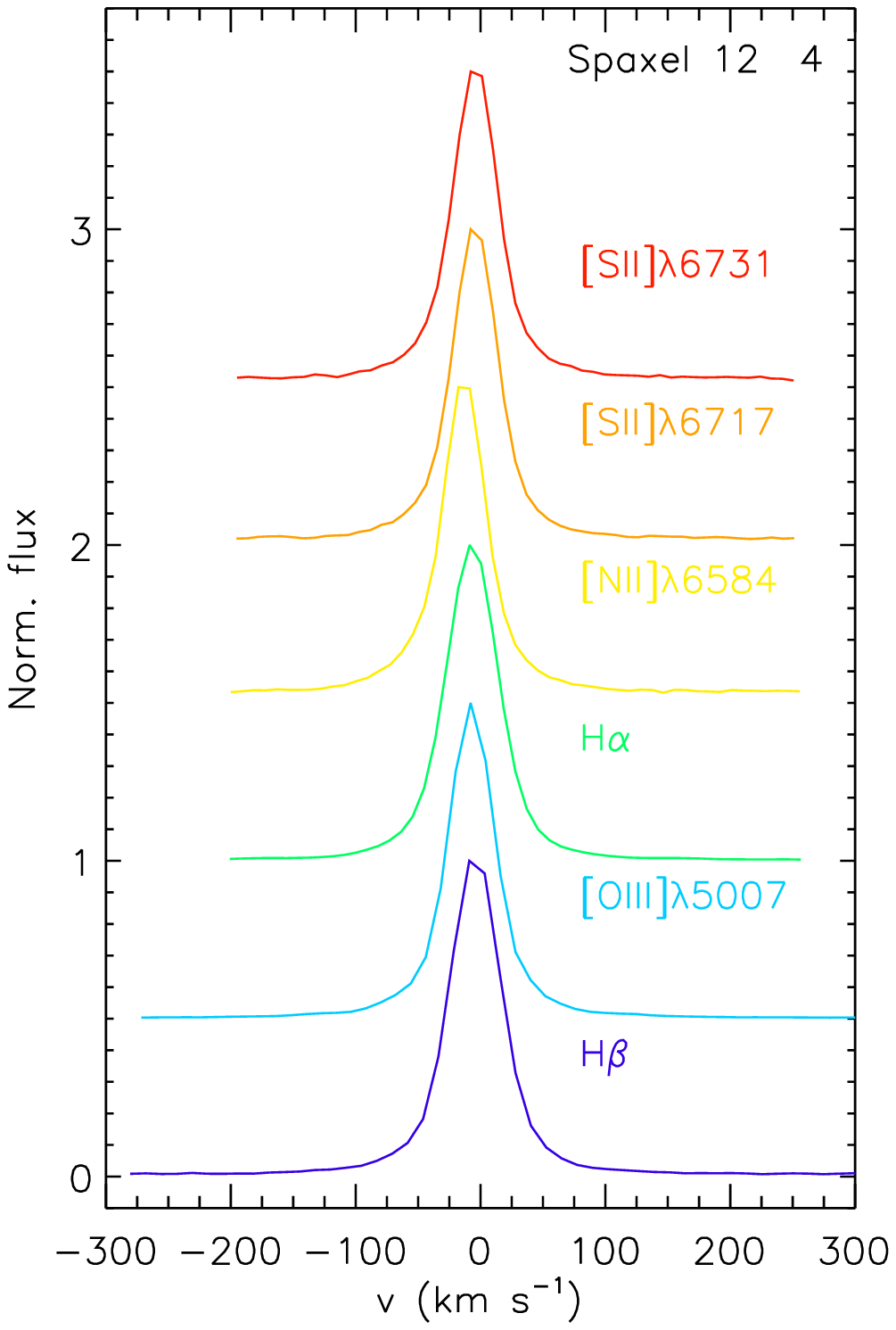}
\includegraphics[angle=0,width=0.27\textwidth, clip=,bbllx=40, bblly=40,
bburx=525, bbury=760]{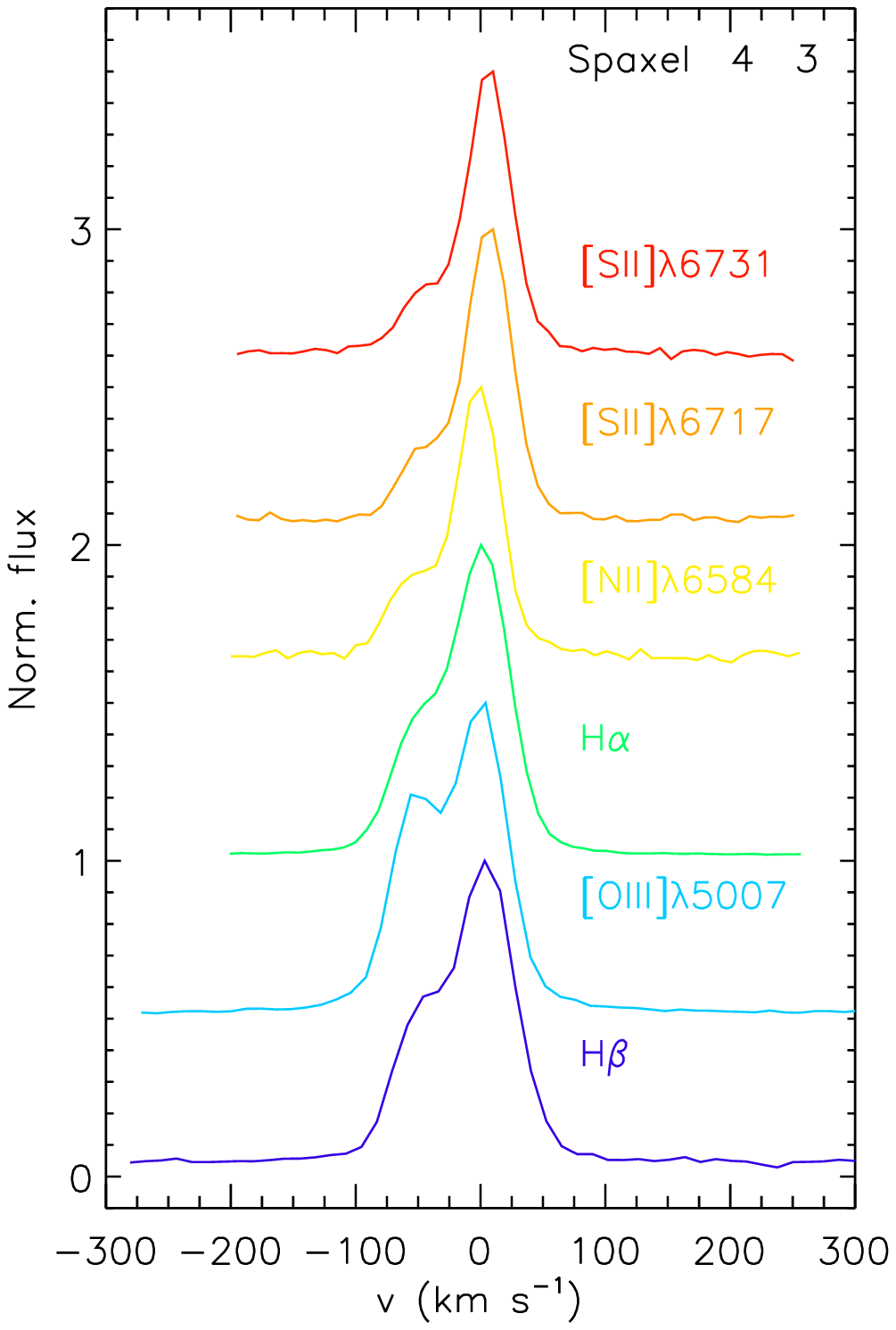}
\includegraphics[angle=0,width=0.27\textwidth, clip=,bbllx=40, bblly=40,
bburx=525, bbury=760]{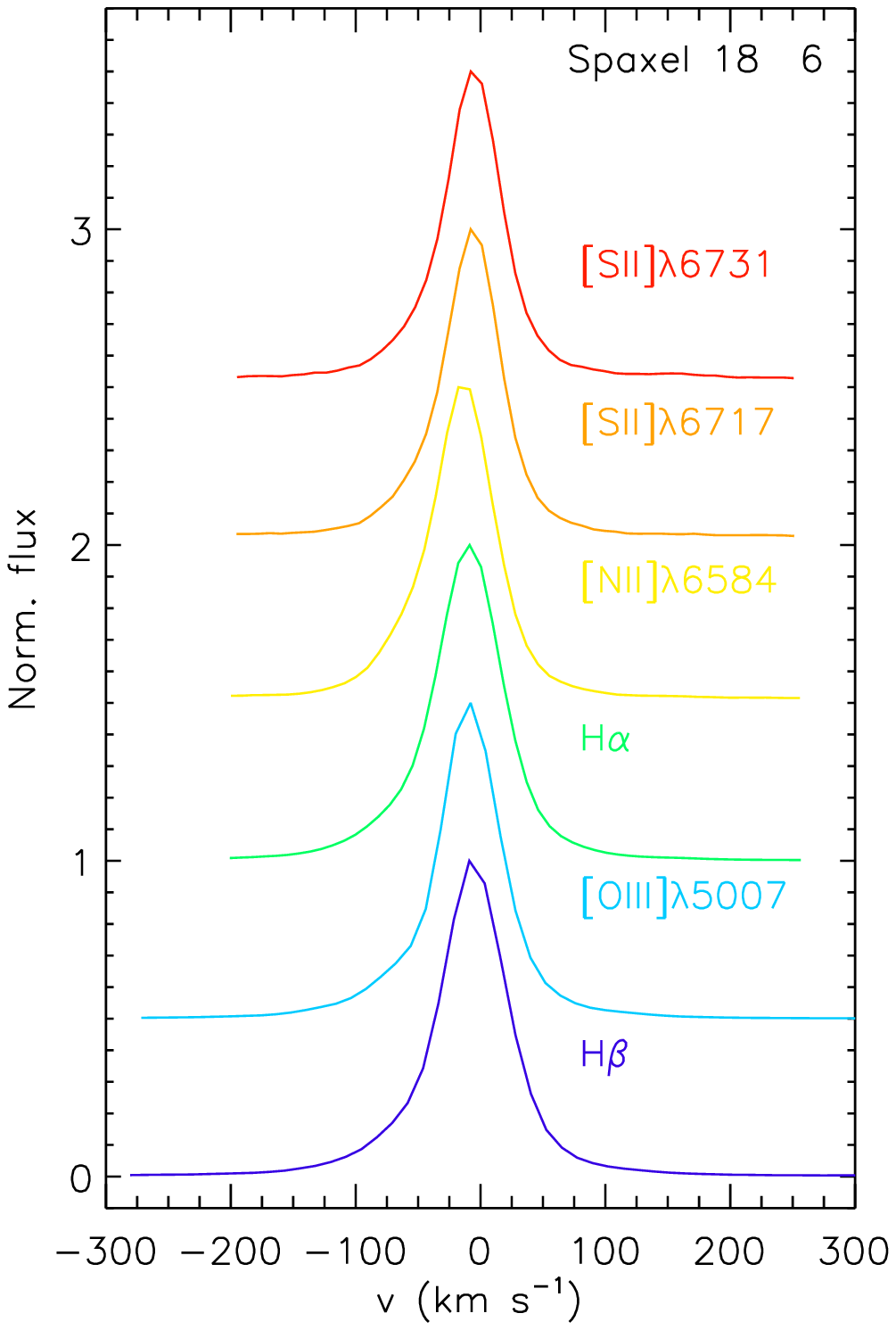}
\includegraphics[angle=0,width=0.27\textwidth, clip=,bbllx=40, bblly=40,
bburx=525, bbury=760]{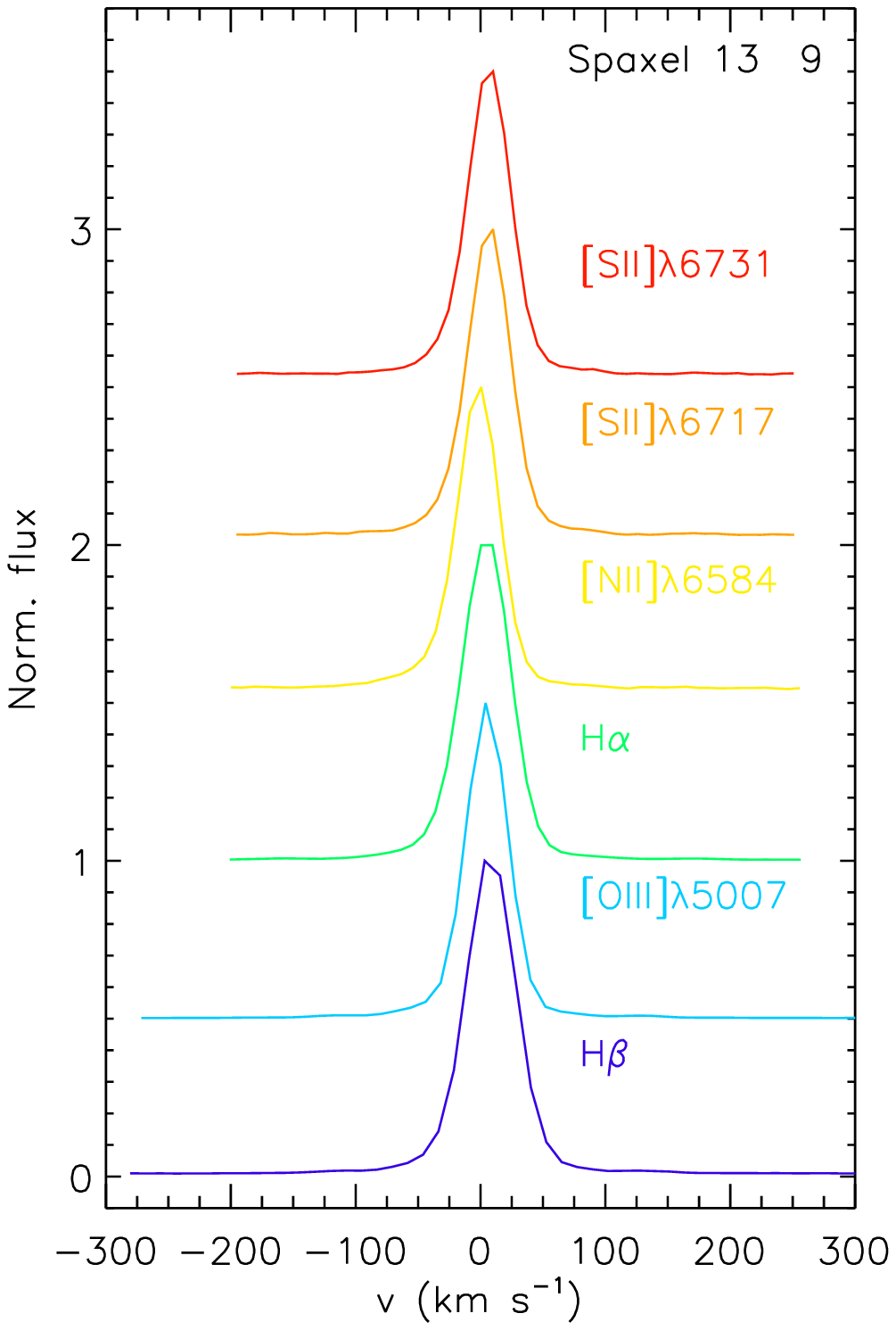}
\includegraphics[angle=0,width=0.27\textwidth, clip=,bbllx=40, bblly=40,
bburx=525, bbury=760]{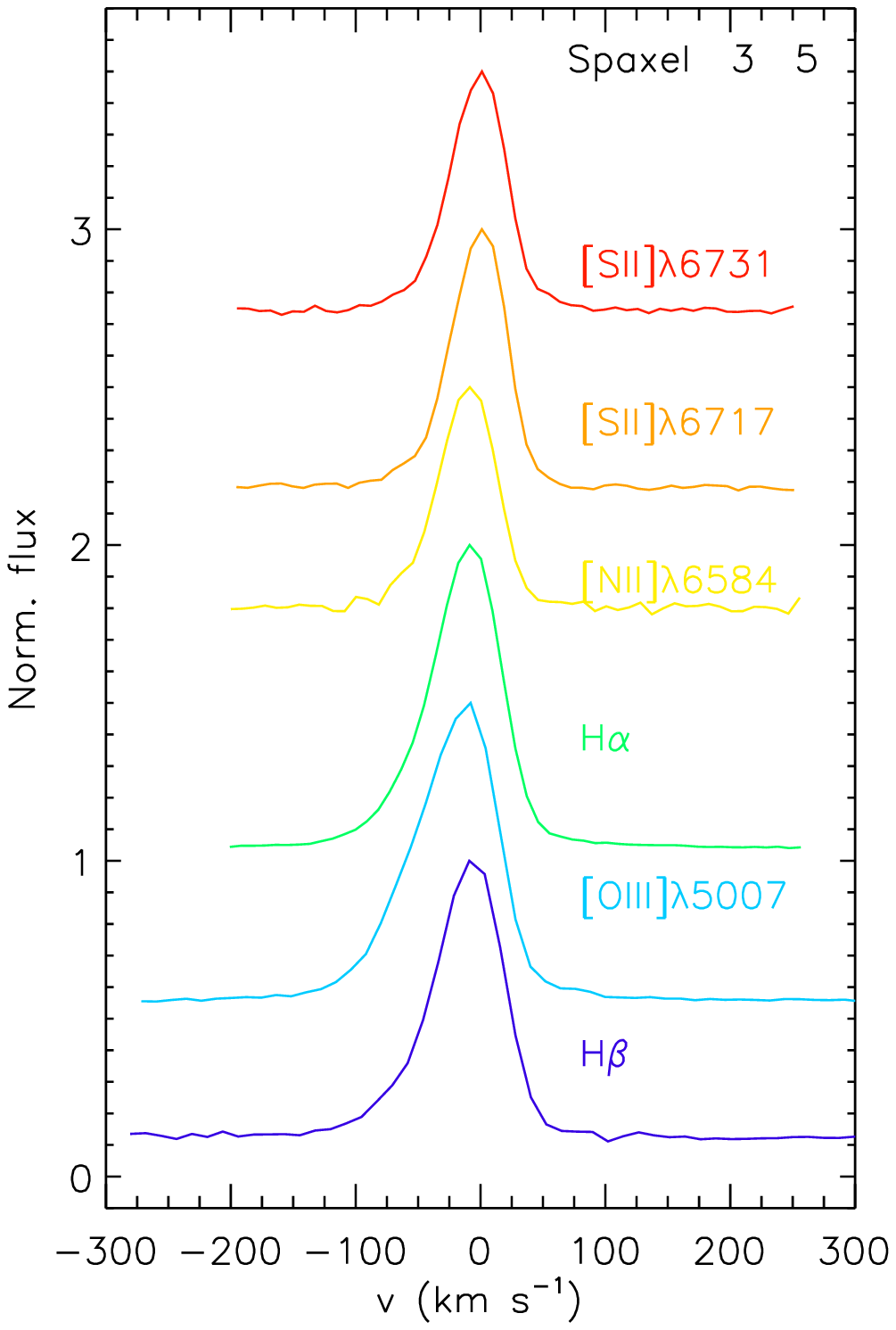}
\includegraphics[angle=0,width=0.27\textwidth, clip=,bbllx=40, bblly=40,
bburx=525, bbury=760]{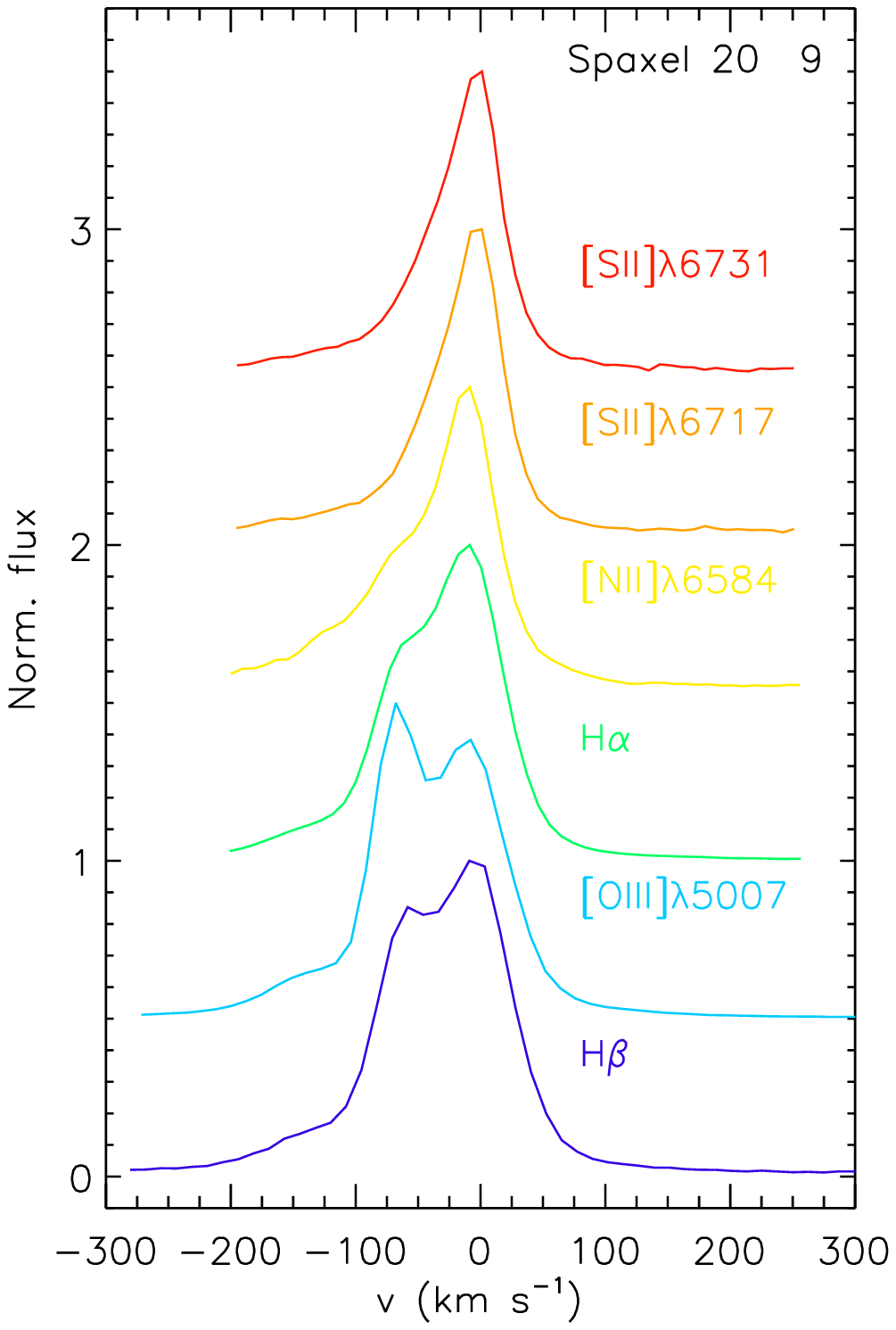}
\includegraphics[angle=0,width=0.27\textwidth, clip=,bbllx=40, bblly=40,
bburx=525, bbury=760]{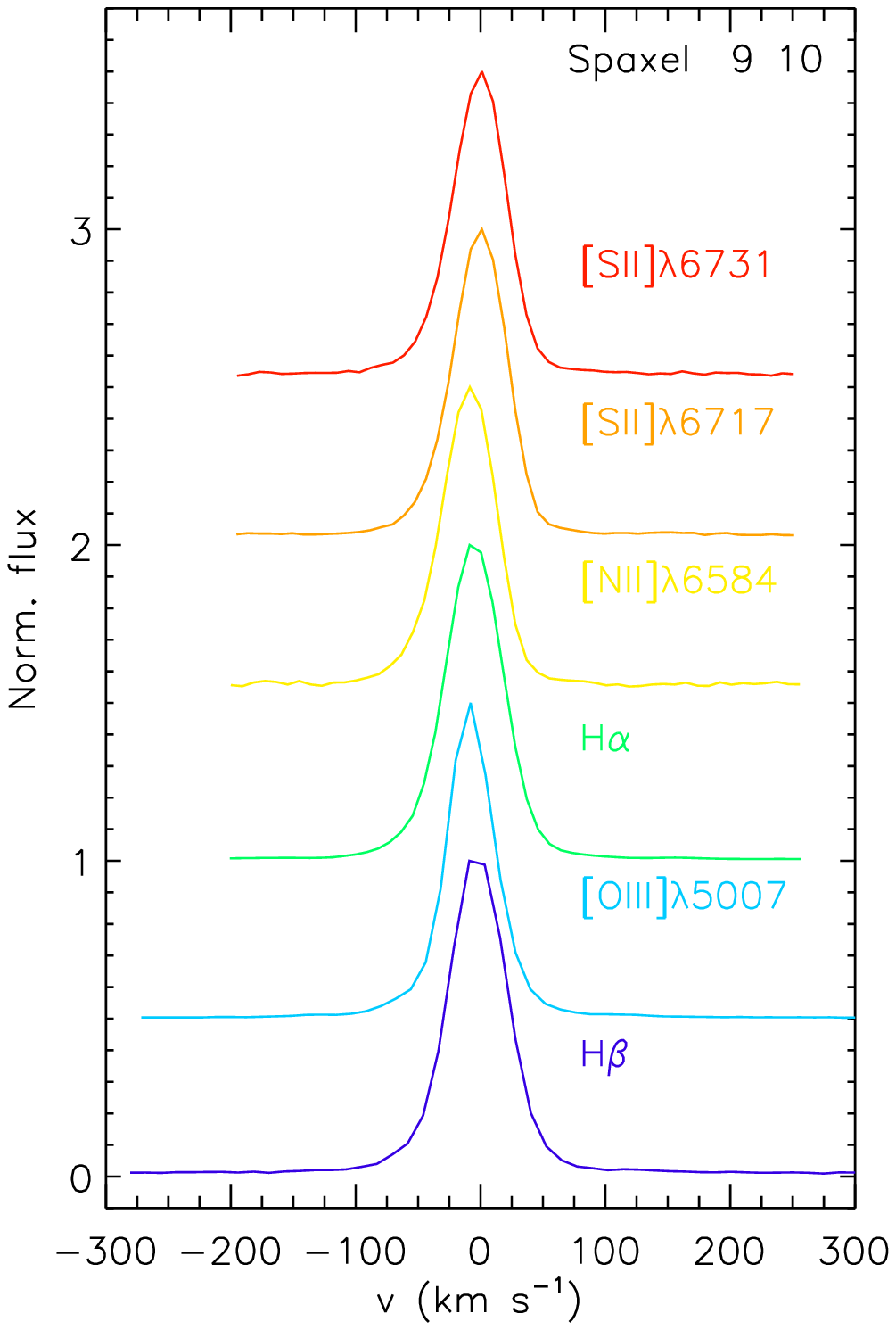}
\includegraphics[angle=0,width=0.27\textwidth, clip=,bbllx=40, bblly=40,
bburx=525, bbury=760]{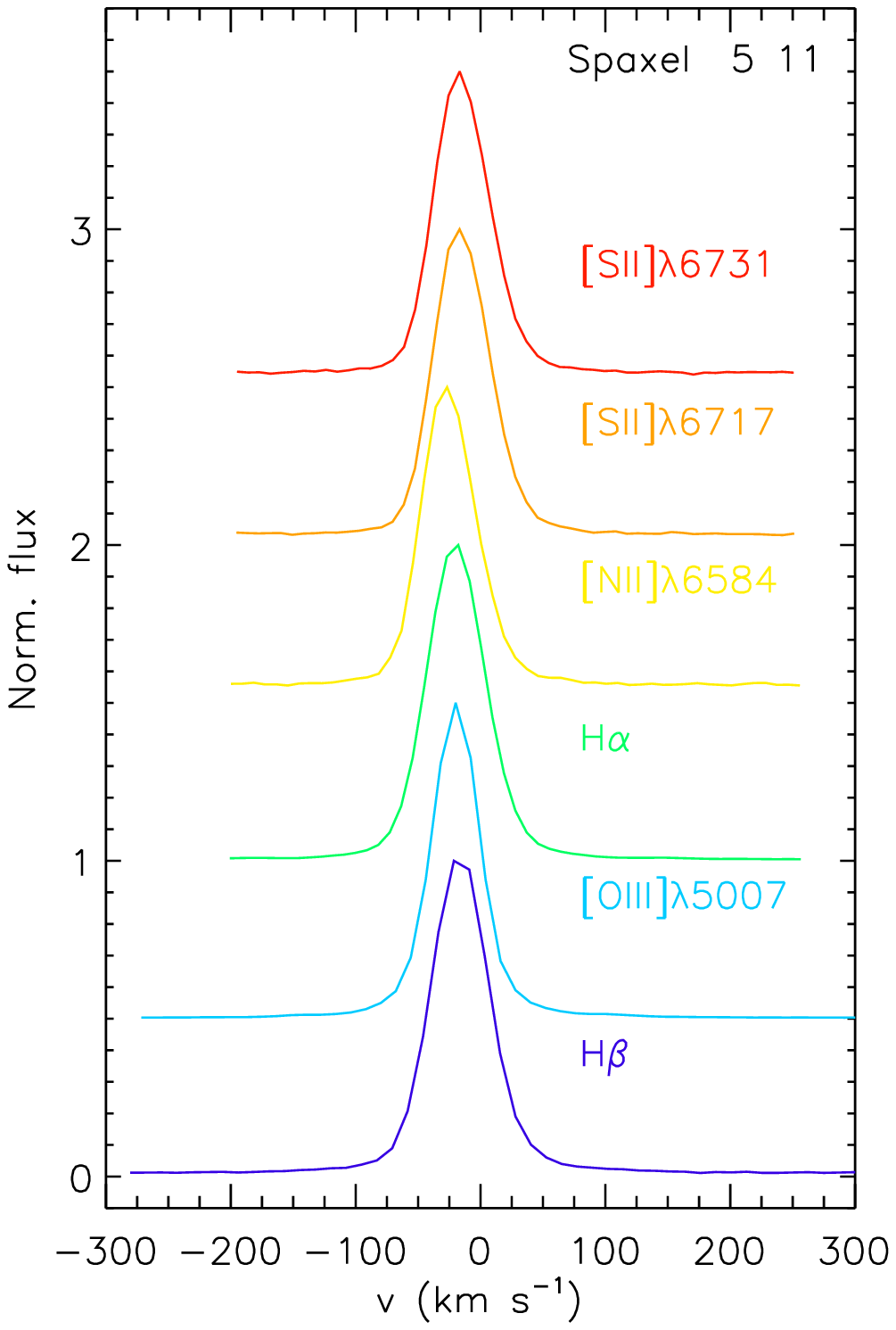}
   \caption[]{Profiles of the main emission lines for representative
     spectra. The distribution within the FLAMES f.o.v. has
      been roughly retained. That is: spaxel (1,1) is located
        at the upper right corner of the FLAMES IFU and numbering
        increases towards the left and bottom. The middle row contains
      the profiles  
      corresponding to the positions of the peaks
     of continuum emission (left: Complex \#1; centre: Complex \#2;
     right: Complex \#3). Spectra in the left column correspond to
     spaxels associated with the giant \ion{H}{ii} region (upper:
     extension towards the north-west; lower: extension towards the
     south-east), those in the 
     central column to areas in the centre of the FLAMES f.o.v. and
     those in the right column to the area associated with Complex
     \#3 (upper: position towards the
     north-west; lower: towards the south-east).
\label{ex_spa}}
 \end{figure*}

\begin{figure*}[htb!]
 \centering
\includegraphics[angle=0,width=0.31\textwidth, clip=,bbllx=60, bblly=61,
bburx=338, bbury=248]{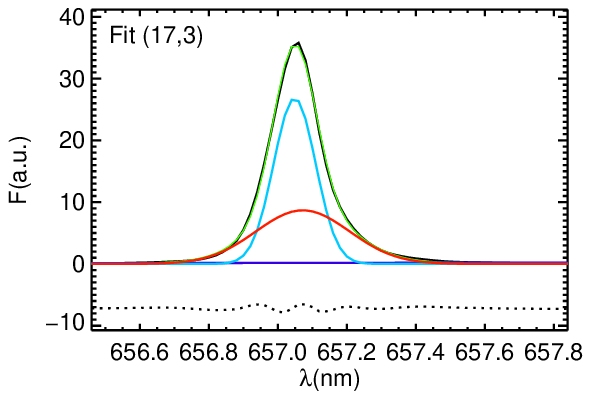}
\includegraphics[angle=0,width=0.31\textwidth, clip=,bbllx=60, bblly=61,
bburx=338, bbury=248]{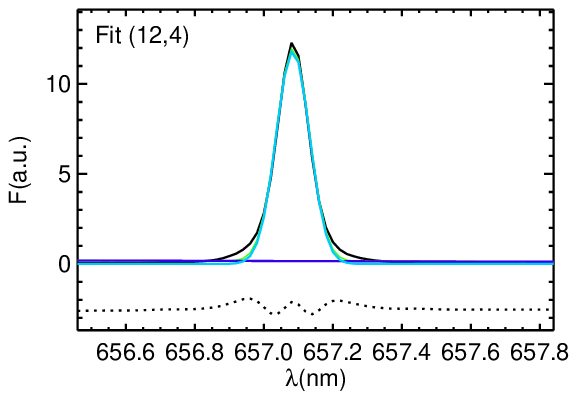}
\includegraphics[angle=0,width=0.31\textwidth, clip=,bbllx=60, bblly=61,
bburx=338, bbury=248]{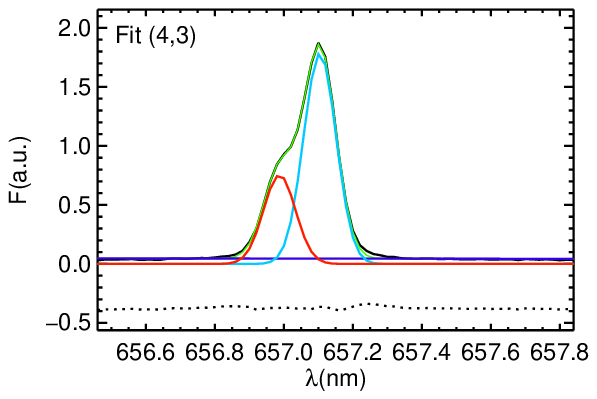}
\includegraphics[angle=0,width=0.31\textwidth, clip=,bbllx=60, bblly=61,
bburx=338, bbury=248]{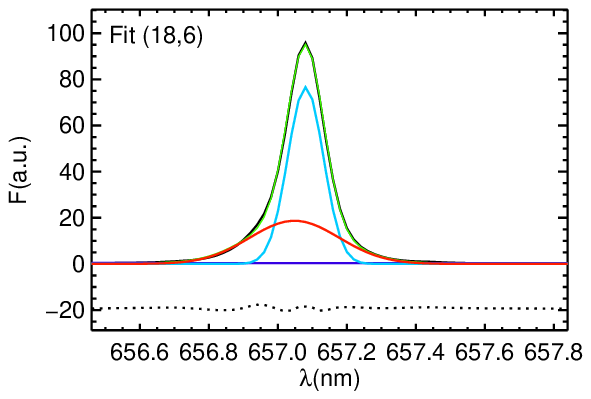}
\includegraphics[angle=0,width=0.31\textwidth, clip=,bbllx=60, bblly=61,
bburx=338, bbury=248]{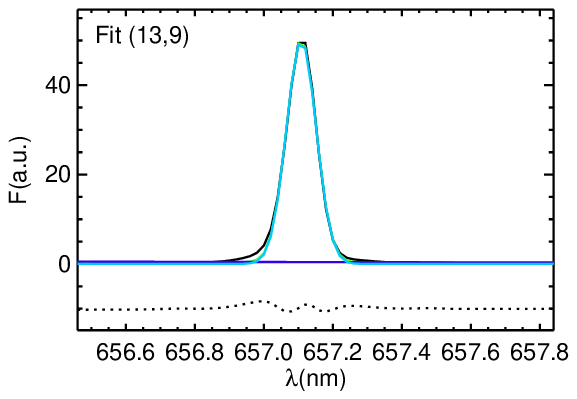}
\includegraphics[angle=0,width=0.31\textwidth, clip=,bbllx=60, bblly=61,
bburx=338, bbury=248]{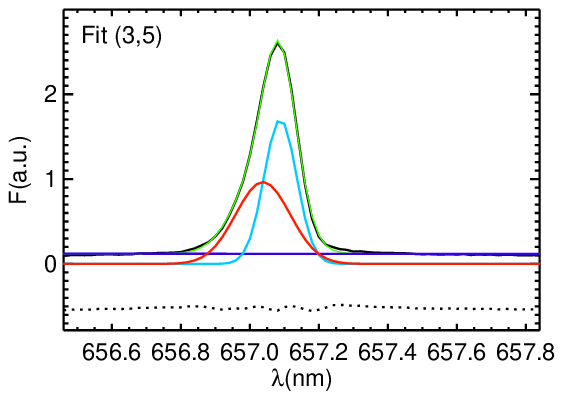}
\includegraphics[angle=0,width=0.31\textwidth, clip=,bbllx=60, bblly=61,
bburx=338, bbury=248]{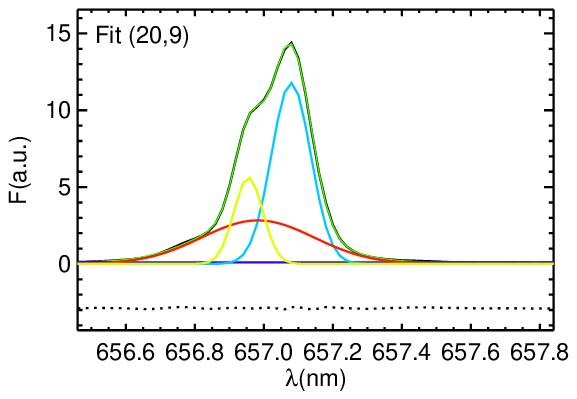}
\includegraphics[angle=0,width=0.31\textwidth, clip=,bbllx=60, bblly=61,
bburx=338, bbury=248]{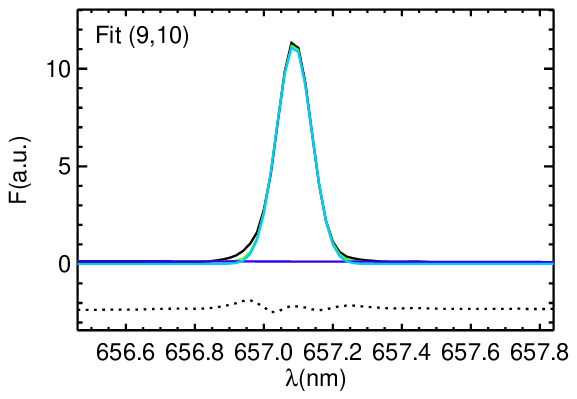}
\includegraphics[angle=0,width=0.31\textwidth, clip=,bbllx=60, bblly=61,
bburx=338, bbury=248]{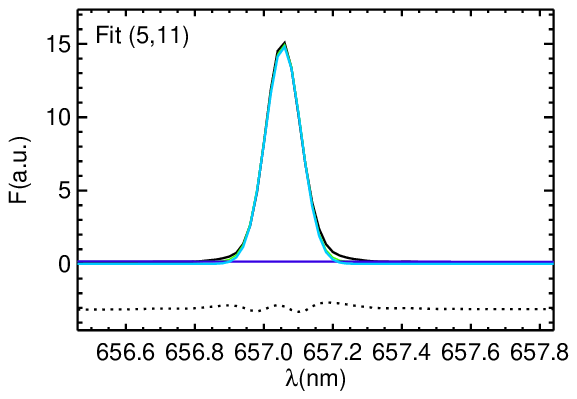}
  \caption[]{Fits in \ha\ for the representative spectra shown
      in Figure \ref{ex_spa}.  Flux is in arbitrary
      units. The observed 
      spectra are presented in black. The first, second and third
      components have been displayed in blue, red, 
      and yellow, respectively, while the underlying continuum and
      total fit are in violet and green. Residuals are shown as a
      dotted line below the profile.
\label{ex_fit}}
 \end{figure*}

\begin{figure*}[htb!]
 \centering
\includegraphics[angle=0,width=0.95\textwidth, clip=,bbllx=30, bblly=150,
bburx=510, bbury=450]{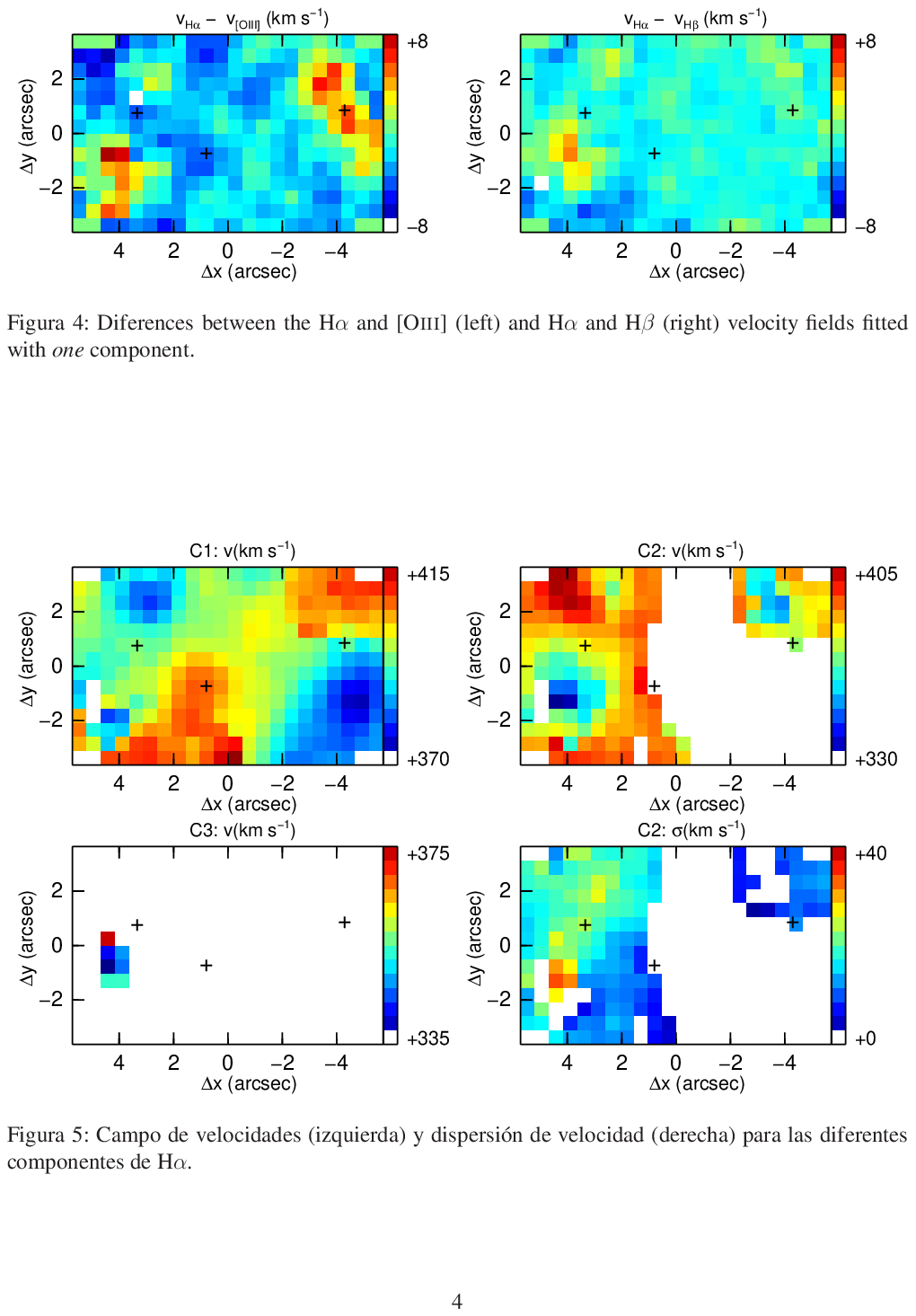}
   \caption[]{Kinematic information derived from the \ha\ emission
     line. \emph{Left:} Velocity fields for the two narrow
     components. \emph{Right:} Velocity field (upper) and velocity
     dispersion map (lower) for the broad component. \label{cinematica}}
\end{figure*}

Slit observations in specific regions of this galaxy have demonstrated
that the kinematics of the ionized gas is rather complex, with
line profiles revealing asymmetric wings
\citep[e.g.][]{mar95,lop07}. These observations usually include the
bright core of the galaxy. However, they might be biased since typically
the long slits only sample specific regions selected by the 
particular slit placement. The present data permit a
2D spatially resolved analysis of the kinematics of the ionized gas in
the central area of the galaxy to be performed, thus overcoming this 
drawback. We based our analysis on the strongest emission lines 
(i.e. mainly \ha, but also \hb, \oiii, \nii, and \sii) where the high
S/N permits the line profiles to be fitted with a high degree of accuracy.

In the following, we will present the results derived from
\ha. Similar results were obtained from the  \hb\  
and \oiii\ emission lines with differences of $|\Delta
v|\lsim2$~km~s$^{-1}$ in most cases and always between -5 and 5 km~s$^{-1}$.
Results for the only emission lines with remarkable differences in the
velocity maps (i.e.  \nii\ and \sii) will be presented, as well.

Typical examples of the different profiles for the main
  emission lines are shown in Figure \ref{ex_spa}. Lines are ordered by
  wavelength from bluer (lower) to red (upper) within each panel. The
  zero point in the 
  abscissa axis corresponds to the measured systemic velocity which is
  defined as the average of the velocities derived from the main
  emission lines for the peak of the continuum emission.
We performed an independent fit for each of the brightest
  emission lines using MPFITEXPR (see section \ref{linefitting} for details).
In general, the line profiles of the individual spectra cannot be
properly reproduced by a single component. A large percentage of them
needed two (and even three) independent components to 
reproduce the observed profile reasonably well. We followed the
approach of keeping the analysis as simple as possible. Thus in those
cases where both fits - the one with one component and the one with several
components - reproduced equally well the line profile, we gave
preference to the fit with one component.
Examples of these fits are shown in Figure \ref{ex_fit}, which
  contains the \ha\ emission line together with the total fit and the
  individual components overplot for the spaxels shown in Figure \ref{ex_spa}. 
Note how the fits for the spaxels in the central area of our
  f.o.v. present relatively larger residuals. These can be attributed
  to a low surface brightness broad extra component which would be the
  subject of a future work.
Central wavelengths were translated into heliocentric velocity
taking into account the radial velocity induced by the Earth's motion
at the time of the observation which was evaluated using the IRAF task
\texttt{rvcorrect}. Velocity dispersions were obtained from the
measured FWHM after correcting for the instrumental width and 
thermal motions. The width of the thermal profile was derived assuming
$T_{\rm e}$=11\,650~K which translates into a $\sigma_{\rm
  ther} = \sqrt{k T_{\rm e}/m_{\rm H}}$ of $\sim $11~km~s$^{-1}$ for
the hydrogen lines. 
The measured systemic velocity  was 392~km~s$^{-1}$. This is slightly lower
  than the one measured from neutral hydrogen
  \citep[407~km~s$^{-1}$,][]{kor04} according to NED.

\begin{figure*}[htb!]
 \centering
\includegraphics[angle=0,width=0.30\textwidth, clip=,bbllx=-10, bblly=10,
bburx=595, bbury=715]{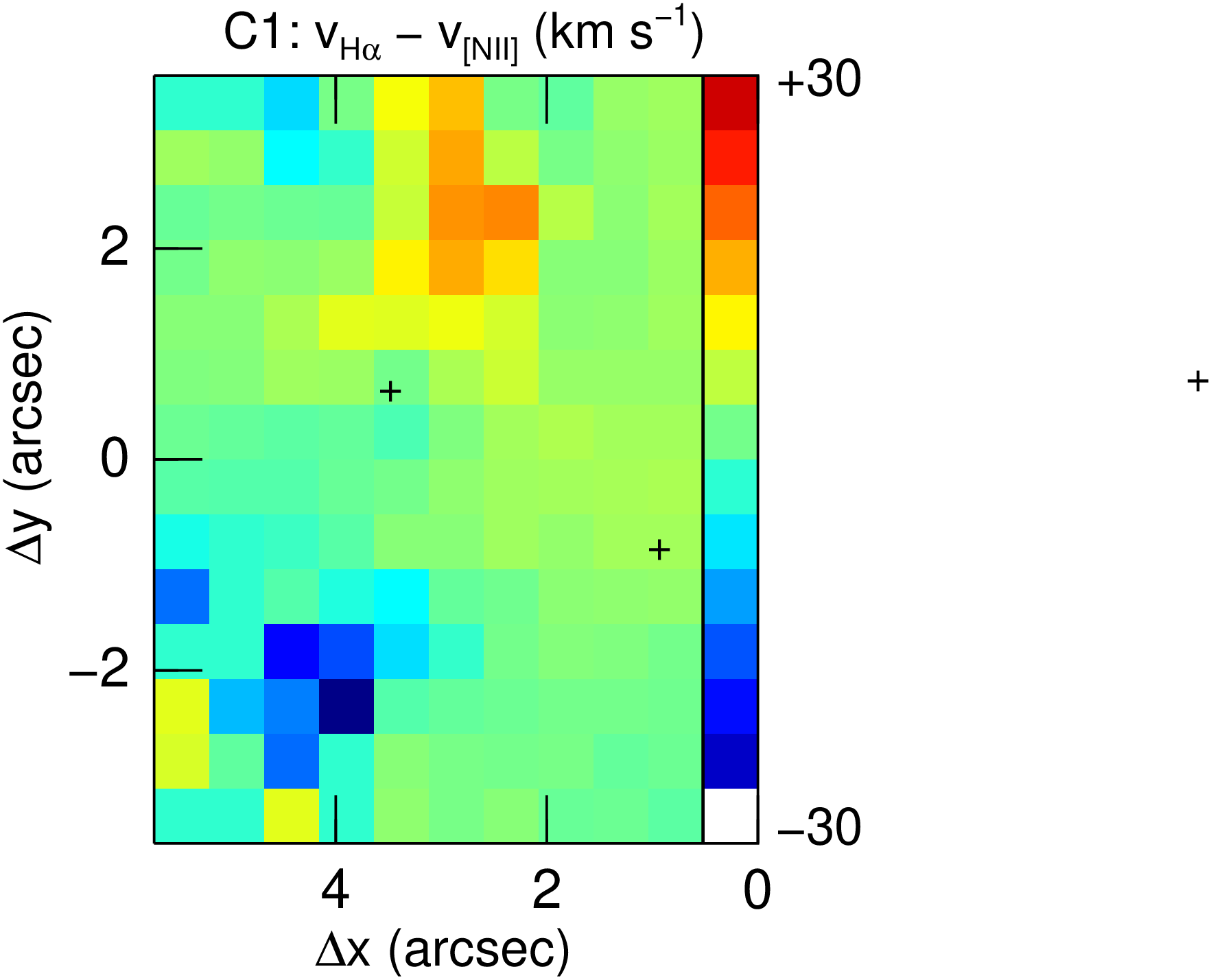}
\includegraphics[angle=0,width=0.30\textwidth, clip=,bbllx=-10, bblly=10,
bburx=595, bbury=715]{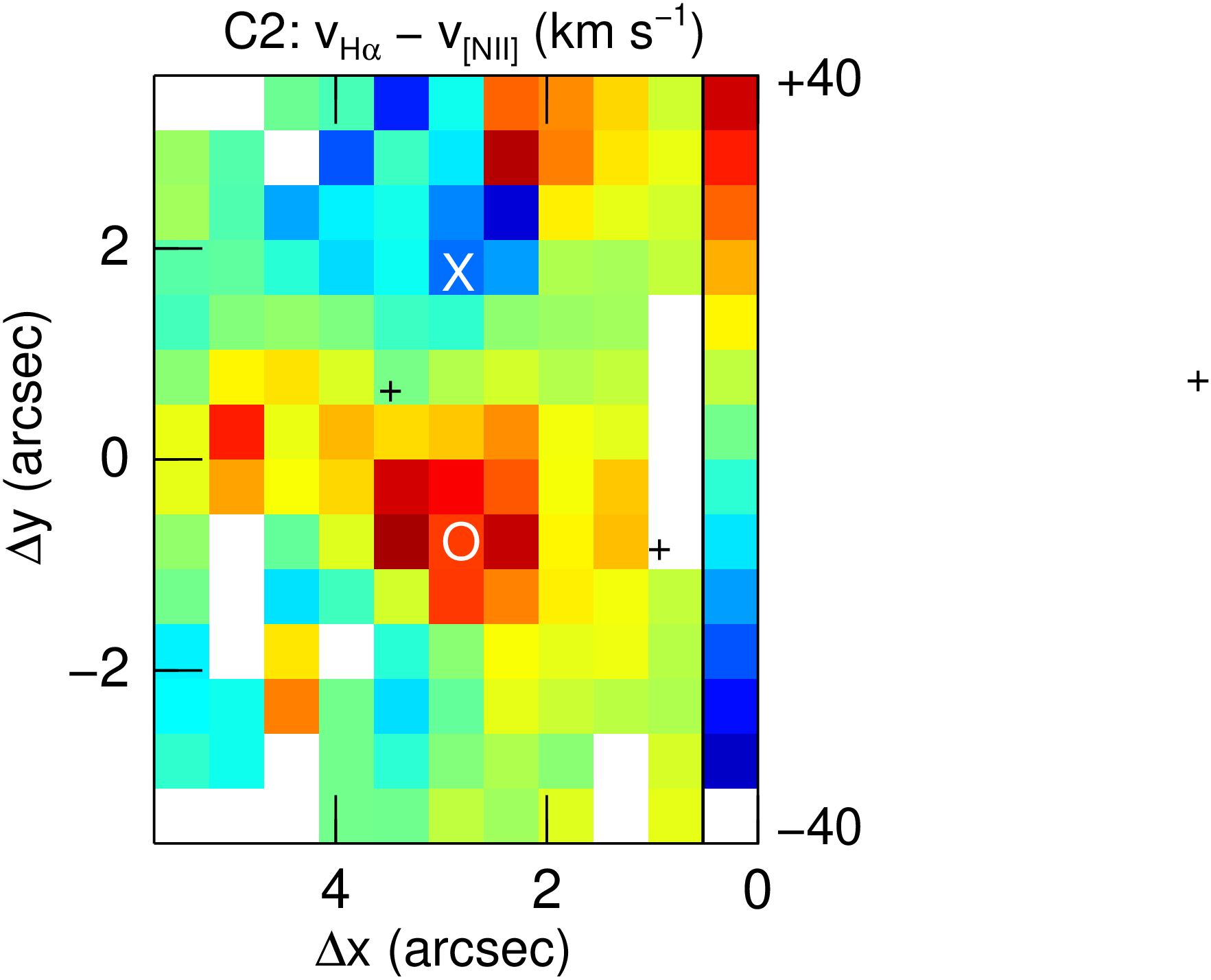}
\includegraphics[angle=0,width=0.30\textwidth, clip=,bbllx=-10, bblly=10,
bburx=595, bbury=715]{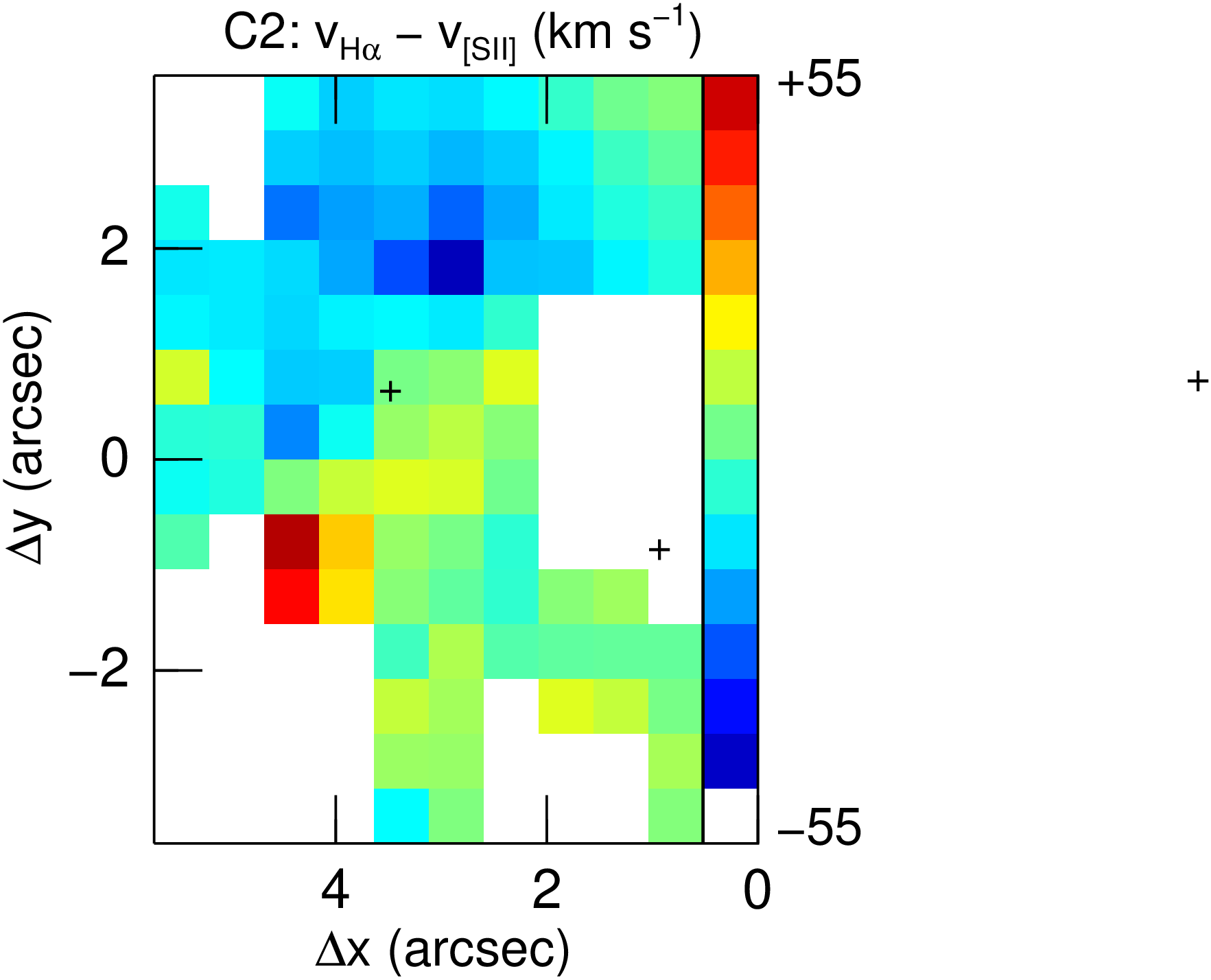}
    \caption[]{Difference between the velocity fields derived from the
     \ha\ and the \nii\ emission lines for the narrow (left) and broad
     (center) components as well as between those derived from the
     \ha\ and \sii\ emission lines for the broad component
     (right). Only the section of the FLAMES field of 
     view associated with the giant \ion{H}{ii} region is
     shown. The ``X'' and the ``O'' symbols in the central 
       panel show the position of the spaxels (17,4) and (17,9),
       repectively (see line fitting in Figure
       \ref{ex_offset}). \label{difvel}} 
 \end{figure*}

\begin{figure*}[htb!]
 \centering
\includegraphics[angle=0,width=0.31\textwidth, clip=,bbllx=60, bblly=61,
bburx=338, bbury=248]{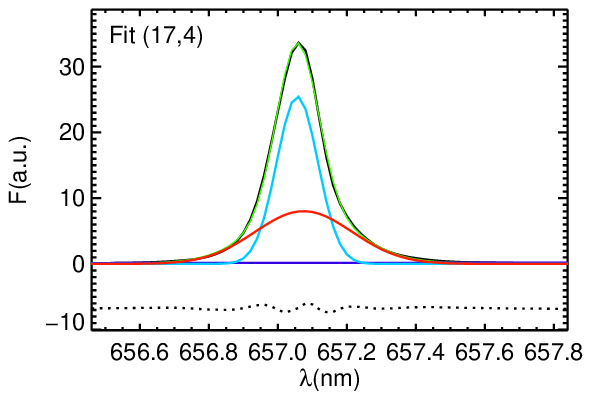}
\includegraphics[angle=0,width=0.31\textwidth, clip=,bbllx=60, bblly=61,
bburx=338, bbury=248]{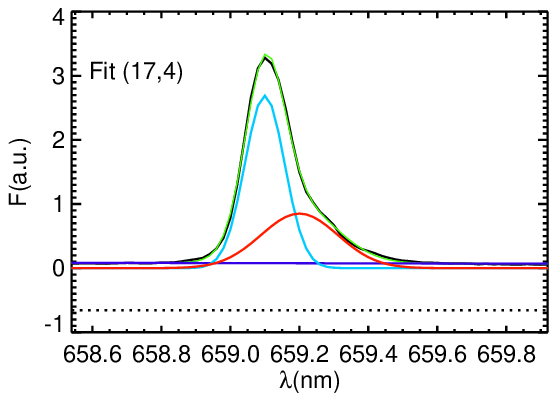}
\includegraphics[angle=0,width=0.31\textwidth, clip=,bbllx=60, bblly=61,
bburx=338, bbury=248]{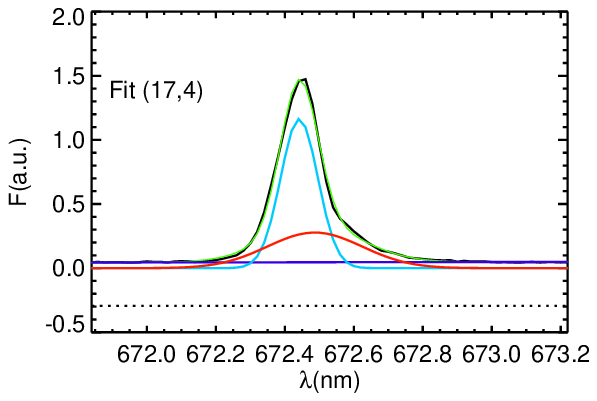}
\includegraphics[angle=0,width=0.31\textwidth, clip=,bbllx=60, bblly=61,
bburx=338, bbury=248]{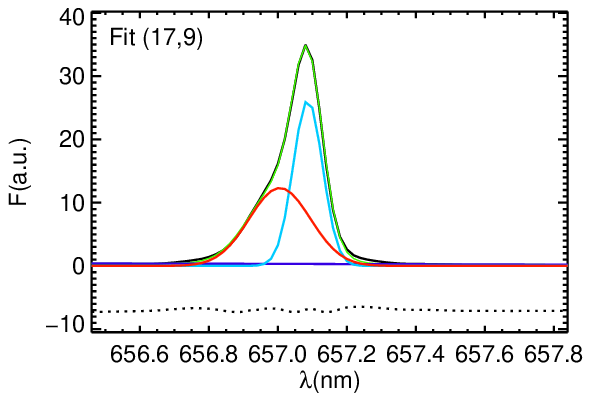}
\includegraphics[angle=0,width=0.31\textwidth, clip=,bbllx=60, bblly=61,
bburx=338, bbury=248]{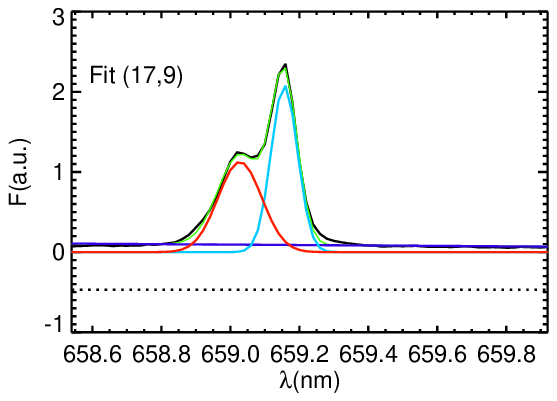}
\includegraphics[angle=0,width=0.31\textwidth, clip=,bbllx=60, bblly=61,
bburx=338, bbury=248]{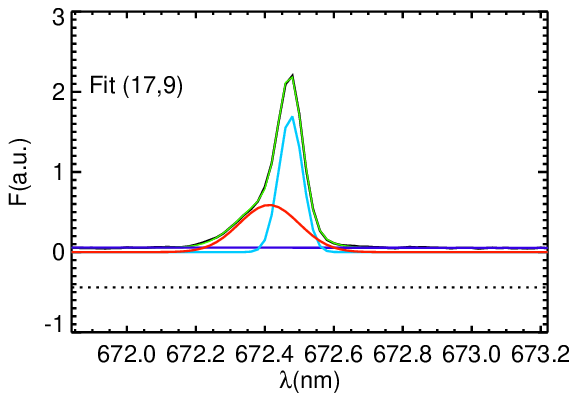}
  \caption[]{Fits in \ha\ (left), \nii\ (middle), and,
    \textsc{[S\,ii]}$\lambda$6717 (right)  for two representative
    spaxels showing offsets in velocity for the different emission
    lines.  Flux is in arbitrary units. The observed 
      spectra are presented in black. The first and second 
      components have been displayed in blue and red, respectively,
      while the underlying continuum and 
      total fit are in violet and green. Residuals are shown as a
      dotted line below the profile.
\label{ex_offset}}
 \end{figure*}

In Figure \ref{cinematica}, we present the velocity fields 
for the three fitted components derived from the \ha\
emission line.
We also included the velocity dispersion map for our
broadest component. Corrected velocity dispersions for the two narrow components
were, in general, subsonic and will not be shown here.
The only exception would be an area at
$\sim[4\farcs0,-2\farcs0]$.  The line profiles in this area show
how the narrow component presents a continuity with the two narrow
components at $\sim[4\farcs0,-1\farcs0]$ as if it were the result of a
strong blending of these two components. However, we were not able to
properly deblend these two components by means of our line fitting technique. 

From Figure \ref{cinematica}, it is clear that the movements of
the ionized gas are far from simple rotation.
For the discussion we will separate the emitting area into the
  zone corresponding to the giant \ion{H}{ii} region and the rest. 
The zone of the giant \ion{H}{ii} region, occupying roughly the left 
part of the FLAMES field of view, shows in the upper
part, line profiles that can be explained by two components while in some
spaxels of the lower part a third component was required.
The area of the giant \ion{H}{ii} region itself, which occupies an area of
  $\sim$120~pc$\times$60~pc, requires up to three
  components to properly reproduce the line profiles. They were named
  C1, C2 and C3, according to their relative fluxes.
The first component (i.e. C1) accounts for the $\sim45-68$\% of the flux in
\ha, depending on the considered spaxel. It is relatively narrow and
constant, with a $\Delta v\sim$10~km~s$^{-1}$ over a distance of
$\sim$6$^{\prime\prime}$ ($\sim$110~pc) with slightly 
bluer velocities in the spaxels associated with the edge of the upper
and lower extensions. 

The second component (i.e. C2) accounts for the $\sim27-55$\% of the flux in
\ha. It is symmetric with respect of an axis that goes
through the Complex \#1 in the north-south direction. In
comparison with the first component, it presents large velocity
variations (i. e. $\Delta v\sim70$~km~s$^{-1}$ over
$\sim4\farcs7$ or $\sim86$~pc) and is relatively broad
($\sigma\sim20-25$~km~s$^{-1}$).  
Low surface brightness broad components have been reported in
  starburst galaxies using a slit since more than a decade and have
  been the subject of several theoretical \citep[e.g.][]{ten97} and
  observational \citep[e.g.][]{cas90,gon94} studies. They usually
  represent a small 
  fraction ($\sim$3-20\%) of the total \ha\ flux and have widths of
  $\sigma\sim$700~km~s$^{-1}$. Recently, 2D spectroscopic analysis of
  very nearby starbursts have shown how \emph{locally}, the line width is
  somewhat smaller \citep[$\sigma\sim$50-170~km$^{-1}$, see][and
    references therein]{wes09b}. This is understood in the context of
  the so-called \emph{Turbulent Mixing Layers}
  \citep[e.g.][]{sla93}. However, the high surface brightness of C2,
   together with the symmetry in the velocity field and its low widths
   made us to explore an alternative 
  explanation for it (see section \ref{reghii}).

The third component (i.e. C3) is present in a small area of
  the field of about 1\farcs0 diameter (i.e. 18~pc) and centered at
  about [4\farcs0,-1\farcs0]. C3 has a width similar to the first
  component (C1) but displaced $\sim$50~km~s$^{-1}$ towards the
blue. This third component appears in a location
about 1\farcs5 south-east to the position of the SSC complexes in the 
tongue-shaped extension described in section \ref{morfologia}. This 
area presents low values of extinction (see Figure \ref{mapa_ebv}) and
surface brightness (see Figure \ref{estructura}). 
This third component shows indications of more extended weaker emission
which was not fitted.

We pointed out before that velocity maps derived from the
  \nii\ and \sii\ emission lines showed areas with important differences with
  respect to the one obtained from \ha.
This is the case for the giant \ion{H}{ii} region. 
Figure \ref{difvel} contains the measured velocity differences for the
two brighter components (C1 and C2) in the subset of the field 
corresponding to the giant \ion{H}{ii} region while Figure
  \ref{ex_offset} presents examples of the independent fits for \ha,
  \nii, and \textsc{[S\,ii]}$\lambda$6717. 
Differences exist for both, the narrow (i.e. C1) and broad (i.e. C2), components
in \nii\ and are relatively symmetric with
respect to the peak of continuum emission Complex \#1, but with
opposite sign and slightly different directions (P.A.$\sim50^\circ$ and
$\sim30^\circ$ for C1 and C2 respectively). Also, the range of
velocity differences, ($v_{\rm H\alpha} - v_{\rm [NII]}$) is larger for
C2 than for C1 ($\sim70$~km~s$^{-1}$ and $\sim30$~km~s$^{-1}$,
respectively). As illustrated in the right hand map of Figure
\ref{difvel}, the broad component for the \sii\ emission lines also
shows velocity differences  with similar range ($v_{\rm H\alpha} - v_{\rm [SII]}\sim
60$~km~s$^{-1}$),~orientation and sign as in the case of the
\nii\ emission~line. For the C1 component, no significant differences 
were found.    
Measured differences in C3, with a mean and standard deviation of 
$-6\pm9$~km~s$^{-1}$ and $5\pm4$~km~s$^{-1}$ for
  $v_{\rm H\alpha} - v_{\rm [NII]}$ and $v_{\rm H\alpha} - v_{\rm
    [SII]}$  respectively, do not appear to be significant. However, since
  C3 was only detected in seven spaxels (see Figure \ref{cinematica}, bottom
  left panel) this result has to be treated
  with caution.

Similar offsets has been detected in galactic \ion{H}{ii}
  regions like Orion \citep{gar08b}, but to our knowledge, this is the
  first time that maps with  
  such offsets in velocity for different emission
lines in starbursts are presented. This can partially be caused by the
fact that 2D-kinematic analysis of starbursts, from dwarfs
\citep[e.g.][]{gar08} to more extreme events like LIRGs
\citep[e.g.][]{alo09}, are usually based on fitting 
techniques that impose restrictions between the \ha\ and \nii\ central
wavelengths and thus, preventing from
detecting such offsets. An example of work where the main emission
lines are fitted independently is presented by \citet{wes07}. However,
since they only analyzed the kinematic for \ha, it is not possible to
assess if they found different kinematics for the other emission lines.
There are however, some works that offer
examples of offsets of this kind using a slit. In particular
\citep{lop07} report also an offset between the \nii\ and the
\ha\ emission line of $\sim$10~km~s$^{-1}$, similar to what we have
measured for C1 in the upper part of our f.o.v. 

The second region of interest is located in the right (south-west) part of the
FLAMES field of view. Emission there shows narrow lines
with velocity dispersion dominated by the thermal width. 
In some areas (the north-east corner in Figure
  \ref{cinematica}), two narrow lines were needed to better reproduce
  the line profile. The primary component (i.e. C1) shows a symmetric
  velocity pattern with respect to the twin clusters associated with the
  peak of emission \#3. A velocity gradient in the 
  north-west to south-east direction is clear with a  $\Delta v \sim
40$~km~$^{-1}$ over about 4\farcs0 (linear scale of $\sim$75~pc). The
secondary component traces a shell blue-shifted 
$\sim$40~km~s$^{-1}$ in the western corner (see Figure
  \ref{cinematica}, upper right panel). Note that C2, although
  relatively narrow, is a bit broader than the thermal width (Figure
  \ref{cinematica}, bottom right panel).  
This component accounts for $\sim30-50$\% of the \ha\
flux in this area. No significant differences in the
velocity fields and the velocity dispersion maps for the main emission
lines have been found in this area.




\section{Discussion \label{discusion}}

\subsection{The giant \ion{H}{ii} region \label{reghii}}

The most interesting area of NGC~5253 in the present data covers
the left (north-east) part of the FLAMES field of view. In previous
sections we have seen that this area is occupied by a giant
\ion{H}{ii} region which: i) harbors two very massive and young SSCs at
its centre
(i.e. Complex \#1); ii) presents high levels of extinction, being
larger in the upper part of the f.o.v.; iii) has high electron
densities as traced by both the sulphur and the argon line ratios, and
again are also larger in the upper part of the f.o.v.; iv) presents an
excess in the \nha\ line ratio with respect to \sha\ which, if
interpreted as N-enrichment, indicates an outward gradient of extra nitrogen
from a point at $\sim0\farcs5-1\farcs0$ towards the north-west of
the peak of continuum emission at Complex \#1; v)
presents W-R features, implying a young age for the harboured stellar
population; vi) displays complex kinematics, not coincident in all
the emission lines, that require a minimum of three components to
reproduce the line profiles. How does all this 
evidence fit together into a coherent picture?  

\begin{figure}[htb!]
 \centering
\includegraphics[angle=270,width=0.43\textwidth, clip=,bbllx=110, bblly=70,
bburx=515, bbury=690]{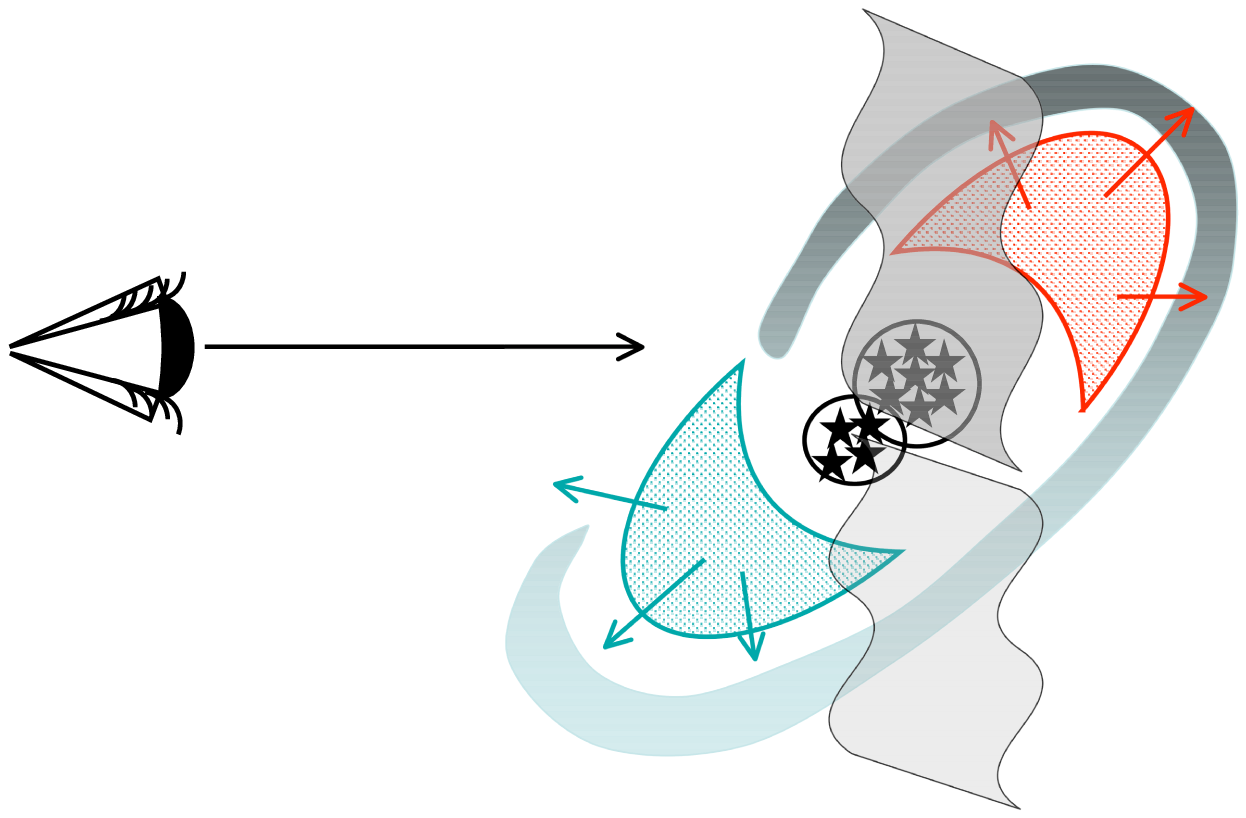}
   \caption[]{Sketch showing the different elements associated with
     the area of the giant \ion{H}{ii} region. The two
       massive SSCs (stars in the circles) expel material (red and
       blue half moons) which encounter quiescent gas (grey
       shell). Regions of dust extinction are represented by the two wavy
       sheets. \label{esquema_reghii}} 
 \end{figure}

In Figure \ref{esquema_reghii}, we sketch a plausible scenario
compatible with all these results. Here, the broad component (C2) would
trace an outflow created by the two SSCs at Complex \#1 while
the two narrow components (C1 and C3) would be caused by a shell of previously
existing quiescent gas that has been reached by the ionization
front. Different grades of grey in Figure \ref{esquema_reghii} in 
the shell represent the 
different densities observed in the upper and lower part of the
FLAMES f.o.v., while two wavy sheets in two grades of grey have been used to
represent the differences in extinction between these two halves. Due
to this extinction distribution, in the upper (i.e. north-western)
half of the \ion{H}{ii}
region, the observer cannot see the further part of the shell, while in
the lower (i.e. south-eastern) half both parts are visible and are
detected as a single broader component when approaching the vertex of the oval.

As in section \ref{secdensidad}, we formed
\textsc{[S\,ii]}$\lambda$6717/\textsc{[S\,ii]}$\lambda$6731 maps for
the individual kinematic components and measured the mean and
standard deviation for the sulphur line ratio in two 4 $\times$ 4 spaxel
squares sampling the upper and lower part of the giant \ion{H}{ii}
region. Although the standard deviations are large ($\sim0.09$),
results support this sketch. While the broad component
presented similar line ratios in both areas ($\sim1.07$ implying
densities of $\sim$470~cm$^{-3}$), the narrow component presented
somewhat lower line ratios in the upper part than in the lower one
($\sim$1.11 vs. $\sim$1.25) implying densities for the
shell of $\sim$390~cm$^{-3}$ and  $\sim$180~cm$^{-3}$, respectively.
Also, we created (noisier) \nha, \sha, and \ohb\ line maps and
  compared the relation between \sha\ and \nha\ for the individual
  components. The three fitted components present extra
  nitrogen in an area coinciding with the one derived
  from one-gaussian fitting. This differs from the findings for
  Mrk~996, a galaxy with several kinematically distinct components
  where only the broad one presented N-enrichment, with an abundance
  $\sim$20 times larger than the one for the narrow component
  \citep[see][]{jam09}. Still, in \object{NGC~5253}, the N-enrichment in the
  broad component is larger than in the narrow one by a factor of
  $\sim$1.7 which is consistent with the scenario sketched above.     
Moreover, while the \ohb\ maps are relatively
  similar for both components\footnote{C3 is not considered here,
    since the so-called \emph{map} would be associated to $\le7$
    spaxels, depending on the line ratio.} (not shown), those associated to
  \nha\ and specially to (the more shock 
  sensitive) \sha\ line ratio display a relatively different ionization
  degree with larger values for C1 than for C2. This is also
  consistent with the presented scenario since a larger contribution
  due to shocks is expected in the area where the outflowing material
  encounters the pre-existent gas. 

An interesting result of the previous section was the offsets derived
for the velocities of the different species, in particular nitrogen
and sulphur. To our knowledge, this is the first time that maps showing
this kind of offsets are reported in an starburst galaxy.
Similar phenomena have already been reported in much closer
regions of star formation. For example, observations in the Galactic Orion
Nebula, a much less extreme event in terms of star formation, show how
\ha\ and \oiii\ display similar velocities while \nii\ and \sii\ are
shifted by $\sim$4-5~km~s$^{-1}$ \citep{gar08b}, an order of magnitude
smaller than the shifts found for NGC~5253. Also, self-consistent
dynamic models of 
steady ionization fronts point towards the detection of such
differences \citep{hen05}. In the context of the scenario sketched in
Figure \ref{esquema_reghii}, the offsets in C2 would fit if \nii\ (and
\sii) traced the outer parts in an outflow which has a Hubble flow
(i.e. velocity proportional to radius).

Finally, an estimate of the time scales
  associated with the pollution process can be determined by using the
  velocity for the 
  outflow derived in section \ref{seccinematica}. Assuming that this
  traces the velocity of nitrogen contamination of the ISM, the
  detected pollution extending up to distances of $\sim$60~pc took
  place over only $\sim$1.3-1.7~Myr. 
This supports the idea that the nitrogen dilution is a relatively
fast process and is 
consistent with the shortage of observed systems presenting this kind
of chemical inhomogeneity.

\subsection{The area associated with the older stellar clusters}


\begin{table}
     \centering
     \caption[]{Integrated line ratios for the rightmost (south-west) part of the
       FLAMES f.o.v. \label{cocientes}}
             \begin{tabular}{ccccccccc}
            \hline
            \noalign{\smallskip}
Component & $\log$(\textsc{[O\,iii]}/\hb) & 
            $\log$(\textsc{[N\,ii]}/\ha) & 
            $\log$(\textsc{[S\,ii]}/\ha) \\
           \noalign{\smallskip}
           \hline
           \noalign{\smallskip}
Upper Blue & 0.69 & $-$1.20 & $-$0.85\\	
Upper Red  & 0.52 & $-$1.09 & $-$0.67\\
Lower      & 0.54 & $-$1.09 & $-$0.75\\
NGC~1952$^{\mathrm{(a)}}$   & 0.92 & 0.15 & 0.18\\
            \noalign{\smallskip}  
            \hline
         \end{tabular}

\begin{list}{}{}
\item[$^{\mathrm{(a)}}$] From \citet{ost06}.
\end{list}
\end{table}


The rightmost (south-west) part of the FLAMES f.o.v. presents a
different picture. We have seen that this region: i) is associated 
with two relatively old ($\sim$70 and $\sim$110~Myr) and massive (3 and
7$\times$10$^4$~M$_\odot$) clusters \citep{har04}; ii) presents
moderate levels of extinction, being higher in the lower part of the
FLAMES field of view; iii) has very low $N_e$ and lower than
100~cm$^{-3}$ in the upper corner; iv) the \ha\ surface brightness is
very low (i.e. one and  two orders of magnitude smaller than in the
giant \ion{H}{ii} region for the upper and lower portions respectively); v)
displays two distinct kinematic components in the upper part of the
f.o.v.. Noteworthy is that two supernova remnant candidates have been
detected in the area \citep{lab06}. One of them (S001) appears very
close in projection to the massive clusters at $\sim$[-5\farcs0,1\farcs0]. The
second one (S002) is located, just outside of the FLAMES f.o.v., at the
right upper corner.

\begin{figure}[htb!]
 \centering
\includegraphics[angle=270,width=0.43\textwidth, clip=,bbllx=110, bblly=70,
bburx=515, bbury=690]{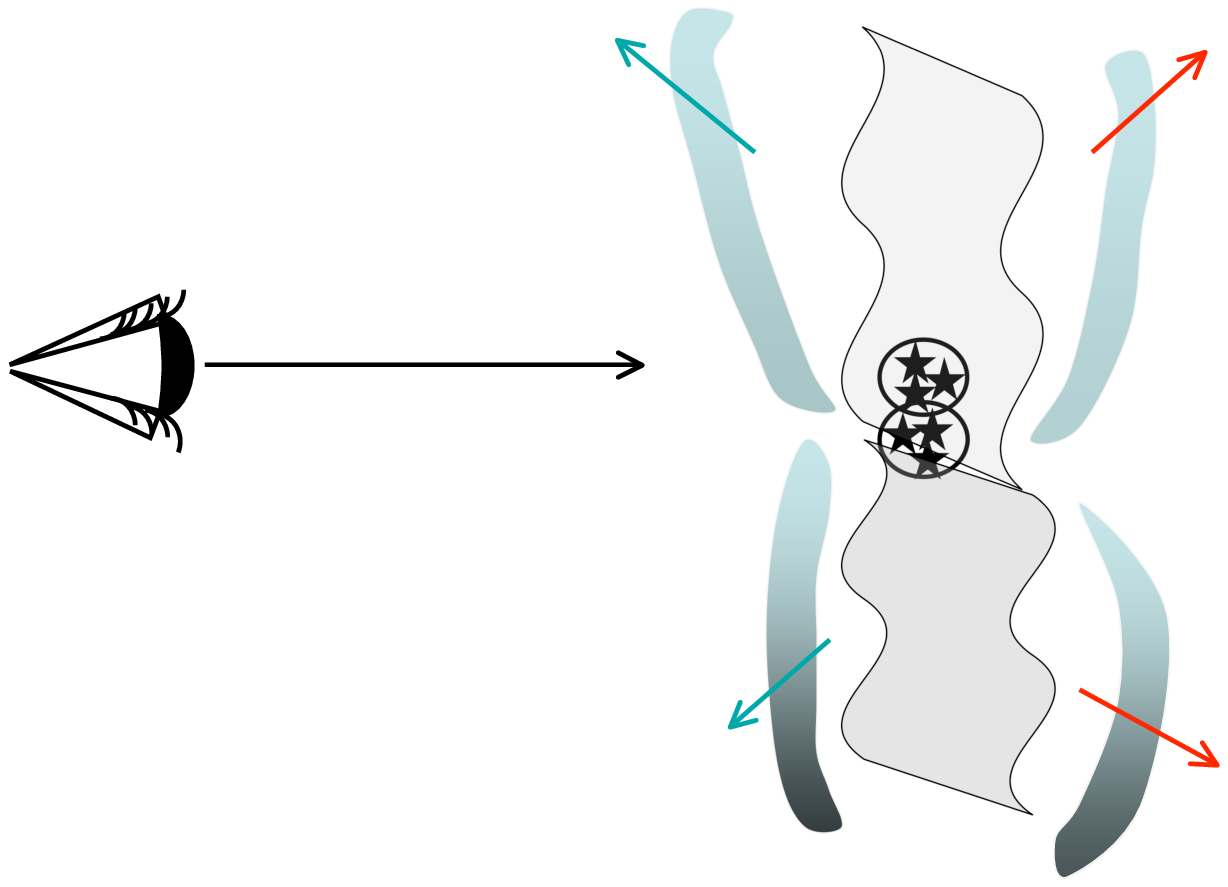}
   \caption[]{Sketch showing the different elements associated with
     the right part of the FLAMES field of view. The two
       massive SSCs (stars in the circles) have already expelled their
       material and are seen as relatively quiescent ionized shells (shown
       in grey). Differences in extinction are represented by two wavy
       sheets. \label{esquema_partedcha}} 
 \end{figure}

The first question to consider is if any of the kinematic
components is related 
to the supernova remnant candidates. However, our measured velocity
dispersions are much more lower than expansion velocities of typical
supernova remnants \citep[i.e. NGC~1952,
  1450~km~s$^{-1}$][]{ost06}. Moreover, we created two integrated 
spectra for the upper and the lower part of the area. Line ratio for 
both components of the upper part and for the lower part were
relatively similar (i.e. within $\sim$0.1~dex, see Table \ref{cocientes}) and
much lower than those expected for a  supernova remnant
\citep[e.g. NGC~1952,][]{ost06}. Thus, supernova remnants are not
obviously the cause of the observed kinematics and physical properties of
the ionized gas in this region.

Instead, given that all the three components present similar line
ratios - consistent with ionization caused by stars - and similar line
widths, a more plausible scenario would be that where all the components
are part of a common picture. Given the age of the clusters, and the
velocity differences between the three components, this area can be viewed
as a snapshot of a more evolved version of what is
happening in the left part of the FLAMES f.o.v. The clusters
have managed to clear out their environment. Only a broken shell made
out of previously quiescent gas remains ionized by the remaining hot
stars and moving away from the
clusters with little evidence for high velocity outflow.  
Figure \ref{esquema_partedcha} presents a sketch of the
different elements associated with this area.  

\section{Summary \label{summary}}

We present a thorough study of the ionized gas and its relation
with the stellar population of NGC~5253 by mapping the central
212~pc$\times$134~pc in a continuous and unbiased manner using with
the ARGUS IFU unit of FLAMES. The analysis of the data have yield the
following results. 

\begin{enumerate}

\item We obtained a 2D detailed map for the extinction suffered by the
  ionized gas, finding an offset of $\sim0\farcs5$ between the
  peak of the optical continuum and the extinction peak, in agreement
  with findings in the infrared.

\item We compared the extinction suffered by gas and stars by defining
  \emph{ad hoc} broad-band colours. We have shown the
  importance of using line-free  
   filters when performing this comparison and found that
   stars suffer less extinction than the ionized gas by a factor
  $\sim0.33$, similar to the findings in other starburst galaxies.

\item We derived $N_e$ sensitive line ratio maps. The one
  involving the sulphur lines shows a gradient from 790~cm$^{-3}$ at the
  peak of emission in the giant \ion{H}{ii} region described
    by \citet{cal97}
  outwards. The argon line ratio is only detected in the  area
  associated with the giant \ion{H}{ii} region and traces the highest
  density ($\sim4200 - 6200$~cm$^{-3}$) regions.

\item We studied the ionization structure by means of the maps of line ratios
  involved in the BPT diagrams. The spatial distribution of the \sha\ and
  \ohb\ line ratios follows that for the flux distribution of the
  ionized gas. On the contrary, the \nha\ map shows a completely
  different structure.

\item We evaluated the possible ionization mechanisms through
  the position of these line ratios in the diagnostic diagrams and
  comparing with the predictions of models. All our line ratios are compatible
  with photoionization caused by stars. The \sha\ indicated a somewhat
  higher ionization degree that might be evidence of some contribution of
  shocks to the measured line ratios. Part of the data in the
  diagram involving the \nha\ line ratio are distributed in a distinct
  cloud. This can be explained 
  within the local N-enrichment scenario proposed for this galaxy.

\item We delimited very precisely the area presenting  local
  N-enrichment. It occupies the whole giant \ion{H}{ii}, including the
  two extensions towards the upper and lower part of the FLAMES
  field of view, peaking at $\sim$1\farcs5 from the peak of emission
  in the continuum and almost coincident (i.e. at $\sim20$~pc) with the
  peak of extinction. 

\item We located the areas that could contain Wolf-Rayet stars by looking
  for the \emph{blue bump}. We confirmed the existence of W-R stars
  associated with the nucleus and the brightest cluster in the 
  ultraviolet. W-R stars are distributed in a wider area than the one
  presenting N-enrichment and in a more irregular manner. We were able
  to identified one (or more) clusters with ages compatible with the
  existence of W-R stars in all but one (i.e. W-R 5) of our
  delineated regions with a W-R signature. 

\item If the scenario of \emph{N-enrichment caused by W-R stars}
  turns out to be applicable, only the W-R detected at the core (Complex 
  \#1), and perhaps in the two extensions of the Giant \ion{H}{ii}
  region, can be considered the cause of the local N-enrichment,
  according to the correlation of the spatial 
  distribution of W-R features and N-enrichment.

\item We measured the He$^+$ and He$^{++}$ abundances. He$^+$/H$^+$ is
  $\sim$0.08-0.09 in most of our field of view except for an area of
  $\sim$2$^{\prime\prime}\times3^{\prime\prime}$ in the upper right
  corner, far away from the main ionization source. We detected
  the nebular \ion{He}{ii}$\lambda$4686 emission line in areas not
  coincident, in general, with those presenting W-R features, nor
  with the one presenting N-enrichment. Abundances in \ion{He}{ii}
  were always $\lsim$0.0005. Given the depth and unbiased mapping of
  the present data, we can exclude the possibility
  of further detections of larger quantities of He$^{++}$ based on
  optical observations in the nuclear region of
  NGC~5253. This result is difficult to reconcile with the scenario of
  \emph{N-enrichment caused by W-R stars} and favours a suggestion
   where the N-enrichment arises during the late O-star
  wind phase. 

\item We studied the kinematics of the ionized gas by using velocity
  fields and velocity dispersion maps for the main emission lines. We
  needed up to three components to properly reproduce the line
  profiles. In particular, one of the components associated with the
  Giant \ion{H}{ii} region presents supersonic widths and \nii\ and
  \sii\ emission lines shifted  up to $40$~km~s$^{-1}$ with respect to
  \ha. Also, one of the narrow components shows velocity offsets in the
  \nii\ line of up to 
  $20$~km~s$^{-1}$. This is the first time that maps providing 
  such offsets for a starburst galaxy have been presented. 

\item We provide a scenario for the event occurring at the Giant
  \ion{H}{ii} region. The two SSCs are producing an outflow that
  encounters previously existing quiescent gas. The scenario is
  consistent with the measured extinction structure, electron
  densities and kinematics. 

\item We explain the different elements in the right (south-west) part of the
  FLAMES field of view as a more evolved stage of a similar scenario
  where the clusters have now cleared their local environment. This is
  supported by the low electron densities and \ha\ surface
  brightness as well as the kinematics in this area.

\end{enumerate}

\begin{acknowledgements}

We thank Peter Weilbacher for his help in the initial stages of this project.
We also thank the anonymous referee for his/her careful and detailed
review of the manuscript.
Based on observations carried out at  the European Southern
Observatory, Paranal (Chile), programme 078.B-0043(A). This paper uses
the plotting package \texttt{jmaplot}, developed by Jes\'us
Ma\'{\i}z-Apell\'aniz,
\texttt{http://dae45.iaa.csic.es:8080/$\sim$jmaiz/software}. This 
research made use of the NASA/IPAC Extragalactic 
Database (NED), which is operated by the Jet Propulsion Laboratory, California
Institute of Technology, under contract with the National Aeronautics and Space
Administration.
AMI is supported by the Spanish Ministry of
Science and Innovation (MICINN) under the program "Specialization in
International Organisations", ref. ES2006-0003. This work has been
partially funded by the Spanish PNAYA, projects AYA2007-67965-C01 and
C02 from the Spanish PNAYA and  CSD2006 - 00070  "1st Science with
GTC"  from the CONSOLIDER 2010 programme of the Spanish MICINN. 

\end{acknowledgements}


\bibliography{14154mybib}{}
\bibliographystyle{./aa}

\end{document}